\documentclass[aps,prl,twocolumn,reprint,amsmath,amssymb,superscriptaddress,floatfix,footinbib,longbibliography]{revtex4-2}

\usepackage{graphicx}
\usepackage{epstopdf}

\usepackage[T1]{fontenc}
\usepackage[applemac]{inputenc}
\usepackage{lmodern}
\usepackage{amsmath}
\usepackage{amssymb}
\usepackage[english]{babel}
\usepackage{natbib}
\usepackage{ae}
\usepackage{units}
\usepackage{ulem}
\usepackage{amsmath,amssymb,natbib,bm}
\usepackage{psfrag}
\usepackage{subfigure}
\usepackage{amsthm}
\usepackage{color}
\usepackage{booktabs}
\usepackage{slashed}

\usepackage[americaninductors]{circuitikz}
\usepackage{tikz}
\usetikzlibrary{arrows}
\usepackage{ulem}
\usepackage{color}
\usepackage{url}

\usepackage[colorlinks]{hyperref}
\hypersetup{%
        plainpages=true,
        breaklinks=true,
        hypertexnames=false,
        pageanchor=true,
        colorlinks=true,
        linkcolor={blue},
        citecolor={magenta},
        urlcolor={blue},
        anchorcolor={black}
      }

\usepackage{mleftright} 

\newcommand{\figref}[1]{\mbox{Fig.~\ref{#1}}}

\renewcommand{\eqref}[1]{\mbox{Eq.~(\ref{#1})}}
\newcommand{\figpanel}[2]{Fig.~\hyperref[#1]{\ref*{#1}(#2)}} 

\newcommand{\figpanels}[3]{Fig.~\hyperref[#1]{\ref*{#1}(#2)--(#3)}} 
\newcommand{\figpanelNoPrefix}[2]{\hyperref[#1]{\ref*{#1}(#2)}} 
\newcommand{\figpanelsNoPrefix}[3]{\hyperref[#1]{\ref*{#1}(#2)--(#3)}} 

\newcommand{\be}{\begin{equation}}
\newcommand{\ee}{\end{equation}}
\newcommand{\bea}{\begin{eqnarray}}
\newcommand{\eea}{\end{eqnarray}}

\AtBeginDocument{%
    \newwrite\bibnotes
    \def\bibnotesext{Notes.bib}
    \immediate\openout\bibnotes=\jobname\bibnotesext
    \immediate\write\bibnotes{@CONTROL{REVTEX42Control}}
    \immediate\write\bibnotes{@CONTROL{%
    apsrev42Control,author="08",editor="1",pages="0",title="0",year="1"}}
     \if@filesw
     \immediate\write\@auxout{\string\citation{apsrev42Control}}%
    \fi
}%

\begin{document}

\title{Realizing on-demand all-to-all selective interactions between distant spin ensembles} 

\author{C.-X.~Run}
\thanks{These authors contributed equally}
\affiliation{Department of Physics, City University of Hong Kong, Kowloon, Hong Kong SAR, China}

\author{K.-T.~Lin}
\thanks{These authors contributed equally}
\affiliation{Trapped-Ion Quantum Computing Laboratory, Hon Hai Research Institute, Taipei 11492, Taiwan}

\author{K.-M.~Hsieh}
\thanks{These authors contributed equally}
\affiliation{Department of Physics, City University of Hong Kong, Kowloon, Hong Kong SAR, China} 

\author{B.-Y.~Wu}
\affiliation{Department of Physics, City University of Hong Kong, Kowloon, Hong Kong SAR, China}

\author{W.-M.~Zhou}
\affiliation{Department of Physics, City University of Hong Kong, Kowloon, Hong Kong SAR, China}

\author{G.-D.~Lin}
\affiliation{Trapped-Ion Quantum Computing Laboratory, Hon Hai Research Institute, Taipei 11492, Taiwan}
\affiliation{Department of Physics and Center for Quantum Science and Engineering, National Taiwan University, Taipei 10617, Taiwan}

\author{A. F.~Kockum}
\affiliation{Department of Microtechnology and Nanoscience, Chalmers University of Technology, 412 96 Gothenburg, Sweden}

\author{I.-C.~Hoi}
\email[e-mail:]{iochoi@cityu.edu.hk}
\affiliation{Department of Physics, City University of Hong Kong, Kowloon, Hong Kong SAR, China}


\date{\today}

\begin{abstract}

Achieving all-to-all coherent networks is critical for the advancement of large-scale coherent computing and communication protocols. By exploiting the resonant dipole-dipole interaction between distant spin ensembles coupled to a one-dimensional coplanar waveguide (CPW) terminated by a mirror, we successfully demonstrate an on-demand all-to-all selective coherent network between four spin ensembles. Furthermore, by repositioning the spin ensembles along the CPW, we achieve collective coupling, and demonstrate coherent energy exchange between multiple spin ensembles in the time domain. These results strongly indicate the potential of this device as a medium-scale all-to-all network structure, which is poised to advance the exploration of many-body physics and enhance coherent information processing capabilities.

\end{abstract}

\maketitle


\textit{Introduction.}---Studies of spin ensembles have highlighted their potential applications in coherent information processing across both classical and quantum regimes~\cite{Lukin2003, Atac2009}, primarily due to their favorable interactions with electromagnetic fields. Recently, cavity magnonics~\cite{ZARERAMESHTI20221} has emerged as a pivotal area for exploring coherent interaction between cavity modes and magnon modes in spin ensembles~\cite{Tabuchi2014, Zhang2014, Goryachev_2014, Zhang2015, Bourhill2016, Kostylev2016, Goryachev_2018, Flower_2019}, facilitating applications such as information transduction~\cite{Xufengscienceadvance2016, Hisatomi2016}, memory~\cite{Zhang2015NC, ShenPRL2021}, nonreciprocity~\cite{Osada2016, Wangprl2019}, bistability~\cite{Yi-Pu2018, Nair_2021, Shen2022}, exceptional points~\cite{ZhangNC2017, ZhangPRB2019}, and entanglement~\cite{Li_PRL_2018, Zhangzhedong_2019}. Additionally, interaction between magnons and superconducting artificial atoms mediated via cavities has attracted significant attention~\cite{TabuchiScience2015, Danysciadv2017, DanyScience2020, XuPRL2023}. Beyond the interaction with single modes within a cavity, the interplay between spin ensembles and a continuum of modes in waveguides has resulted in numerous intriguing physical phenomena~\cite{Rao2020, Wang2022, Qian2023, Raoprl2023, Wang_NP_2024, wu2024arxiv}, including indirect magnon-cavity mode interactions~\cite{Rao2020}, giant spin ensemble physics~\cite{Wang2022}, and microwave interference due to the coupling between a spin ensemble and its mirror image~\cite{wu2024arxiv}.

One area in the development of scalable coherent information processing and communication protocols that presents a significant challenge is achieving all-to-all coherent networks through remote coupling. Various schemes to facilitate long-distance coupling have been investigated, including central bus resonators~\cite{songchaosci2019, stassi2020scalable, bhattacharjee2025}, adjustable couplers~\cite{Arute2019}, waveguide mirrors~\cite{Mir2019nat, Wen2019, Lin2019}, hybrid magnonic circuits~\cite{li2022coherent}, giant atoms and molecules~\cite{kockprl2018, Kannan2020, Wang2022, Almanakly2025}, light-mediated interactions~\cite{karg2019, Thomssci2020, zhong2021deterministic, Liprxq2021, HOLLER2022, cheung2024photon}, microwave quantum state routers~\cite{Zhou2023}, gain-driven cavities~\cite{RAO2023PRL}, and critical phenomena~\cite{YANG2024PRL}. However, these strategies often encounter considerable obstacles such as fabrication difficulties~\cite{songchaosci2019, stassi2020scalable, bhattacharjee2025, Arute2019, Mir2019nat, Wen2019, Lin2019}, stability concerns~\cite{Thomssci2020}, loss issues~\cite{Liprxq2021, HOLLER2022}, and bandwidth limitations~\cite{Zhou2023, RAO2023PRL} that hinder their effective implementation. Notably, research on employing spin ensembles to establish all-to-all coherent networks with long-range interactions remains limited. Addressing this gap is one possible avenue for developing robust and versatile coherent networks capable of supporting advanced coherent information processing applications.

In this Letter, we demonstrate how selective all-to-all interaction between spin ensembles can be realized. We utilize yttrium iron garnet (YIG, $\rm Y_{3}Fe_{5}O_{12}$) spheres as spin ensembles due to their high spin density of $\unit[2.1 \times 10^{22} \mu_B]{cm^{-3}}$ ($\mu_B$: Bohr magneton)~\cite{Gilleo1958, Huebl2013} and low damping rate~\cite{Zhang2014, Tabuchi2014}. Our experiments are conducted at room temperature. Upon application of an external magnetic field, the collective excitation mode of the spin ensemble, commonly referred to as the magnon mode, can interact with propagating photons in a coplanar waveguide (CPW), resulting in a resonant dip observable through spectroscopic measurement. This magnon mode can be conceptualized as a harmonic oscillator characterized by sufficiently large excitation at room temperature. When the spin ensemble is uniformly magnetized, the frequency of the magnon mode is directly proportional to the magnitude of the applied magnetic field~\cite{Zhang2014, Tabuchi2014}. In this study, we focus on a specific excitation, the Kittel mode (KM), also known as the uniform ferromagnetic resonance (FMR) mode, which behaves akin to a large magnetic dipole moment~\cite{Zhang2014, Tabuchi2014}, facilitating strong magnon-photon interactions. This characteristic provides an ideal platform for exploring resonant dipole-dipole interaction (RDDI) between spin ensembles, mediated by traveling microwave photons.

\begin{figure}
\includegraphics[width=\linewidth]{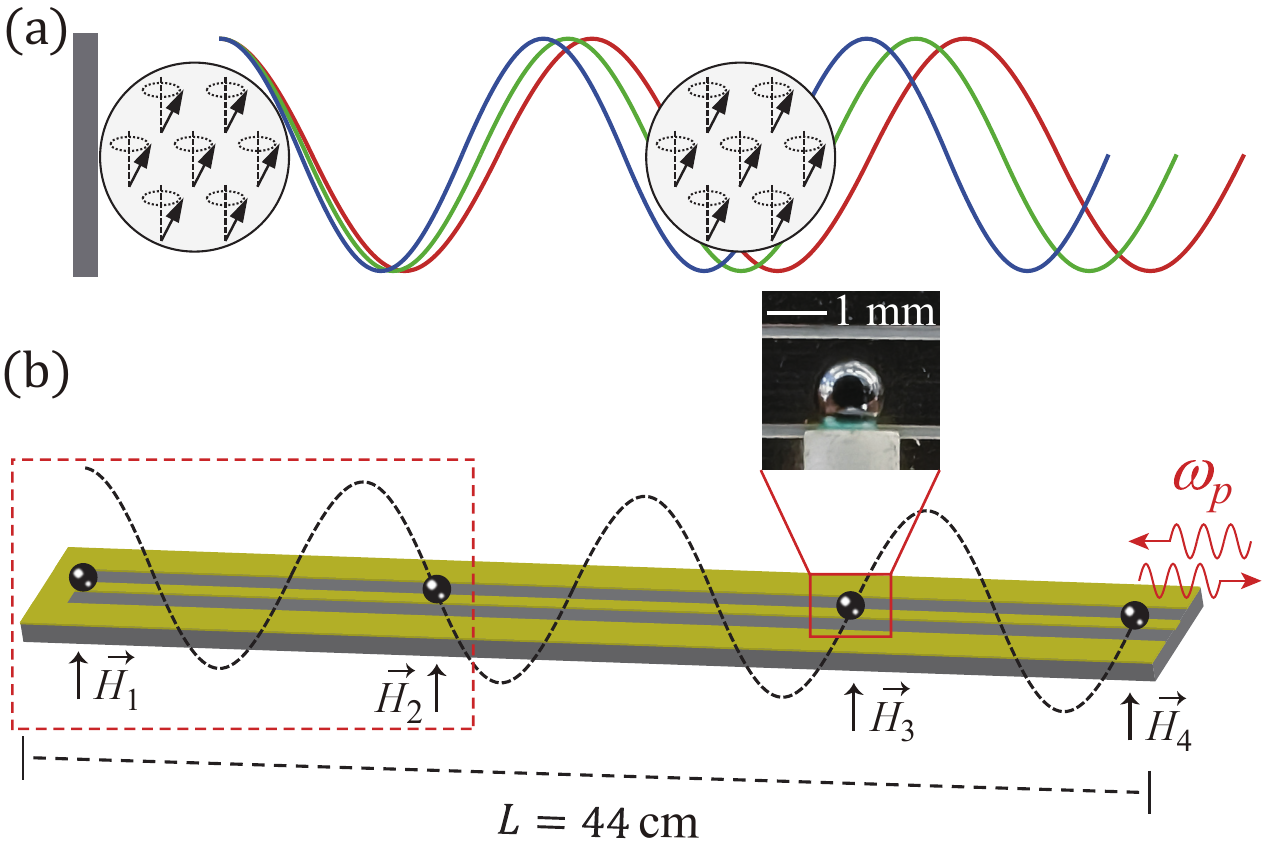}
\caption{Sketch of the system. 
(a) Two spin ensembles positioned in front of a mirror interact via a continuum of microwaves (curves in different colors) in one dimension.
(b) The experimental configuration consists of a $\unit[44]{cm}$ long CPW, with its left end shorted to ground, acting as a magnetic antinode mirror. Four YIG spheres (black spheres; an enlarged photograph of one YIG sphere is shown as an inset) are strategically placed along the CPW, with their Kittel mode (KM) transition frequencies independently tunable via corresponding local magnetic fields ($\vec{H_i}, i = 1, 2, 3, 4$), which are set by an applied current $I$. The black dashed sinusoidal curve indicates that at a specific transition frequency, all YIG spheres are positioned at the magnetic nodes of the microwave, except for the one at the mirror, which consistently resides at the magnetic antinode. The dashed red box shows the part of the system sketched in more detail in panel (a). The downscaling and scalability of this device are discussed in Sec.~S9 of the Supplementary Material~\cite{SupMat}.}
\label{fig:setup}
\end{figure}

In this study, RDDI is constructed between distant spin ensembles via a continuum of microwaves confined in a one-dimensional CPW shorted to a mirror [see~\figref{fig:setup}]. Thereby, this setup~\cite{Astafiev2010, You2011, Hoi2011, Hoi2012, Hoi2013, GU20171, Kockum_2019, Wen_PRL_2018, Wen2019} eliminates the need for direct contact or additional physical mediators, and enables interactions over significant physical distances. By leveraging this interaction, we successfully demonstrate an all-to-all coherent network with selective interactions through strategic design of the spatial one-dimensional arrangement and the tunability of the resonance frequencies of the YIG spheres. In contrast to other approaches~\cite{songchaosci2019, bhattacharjee2025, Arute2019, Zhou2023}, our device operates at room temperature, mitigates fabrication challenges and crosstalk issues, and presents additional advantages of flexibility and straightforward extensibility for large-scale networks. Furthermore, by repositioning the YIG spheres on the CPW, we realize collective coupling and demonstrate coherent energy exchange following resonant excitation. Additionally, we validate the relationship between the strength of the RDDI and the number of YIG spheres located at the nodes.


\textit{System and model.}---Our experimental setup, shown in~\figpanel{fig:setup}{b}, comprises four YIG spheres, each with a diameter of $\unit[1.2]{mm}$, positioned on top of a $\unit[44]{cm}$ long CPW, with the left end shorted to ground, effectively acting as a magnetic antinode mirror. The KM transition frequency $\omega_m$ of each YIG sphere is independently tunable by the local magnetic field ($\vec{H_i},  i = 1, 2, 3, 4$) perpendicular to the CPW and the YIG's crystal axis $\langle 110 \rangle$, controlled by the applied current $I$ and a permanent local magnet. Figure~S1 in the Supplementary Material~\cite{SupMat} illustrates the measurement apparatus for both frequency- and time-domain experiments.

In the presence of the mirror, when a YIG sphere is positioned at the node of the standing wave formed by incident and reflected waves at its resonance frequency, it is considered completely decoupled from the field~\cite{Hoi2015}. However, RDDI between two YIG spheres can in this situation still be induced by the continuum of propagating microwave modes. Theoretically, when two YIG spheres resonate at $\omega_r$, their interaction strength is given by [Eq.~(S11)~\cite{Wen2019, Lin2019, SupMat}]
\begin{equation}
\Delta_{ij} = \frac{\kappa_{ij}}{2} \mleft[ \sin \mleft( \frac{\omega_r}{v} \mleft( x_i + x_j \mright) \mright) + \sin \mleft( \frac{\omega_r}{v} \mleft| x_i - x_j \mright| \mright) \mright],
\label{Eq:strength}
\end{equation}
where $x_i$ denotes the distance of YIG sphere $i$ from the mirror located at $x = 0$, $\kappa_{ij} \equiv \sqrt{\kappa_i (\omega_r) \kappa_j (\omega_r)}$ with $\kappa_i (\omega_r)$ the bare radiative damping rate of YIG sphere $i$ into an open waveguide at the KM frequency $\omega_m = \omega_r$, and $v$ is the velocity of microwaves propagating in the CPW. According to~\eqref{Eq:strength}, when YIG sphere $i$ is at an antinode ($\omega_r x_i / v = n_i \pi$, $n_i = 0, 1, 2, \ldots$) and YIG sphere $j$ is at a node ($\omega_r x_j / v = n_j \pi + \frac{\pi}{2}$, $n_j = 0, 1, 2, \ldots$, $n_j \geq n_i$), the interaction reaches its maximum, given by $\Delta_{ij} = \kappa_{ij}$. This interaction leads an avoided level crossing between the two YIG spheres, and the reflection spectrum is anticipated to exhibit two resonance modes with a frequency splitting of $2 \Delta$. Compared to the scenario in which the YIG spheres are placed on an open CPW, the presence of the mirror in our setup enhances the interaction and circumvents the challenge posed by the linewidths of the two resonance modes obscuring the interaction~\cite{Wen2019, Lin2019}. In addition, this mirror technique enables on-demand magnon-magnon interactions and complete suppression of both coherent [\eqref{Eq:strength}] and dissipative interactions [Eq.~(S10)] (see Sec.~S2~\cite{SupMat}). The further detailed advantages of this mirror-terminated waveguide system are summarized in Table~S4~\cite{SupMat}.


\begin{figure*}
\includegraphics[width=\linewidth]{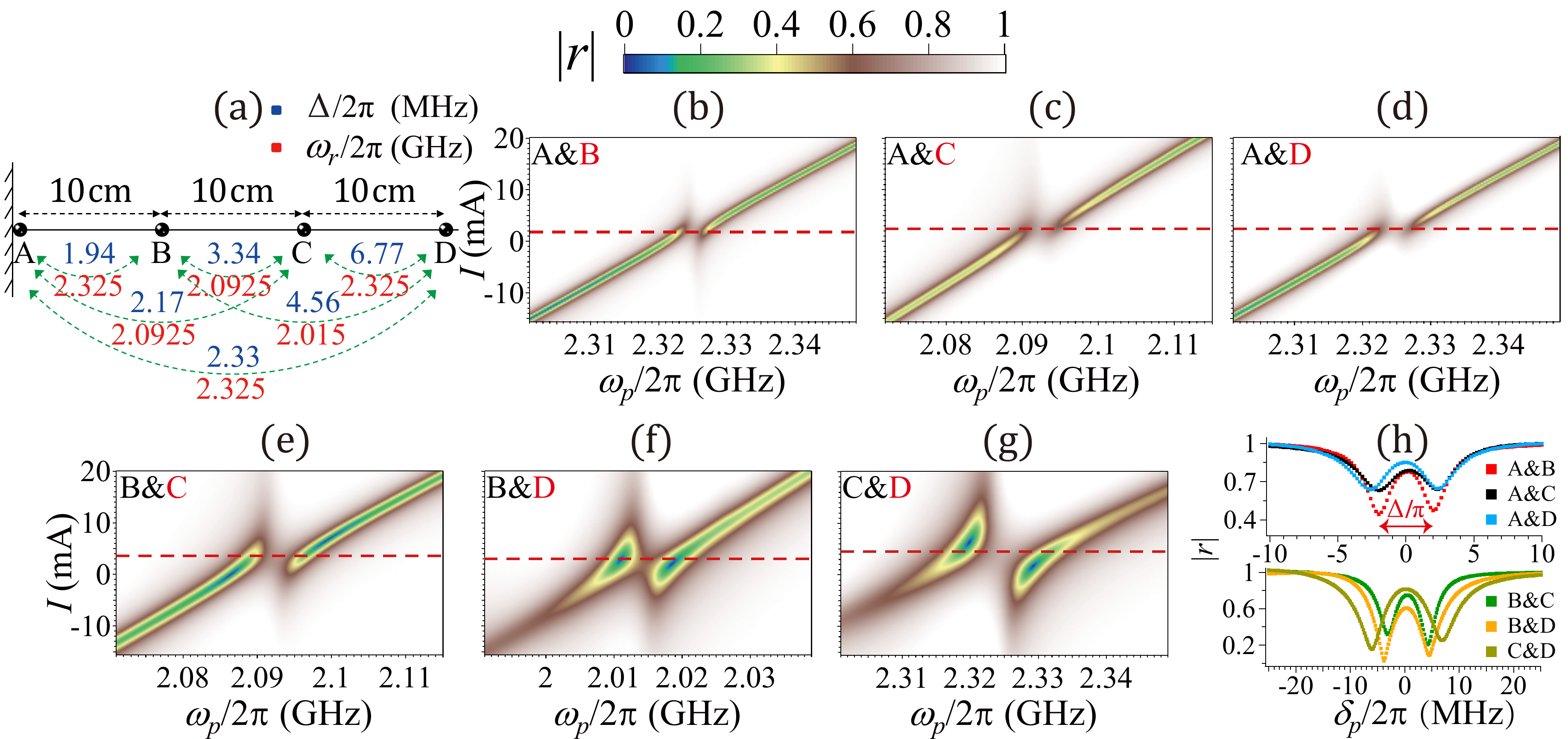}
\caption{All-to-all selective interaction between distant YIG-sphere pairs.
(a) Schematic diagram of the setup. Four YIG spheres are placed equidistantly in front of a mirror along the CPW, with a spacing of $\unit[10]{cm}$ between them, labeled sequentially as A, B, C, and D. Green dashed curves with arrows below the YIG spheres indicate the interacting pairs. The blue text denotes the coherent interaction $\Delta / 2\pi$ between corresponding YIG spheres, while the red text represents the designed resonance frequency $\omega_r / 2\pi$. 
(b)--(g) Amplitude reflection coefficient $|r|$ for a weak probe as a function of the probe frequency $\omega_p$ and the current $I$ controlling the KM frequency of one YIG sphere in a pair. The six panels correspond to the pairs A\&B, A\&C, A\&D, B\&C, B\&D, and C\&D, respectively. In each panel, the KM frequency of the YIG sphere at the node is fixed (highlighted in red font), while the KM frequency of the other YIG sphere is tuned through resonance with this frequency. Each panel shows a clear avoided level crossing, demonstrating interactions between the selected pairs of YIG spheres. 
(h) Line cuts of the data [dashed lines in panels (b)--(g)] at the point where the corresponding selected YIG pair is on resonance. Here, $\delta_p = \omega_p - \omega_r$ is the detuning between the probe and resonance frequencies. The simulation results corresponding to panels (b)--(h) are shown in Fig.~S2~\cite{SupMat}.}
\label{fig:fig2}
\end{figure*}

\textit{Results and discussion.}---To establish an all-to-all selective network, we position four YIG spheres on the CPW at equal intervals of \unit[10]{cm}, designated sequentially as A, B, C, and D, as shown in~\figpanel{fig:fig2}{a}. We begin by characterizing each YIG sphere through reflection spectroscopy. From the results, we derive the KM frequency $\omega_m$, the radiative damping rate $\kappa_r$, the overall damping rate $\Gamma$, the non-radiative decay rate $\alpha$, and the propagation velocity of microwaves within the CPW ($v \approx 1.86 \times \unit[10^8]{m/s}$). All extracted and calculated parameters are detailed in Table~S1~\cite{SupMat}.

\begin{figure*}
\centering
\includegraphics[width=\linewidth]{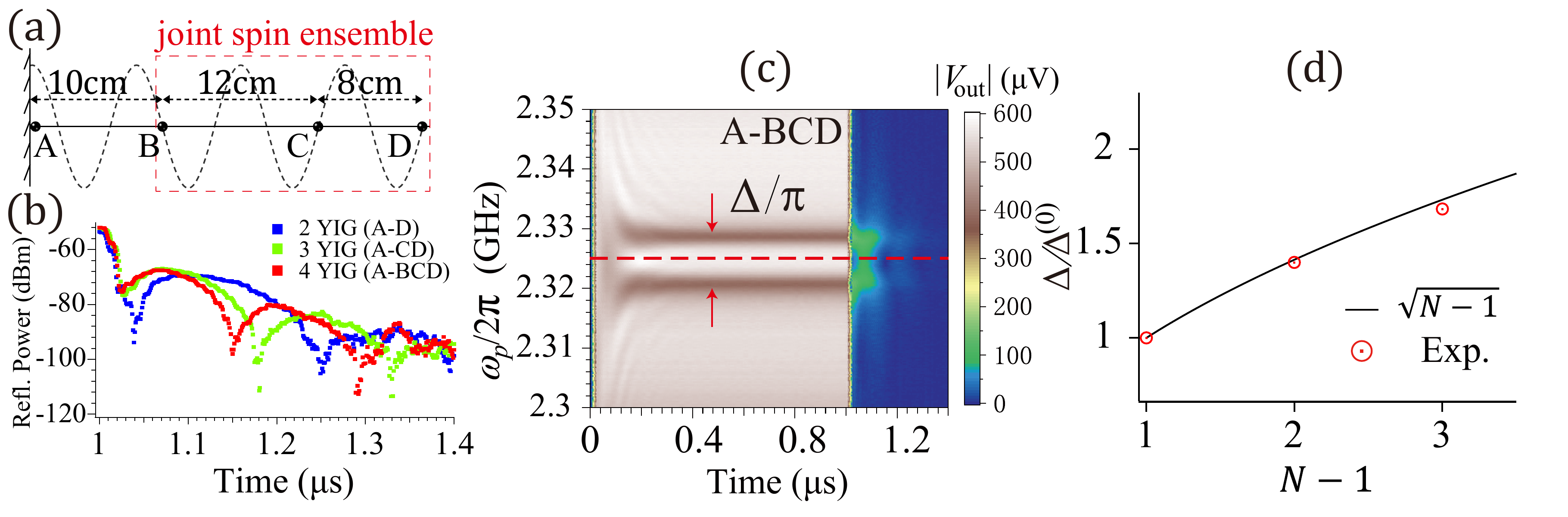}
\caption{Time-resolved energy oscillations between YIG sphere A and joint spin ensemble.
(a) Sketch of the setup. Four YIG spheres are positioned in front of a mirror along the CPW, with distances from the mirror set at $x_{\rm A} = 0$, $x_{\rm B} = \unit[10]{cm}$, $x_{\rm C} = \unit[22]{cm}$, and $x_{\rm D} = \unit[30]{cm}$. The black dashed sinusoidal curve indicates that when $\omega_p / 2\pi = \unit[2.325]{GHz}$, YIG A is at an antinode, while B, C, and D are at nodes.
(b) Time evolution of the system for configurations with two (blue), three (green), and four (red) YIG spheres, each tuned on resonance at $\omega_r / 2\pi = \unit[2.325]{GHz}$, following the termination of a long ($\sim \unit[1]{\mu s}$) resonant rectangular microwave pulse. The time traces correspond to line cuts of more extensive data [dashed lines in panel (c) and Fig.~S5(a)--(b)~\cite{SupMat}]. To enhance visibility of the oscillations, $\mleft| V_{\rm out} \mright|$ is converted to reflected power in dBm. We observe that the oscillation period decreases with an increasing number of YIG spheres. The corresponding simulation results are shown in Fig.~S5(f)~\cite{SupMat}.
(c) Time evolution of the reflected signal $\mleft| V_{\rm out} \mright|$ for the configuration involving four YIG spheres. The plot consists of two distinct regions: from 0 to $\unit[1]{\mu s}$, during which the pulse is applied, and the remaining duration when the pulse is off. Shortly after the pulse is activated and deactivated, energy oscillations can be observed. When the pulse is applied for a long time, the system reaches a steady state, revealing mode splitting, consistent with the frequency-domain measurement in Fig.~S3(b)~\cite{SupMat}. The corresponding time-domain simulation results are shown in Fig.~S5(e)~\cite{SupMat}.
(d) Based on the oscillation periods in panel (b), $\Delta$ as a function of the number of YIG spheres located at the nodes ($N-1$). Here, $\Delta^{(0)}$ is the coherent interaction for the case of two YIG spheres (A-D). The agreement between the experimental data and the prediction ($\sqrt{N-1}$) is good.}
\label{fig:fig3}
\end{figure*}

We then focus on the YIG pair A\&B, while ensuring that the other YIG spheres are significantly detuned. The YIG sphere A is consistently positioned at the antinode of the microwave field within the CPW, as it is located at the reflecting mirror. Tuning $\omega_m$ of YIG B enables continuous adjustment of $\omega_m x_{\rm B} / v$, where $x_{\rm B} = \unit[10]{cm}$. With the KM frequency of YIG B fixed at \unit[2.325]{GHz}, such that $\omega_m x_{\rm B} / v = 2.5 \pi$, placing it at a node of the resonant mode, we bias the KM of YIG A close to this value and tune it across this region. The resulting reflection spectrum is presented in~\figpanel{fig:fig2}{b}, where the resonance observed along the diagonal corresponds to YIG A. The resonance of YIG B is hidden from detection due to its position at the node. However, the presence of an anticrossing indicates that when the two YIG spheres are on resonance, they exhibit coherent coupling characterized by the interaction strength $\Delta$, which must arise from the microwave continuum in the CPW, as the two YIG spheres are spatially separated by a large distance. Through the line cut on resonance [red dashed line in~\figpanel{fig:fig2}{b}], shown in the red curve in~\figpanel{fig:fig2}{h}, the coupling strength is found to be $\Delta / 2\pi \approx \unit[1.94]{MHz}$.

We conduct the same experiment for the five other possible pairs among the four YIG spheres: A\&C, A\&D, B\&C, B\&D, and C\&D. The corresponding results are shown in~\figpanels{fig:fig2}{c}{g}, with their resonance line cuts summarized in~\figpanel{fig:fig2}{h}. From these results, we extract the corresponding resonance frequencies and coherent interactions given in~\figpanel{fig:fig2}{a}. Notably, in the configurations A\&B, A\&C, A\&D, and C\&D, the YIG sphere positioned closer to the mirror is consistently at an antinode, while the other is at a node. In contrast, in the cases B\&C and B\&D, although the YIG sphere further from the mirror remains at a node, YIG B is not at an antinode; the values of $\omega_m x_{\rm B} / v$ are $2.25 \pi$ and $2.166 \pi$, respectively, for the corresponding resonance frequencies. 

We make several observations here. (1) When the YIG sphere distant from the mirror is at a node and the YIG sphere close to the mirror is not at an antinode, coherent interaction can still occur. (2) Different pairs of YIG spheres exhibit different coupling strengths, which correlate with their bare radiative damping rates according to~\eqref{Eq:strength} (see Table~S2~\cite{SupMat}). The strongest coupling strength we measured was $\Delta / 2\pi \approx \unit[6.77]{MHz}$ and the weakest was $\Delta / 2\pi \approx \unit[1.94]{MHz}$, with an average strength of $\Delta / 2\pi \approx \unit[3.52]{MHz}$. (3) In this setup, by fixing the selected pair to a strategically designed frequency and ensuring that all other YIG spheres are far detuned, we have successfully established an all-to-all coherent network with selective interactions between distant YIG pairs. A more detailed demonstration is shown in Fig.~S4~\cite{SupMat}.

We proceed to couple multiple YIG spheres together to investigate the dependence of coupling strength on the number of YIG spheres and to demonstrate coherent energy exchange between them. In our previous experiment, at the resonance frequency of \unit[2.325]{GHz}, YIG spheres B and D were positioned at nodes, while YIG A was consistently located at a mirror antinode. To facilitate our analysis, we relocate YIG C \unit[2]{cm} further away from the mirror than its original position, ensuring that when $\omega_r / 2\pi = \unit[2.325]{GHz}$, YIG spheres B, C, and D all reside at nodes, as depicted in~\figpanel{fig:fig3}{a}. Consequently, we can treat these three spheres together as a `joint spin ensemble' [red dashed box in~\figpanel{fig:fig3}{a}] to explore the coupling dynamics between them and YIG A. 

Initially, in the case of two YIG spheres, we focus on the YIG pair A\&D, which are the most distantly spaced. We fix both YIG spheres at \unit[2.325]{GHz} and then apply a long ($\sim \unit[1]{\mu s}$) microwave square pulse while sweeping the carrier frequency around this value. The reflected signal $\mleft| V_{\rm out} \mright|$, which indicates the stored field in YIG A, is recorded using a digitizer (see the setup in Fig.~S1~\cite{SupMat}). After the excitation resonant pulse ends, we observe the coherent energy exchange between the two YIG spheres, as illustrated by the blue trace in~\figpanel{fig:fig3}{b}. Due to the inherent lifetime limitation of the YIG spheres, only one oscillation is observed with a period of $T^{(0)} \approx \unit[220]{ns}$, resulting in a calculated coupling strength of $\Delta^{(0)} / 2\pi = 1 / 2 T^{(0)} \approx \unit[2.27]{MHz}$, consistent with the frequency-domain measurement in~\figpanel{fig:fig2}{d}.

For three YIG spheres, we select the configuration comprising YIGs A, C, and D. During resonance, we observe two oscillations with a period of $T^{(1)} \approx \unit[157]{ns}$, as shown by the green trace in~\figpanel{fig:fig3}{b}, leading to a coupling strength of $\Delta^{(1)} / 2\pi = 1 / 2 T^{(1)} \approx \unit[3.18]{MHz}$.

For the case of four YIG spheres, a 2D plot of $\mleft| V_{\rm out} \mright|$ as a function of $\omega_p$ and evolution time is shown in~\figpanel{fig:fig3}{c}. This plot reveals oscillatory energy transfer between YIG A and the joint spin ensemble consisting of YIG spheres B, C, and D for a brief duration following the initiation and cessation of the pulse. Additionally, we observe that when the pulse is applied for a long time (much longer than the lifetime of the magnon, which is on the order of tens of nanoseconds~\cite{Zhang2014, wu2024arxiv}), the system evolves into a steady state characterized by two split modes (indicated by red arrows in the brown region). The red trace in~\figpanel{fig:fig3}{b} indicates two post-pulse oscillations, with $T^{(2)} \approx \unit[131]{ns}$ and $\Delta^{(2)} / 2\pi = 1 / 2 T^{(2)} \approx \unit[3.82]{MHz}$. 

Further experimental details regarding time-domain measurements of two and three YIG spheres, as well as frequency-domain measurements involving three and four YIG spheres, along with the corresponding simulation results, are provided in Figs.~S3, S5, and Table~S3~\cite{SupMat}. Finally, we calculate the normalized coupling strengths ($\Delta^{(i)} / \Delta^{(0)}, i = 0, 1, 2$), as shown in~\figpanel{fig:fig3}{d}. Our results indicate a proportional relationship $\Delta^{(i)} / \Delta^{(0)} \propto \sqrt{N-1}$, where $N$ ($N-1$) is the number of YIG spheres (at nodes). This finding substantiates the scaling law presented in Ref.~\cite{Lin2019}.


\textit{Summary and outlook.}---We have demonstrated resonant dipole-dipole interaction between distant spin ensembles coupled to a one-dimensional CPW with a mirror. Our experimental findings indicate that when a spin ensemble is positioned at the node of the standing wave created by incident and reflected waves at its resonance frequency, it can significantly interact with another resonant spin ensemble, despite being effectively decoupled from the CPW. Leveraging this effect, we constructed an all-to-all coherent network comprising four detachable and flexibly controllable spin ensembles, achieved through strategic design of their interensemble spacing and corresponding resonance frequencies. We demonstrated coherent energy exchange through time-domain measurements, highlighting the substantial potential of magnons for information exchange within the network. We further demonstrate the ability to rapidly ($\sim$$\unit[200]{ns}$) decouple two YIG spheres on demand (Fig.~S6)~\cite{SupMat}. This device serves as a prototype demonstration of a modular structure for an all-to-all connected network, which holds great promise for advancing our understanding of many-body physics and enhancing coherent information processing capabilities. Furthermore, as the underlying principle relies entirely on linear interactions, this circuit architecture can be readily extended to the quantum regime at millikelvin temperatures~\cite{Wen2019}.


\begin{acknowledgments}

\textit{Acknowledgements.}---I.-C.H.~acknowledges financial support from City University of Hong Kong through the start-up project 9610569, from the Research Grants Council of Hong Kong (Grant No.~11312322) and from Guangdong Provincial Quantum Science Strategic Initiative (GDZX2303005, GDZX2203001, and GDZX2403001). A.F.K.~acknowledges support from the Swedish Foundation for Strategic Research (grant numbers FFL21-0279 and FUS21-0063), the Horizon Europe programme HORIZON-CL4-2022-QUANTUM-01-SGA via the project 101113946 OpenSuperQPlus100, and from the Knut and Alice Wallenberg Foundation through the Wallenberg Centre for Quantum Technology (WACQT). G.-D.L.~acknowledges support from Grant No.~NSTC-113-2112-M-002-025 and No.~NSTC-112-2112-M-002-001.

\end{acknowledgments}


\normalem
\bibliography{References}

\begin{thebibliography}{16}%
\makeatletter
\providecommand \@ifxundefined [1]{%
 \@ifx{#1\undefined}
}%
\providecommand \@ifnum [1]{%
 \ifnum #1\expandafter \@firstoftwo
 \else \expandafter \@secondoftwo
 \fi
}%
\providecommand \@ifx [1]{%
 \ifx #1\expandafter \@firstoftwo
 \else \expandafter \@secondoftwo
 \fi
}%
\providecommand \natexlab [1]{#1}%
\providecommand \enquote  [1]{``#1''}%
\providecommand \bibnamefont  [1]{#1}%
\providecommand \bibfnamefont [1]{#1}%
\providecommand \citenamefont [1]{#1}%
\providecommand \href@noop [0]{\@secondoftwo}%
\providecommand \href [0]{\begingroup \@sanitize@url \@href}%
\providecommand \@href[1]{\@@startlink{#1}\@@href}%
\providecommand \@@href[1]{\endgroup#1\@@endlink}%
\providecommand \@sanitize@url [0]{\catcode `\\12\catcode `\$12\catcode `\&12\catcode `\#12\catcode `\^12\catcode `\_12\catcode `\%12\relax}%
\providecommand \@@startlink[1]{}%
\providecommand \@@endlink[0]{}%
\providecommand \url  [0]{\begingroup\@sanitize@url \@url }%
\providecommand \@url [1]{\endgroup\@href {#1}{\urlprefix }}%
\providecommand \urlprefix  [0]{URL }%
\providecommand \Eprint [0]{\href }%
\providecommand \doibase [0]{https://doi.org/}%
\providecommand \selectlanguage [0]{\@gobble}%
\providecommand \bibinfo  [0]{\@secondoftwo}%
\providecommand \bibfield  [0]{\@secondoftwo}%
\providecommand \translation [1]{[#1]}%
\providecommand \BibitemOpen [0]{}%
\providecommand \bibitemStop [0]{}%
\providecommand \bibitemNoStop [0]{.\EOS\space}%
\providecommand \EOS [0]{\spacefactor3000\relax}%
\providecommand \BibitemShut  [1]{\csname bibitem#1\endcsname}%
\let\auto@bib@innerbib\@empty
\bibitem [{\citenamefont {Wen}\ \emph {et~al.}(2019)\citenamefont {Wen}, \citenamefont {Lin}, \citenamefont {Kockum}, \citenamefont {Suri}, \citenamefont {Ian}, \citenamefont {Chen}, \citenamefont {Mao}, \citenamefont {Chiu}, \citenamefont {Delsing}, \citenamefont {Nori}, \citenamefont {Lin},\ and\ \citenamefont {Hoi}}]{Wen2019}%
  \BibitemOpen
  \bibfield  {author} {\bibinfo {author} {\bibfnamefont {P.~Y.}\ \bibnamefont {Wen}}, \bibinfo {author} {\bibfnamefont {K.-T.}\ \bibnamefont {Lin}}, \bibinfo {author} {\bibfnamefont {A.~F.}\ \bibnamefont {Kockum}}, \bibinfo {author} {\bibfnamefont {B.}~\bibnamefont {Suri}}, \bibinfo {author} {\bibfnamefont {H.}~\bibnamefont {Ian}}, \bibinfo {author} {\bibfnamefont {J.~C.}\ \bibnamefont {Chen}}, \bibinfo {author} {\bibfnamefont {S.~Y.}\ \bibnamefont {Mao}}, \bibinfo {author} {\bibfnamefont {C.~C.}\ \bibnamefont {Chiu}}, \bibinfo {author} {\bibfnamefont {P.}~\bibnamefont {Delsing}}, \bibinfo {author} {\bibfnamefont {F.}~\bibnamefont {Nori}}, \bibinfo {author} {\bibfnamefont {G.-D.}\ \bibnamefont {Lin}},\ and\ \bibinfo {author} {\bibfnamefont {I.-C.}\ \bibnamefont {Hoi}},\ }\bibfield  {title} {\bibinfo {title} {{Large Collective Lamb Shift of Two Distant Superconducting Artificial Atoms}},\ }\href {https://doi.org/10.1103/PhysRevLett.123.233602} {\bibfield  {journal} {\bibinfo  {journal} {Physical Review
  Letters}\ }\textbf {\bibinfo {volume} {123}},\ \bibinfo {pages} {233602} (\bibinfo {year} {2019})}\BibitemShut {NoStop}%
\bibitem [{\citenamefont {Wu}\ \emph {et~al.}(2024)\citenamefont {Wu}, \citenamefont {Cheng}, \citenamefont {Lin}, \citenamefont {Aziz}, \citenamefont {Liu}, \citenamefont {Rangdhol}, \citenamefont {Yeung}, \citenamefont {Yang}, \citenamefont {Shao}, \citenamefont {Wang}, \citenamefont {Lin}, \citenamefont {Nori},\ and\ \citenamefont {Hoi}}]{Wu2024}%
  \BibitemOpen
  \bibfield  {author} {\bibinfo {author} {\bibfnamefont {B.-Y.}\ \bibnamefont {Wu}}, \bibinfo {author} {\bibfnamefont {Y.-T.}\ \bibnamefont {Cheng}}, \bibinfo {author} {\bibfnamefont {K.-T.}\ \bibnamefont {Lin}}, \bibinfo {author} {\bibfnamefont {F.}~\bibnamefont {Aziz}}, \bibinfo {author} {\bibfnamefont {J.-C.}\ \bibnamefont {Liu}}, \bibinfo {author} {\bibfnamefont {K.-V.}\ \bibnamefont {Rangdhol}}, \bibinfo {author} {\bibfnamefont {Y.-Y.}\ \bibnamefont {Yeung}}, \bibinfo {author} {\bibfnamefont {S.}~\bibnamefont {Yang}}, \bibinfo {author} {\bibfnamefont {Q.}~\bibnamefont {Shao}}, \bibinfo {author} {\bibfnamefont {X.}~\bibnamefont {Wang}}, \bibinfo {author} {\bibfnamefont {G.-D.}\ \bibnamefont {Lin}}, \bibinfo {author} {\bibfnamefont {F.}~\bibnamefont {Nori}},\ and\ \bibinfo {author} {\bibfnamefont {I.-C.}\ \bibnamefont {Hoi}},\ }\href@noop {} {\bibinfo {title} {Microwave interference from a spin ensemble and its mirror image in waveguide magnonics}} (\bibinfo {year} {2024}),\ \Eprint
  {https://arxiv.org/abs/2409.17867} {arXiv:2409.17867} \BibitemShut {NoStop}%
\bibitem [{\citenamefont {Wang}\ \emph {et~al.}(2022)\citenamefont {Wang}, \citenamefont {Wang}, \citenamefont {Yao}, \citenamefont {Shen}, \citenamefont {Wu}, \citenamefont {Qian}, \citenamefont {Li}, \citenamefont {Zhu},\ and\ \citenamefont {You}}]{Wang2022}%
  \BibitemOpen
  \bibfield  {author} {\bibinfo {author} {\bibfnamefont {Z.-Q.}\ \bibnamefont {Wang}}, \bibinfo {author} {\bibfnamefont {Y.-P.}\ \bibnamefont {Wang}}, \bibinfo {author} {\bibfnamefont {J.}~\bibnamefont {Yao}}, \bibinfo {author} {\bibfnamefont {R.-C.}\ \bibnamefont {Shen}}, \bibinfo {author} {\bibfnamefont {W.-J.}\ \bibnamefont {Wu}}, \bibinfo {author} {\bibfnamefont {J.}~\bibnamefont {Qian}}, \bibinfo {author} {\bibfnamefont {J.}~\bibnamefont {Li}}, \bibinfo {author} {\bibfnamefont {S.-Y.}\ \bibnamefont {Zhu}},\ and\ \bibinfo {author} {\bibfnamefont {J.~Q.}\ \bibnamefont {You}},\ }\bibfield  {title} {\bibinfo {title} {Giant spin ensembles in waveguide magnonics},\ }\href {https://doi.org/10.1038/s41467-022-35174-9} {\bibfield  {journal} {\bibinfo  {journal} {Nature Communications}\ }\textbf {\bibinfo {volume} {13}},\ \bibinfo {pages} {7580} (\bibinfo {year} {2022})}\BibitemShut {NoStop}%
\bibitem [{\citenamefont {Zhan}\ \emph {et~al.}(2022)\citenamefont {Zhan}, \citenamefont {Sun},\ and\ \citenamefont {Tan}}]{Zhan2022}%
  \BibitemOpen
  \bibfield  {author} {\bibinfo {author} {\bibfnamefont {H.}~\bibnamefont {Zhan}}, \bibinfo {author} {\bibfnamefont {L.}~\bibnamefont {Sun}},\ and\ \bibinfo {author} {\bibfnamefont {H.}~\bibnamefont {Tan}},\ }\bibfield  {title} {\bibinfo {title} {Chirality-induced one-way quantum steering between two waveguide-mediated ferrimagnetic microspheres},\ }\href {https://doi.org/10.1103/PhysRevB.106.104432} {\bibfield  {journal} {\bibinfo  {journal} {Physical Review B}\ }\textbf {\bibinfo {volume} {106}},\ \bibinfo {pages} {104432} (\bibinfo {year} {2022})}\BibitemShut {NoStop}%
\bibitem [{\citenamefont {Lehmberg}(1970)}]{Lehmberg1970}%
  \BibitemOpen
  \bibfield  {author} {\bibinfo {author} {\bibfnamefont {R.~H.}\ \bibnamefont {Lehmberg}},\ }\bibfield  {title} {\bibinfo {title} {{Radiation from an N-Atom System. I. General Formalism}},\ }\href {https://doi.org/10.1103/PhysRevA.2.883} {\bibfield  {journal} {\bibinfo  {journal} {Physical Review A}\ }\textbf {\bibinfo {volume} {2}},\ \bibinfo {pages} {883} (\bibinfo {year} {1970})}\BibitemShut {NoStop}%
\bibitem [{\citenamefont {Meystre}(2021)}]{Meystre2021}%
  \BibitemOpen
  \bibfield  {author} {\bibinfo {author} {\bibfnamefont {P.}~\bibnamefont {Meystre}},\ }\href {https://doi.org/10.1007/978-3-030-76183-7} {\emph {\bibinfo {title} {{Quantum Optics: Taming the Quantum}}}}\ (\bibinfo  {publisher} {Springer International Publishing},\ \bibinfo {year} {2021})\BibitemShut {NoStop}%
\bibitem [{\citenamefont {Lin}\ \emph {et~al.}(2019)\citenamefont {Lin}, \citenamefont {Hsu}, \citenamefont {Lee}, \citenamefont {Hoi},\ and\ \citenamefont {Lin}}]{Lin2019}%
  \BibitemOpen
  \bibfield  {author} {\bibinfo {author} {\bibfnamefont {K.-T.}\ \bibnamefont {Lin}}, \bibinfo {author} {\bibfnamefont {T.}~\bibnamefont {Hsu}}, \bibinfo {author} {\bibfnamefont {C.-Y.}\ \bibnamefont {Lee}}, \bibinfo {author} {\bibfnamefont {I.-C.}\ \bibnamefont {Hoi}},\ and\ \bibinfo {author} {\bibfnamefont {G.-D.}\ \bibnamefont {Lin}},\ }\bibfield  {title} {\bibinfo {title} {{Scalable collective Lamb shift of a 1D superconducting qubit array in front of a mirror}},\ }\href {https://doi.org/10.1038/s41598-019-55545-5} {\bibfield  {journal} {\bibinfo  {journal} {Scientific Reports}\ }\textbf {\bibinfo {volume} {9}},\ \bibinfo {pages} {19175} (\bibinfo {year} {2019})}\BibitemShut {NoStop}%
\bibitem [{\citenamefont {Li}\ \emph {et~al.}(2022)\citenamefont {Li}, \citenamefont {Yefremenko}, \citenamefont {Lisovenko}, \citenamefont {Trevillian}, \citenamefont {Polakovic}, \citenamefont {Cecil}, \citenamefont {Barry}, \citenamefont {Pearson}, \citenamefont {Divan}, \citenamefont {Tyberkevych}, \citenamefont {Chang}, \citenamefont {Welp}, \citenamefont {Kwok},\ and\ \citenamefont {Novosad}}]{Li2022}%
  \BibitemOpen
  \bibfield  {author} {\bibinfo {author} {\bibfnamefont {Y.}~\bibnamefont {Li}}, \bibinfo {author} {\bibfnamefont {V.~G.}\ \bibnamefont {Yefremenko}}, \bibinfo {author} {\bibfnamefont {M.}~\bibnamefont {Lisovenko}}, \bibinfo {author} {\bibfnamefont {C.}~\bibnamefont {Trevillian}}, \bibinfo {author} {\bibfnamefont {T.}~\bibnamefont {Polakovic}}, \bibinfo {author} {\bibfnamefont {T.~W.}\ \bibnamefont {Cecil}}, \bibinfo {author} {\bibfnamefont {P.~S.}\ \bibnamefont {Barry}}, \bibinfo {author} {\bibfnamefont {J.}~\bibnamefont {Pearson}}, \bibinfo {author} {\bibfnamefont {R.}~\bibnamefont {Divan}}, \bibinfo {author} {\bibfnamefont {V.}~\bibnamefont {Tyberkevych}}, \bibinfo {author} {\bibfnamefont {C.~L.}\ \bibnamefont {Chang}}, \bibinfo {author} {\bibfnamefont {U.}~\bibnamefont {Welp}}, \bibinfo {author} {\bibfnamefont {W.-K.}\ \bibnamefont {Kwok}},\ and\ \bibinfo {author} {\bibfnamefont {V.}~\bibnamefont {Novosad}},\ }\bibfield  {title} {\bibinfo {title} {Coherent coupling of two remote magnonic resonators
  mediated by superconducting circuits},\ }\href {https://doi.org/10.1103/PhysRevLett.128.047701} {\bibfield  {journal} {\bibinfo  {journal} {Physical Review Letters}\ }\textbf {\bibinfo {volume} {128}},\ \bibinfo {pages} {047701} (\bibinfo {year} {2022})}\BibitemShut {NoStop}%
\bibitem [{\citenamefont {Gardiner}\ and\ \citenamefont {Collett}(1985)}]{Gardiner1985}%
  \BibitemOpen
  \bibfield  {author} {\bibinfo {author} {\bibfnamefont {C.~W.}\ \bibnamefont {Gardiner}}\ and\ \bibinfo {author} {\bibfnamefont {M.~J.}\ \bibnamefont {Collett}},\ }\bibfield  {title} {\bibinfo {title} {{Input and output in damped quantum systems: Quantum stochastic differential equations and the master equation}},\ }\href {https://doi.org/10.1103/PhysRevA.31.3761} {\bibfield  {journal} {\bibinfo  {journal} {Physical Review A}\ }\textbf {\bibinfo {volume} {31}},\ \bibinfo {pages} {3761} (\bibinfo {year} {1985})}\BibitemShut {NoStop}%
\bibitem [{\citenamefont {Lin}\ \emph {et~al.}(2025)\citenamefont {Lin}, \citenamefont {Hsu}, \citenamefont {Aziz}, \citenamefont {Lin}, \citenamefont {Wen}, \citenamefont {Hoi},\ and\ \citenamefont {Lin}}]{Lin2025}%
  \BibitemOpen
  \bibfield  {author} {\bibinfo {author} {\bibfnamefont {K.-T.}\ \bibnamefont {Lin}}, \bibinfo {author} {\bibfnamefont {T.}~\bibnamefont {Hsu}}, \bibinfo {author} {\bibfnamefont {F.}~\bibnamefont {Aziz}}, \bibinfo {author} {\bibfnamefont {Y.-C.}\ \bibnamefont {Lin}}, \bibinfo {author} {\bibfnamefont {P.-Y.}\ \bibnamefont {Wen}}, \bibinfo {author} {\bibfnamefont {I.-C.}\ \bibnamefont {Hoi}},\ and\ \bibinfo {author} {\bibfnamefont {G.-D.}\ \bibnamefont {Lin}},\ }\bibfield  {title} {\bibinfo {title} {Single-atom amplification assisted by multiple sideband interference in waveguide {QED} systems},\ }\href {https://doi.org/10.1088/1367-2630/add7fe} {\bibfield  {journal} {\bibinfo  {journal} {New Journal of Physics}\ }\textbf {\bibinfo {volume} {27}},\ \bibinfo {pages} {064108} (\bibinfo {year} {2025})}\BibitemShut {NoStop}%
\bibitem [{\citenamefont {Koshino}\ and\ \citenamefont {Nakamura}(2012)}]{Koshino2012}%
  \BibitemOpen
  \bibfield  {author} {\bibinfo {author} {\bibfnamefont {K.}~\bibnamefont {Koshino}}\ and\ \bibinfo {author} {\bibfnamefont {Y.}~\bibnamefont {Nakamura}},\ }\bibfield  {title} {\bibinfo {title} {Control of the radiative level shift and linewidth of a superconducting artificial atom through a variable boundary condition},\ }\href {https://doi.org/10.1088/1367-2630/14/4/043005} {\bibfield  {journal} {\bibinfo  {journal} {New Journal of Physics}\ }\textbf {\bibinfo {volume} {14}},\ \bibinfo {pages} {043005} (\bibinfo {year} {2012})}\BibitemShut {NoStop}%
\bibitem [{\citenamefont {Zhang}\ \emph {et~al.}(2014)\citenamefont {Zhang}, \citenamefont {Zou}, \citenamefont {Jiang},\ and\ \citenamefont {Tang}}]{Zhang2014}%
  \BibitemOpen
  \bibfield  {author} {\bibinfo {author} {\bibfnamefont {X.}~\bibnamefont {Zhang}}, \bibinfo {author} {\bibfnamefont {C.-L.}\ \bibnamefont {Zou}}, \bibinfo {author} {\bibfnamefont {L.}~\bibnamefont {Jiang}},\ and\ \bibinfo {author} {\bibfnamefont {H.~X.}\ \bibnamefont {Tang}},\ }\bibfield  {title} {\bibinfo {title} {{Strongly Coupled Magnons and Cavity Microwave Photons}},\ }\href {https://doi.org/10.1103/PhysRevLett.113.156401} {\bibfield  {journal} {\bibinfo  {journal} {Physical Review Letters}\ }\textbf {\bibinfo {volume} {113}},\ \bibinfo {pages} {156401} (\bibinfo {year} {2014})}\BibitemShut {NoStop}%
\bibitem [{\citenamefont {Zhang}\ \emph {et~al.}(2015)\citenamefont {Zhang}, \citenamefont {Zou}, \citenamefont {Zhu}, \citenamefont {Marquardt}, \citenamefont {Jiang},\ and\ \citenamefont {Tang}}]{Zhang2015NC}%
  \BibitemOpen
  \bibfield  {author} {\bibinfo {author} {\bibfnamefont {X.}~\bibnamefont {Zhang}}, \bibinfo {author} {\bibfnamefont {C.-L.}\ \bibnamefont {Zou}}, \bibinfo {author} {\bibfnamefont {N.}~\bibnamefont {Zhu}}, \bibinfo {author} {\bibfnamefont {F.}~\bibnamefont {Marquardt}}, \bibinfo {author} {\bibfnamefont {L.}~\bibnamefont {Jiang}},\ and\ \bibinfo {author} {\bibfnamefont {H.~X.}\ \bibnamefont {Tang}},\ }\bibfield  {title} {\bibinfo {title} {{Magnon dark modes and gradient memory}},\ }\href {https://doi.org/10.1038/ncomms9914} {\bibfield  {journal} {\bibinfo  {journal} {Nature Communications}\ }\textbf {\bibinfo {volume} {6}},\ \bibinfo {pages} {8914} (\bibinfo {year} {2015})}\BibitemShut {NoStop}%
\bibitem [{\citenamefont {Lalumi\`ere}\ \emph {et~al.}(2013)\citenamefont {Lalumi\`ere}, \citenamefont {Sanders}, \citenamefont {van Loo}, \citenamefont {Fedorov}, \citenamefont {Wallraff},\ and\ \citenamefont {Blais}}]{LalumiPRA2013}%
  \BibitemOpen
  \bibfield  {author} {\bibinfo {author} {\bibfnamefont {K.}~\bibnamefont {Lalumi\`ere}}, \bibinfo {author} {\bibfnamefont {B.~C.}\ \bibnamefont {Sanders}}, \bibinfo {author} {\bibfnamefont {A.~F.}\ \bibnamefont {van Loo}}, \bibinfo {author} {\bibfnamefont {A.}~\bibnamefont {Fedorov}}, \bibinfo {author} {\bibfnamefont {A.}~\bibnamefont {Wallraff}},\ and\ \bibinfo {author} {\bibfnamefont {A.}~\bibnamefont {Blais}},\ }\bibfield  {title} {\bibinfo {title} {Input-output theory for waveguide qed with an ensemble of inhomogeneous atoms},\ }\href {https://doi.org/10.1103/PhysRevA.88.043806} {\bibfield  {journal} {\bibinfo  {journal} {Physical Review A}\ }\textbf {\bibinfo {volume} {88}},\ \bibinfo {pages} {043806} (\bibinfo {year} {2013})}\BibitemShut {NoStop}%
\bibitem [{\citenamefont {Zhang}(2016)}]{zhang2016magnon}%
  \BibitemOpen
  \bibfield  {author} {\bibinfo {author} {\bibfnamefont {X.}~\bibnamefont {Zhang}},\ }\href@noop {} {\emph {\bibinfo {title} {Magnon-based information transduction in ferrimagnetic insulators}}}\ (\bibinfo  {publisher} {Yale University},\ \bibinfo {year} {2016})\BibitemShut {NoStop}%
\bibitem [{\citenamefont {Xu}\ \emph {et~al.}(2023)\citenamefont {Xu}, \citenamefont {Wang}, \citenamefont {Kong},\ and\ \citenamefont {Hu}}]{Jun2023}%
  \BibitemOpen
  \bibfield  {author} {\bibinfo {author} {\bibfnamefont {J.}~\bibnamefont {Xu}}, \bibinfo {author} {\bibfnamefont {F.}~\bibnamefont {Wang}}, \bibinfo {author} {\bibfnamefont {D.}~\bibnamefont {Kong}},\ and\ \bibinfo {author} {\bibfnamefont {X.}~\bibnamefont {Hu}},\ }\bibfield  {title} {\bibinfo {title} {Microwave-mediated magnon–atom interactions: Two-mode higher-order squeezing of two yig spheres},\ }\href {https://doi.org/https://doi.org/10.1016/j.rinp.2023.106818} {\bibfield  {journal} {\bibinfo  {journal} {Results in Physics}\ }\textbf {\bibinfo {volume} {52}},\ \bibinfo {pages} {106818} (\bibinfo {year} {2023})}\BibitemShut {NoStop}%
\end{thebibliography}%


\begin{thebibliography}{66}%
\makeatletter
\providecommand \@ifxundefined [1]{%
 \@ifx{#1\undefined}
}%
\providecommand \@ifnum [1]{%
 \ifnum #1\expandafter \@firstoftwo
 \else \expandafter \@secondoftwo
 \fi
}%
\providecommand \@ifx [1]{%
 \ifx #1\expandafter \@firstoftwo
 \else \expandafter \@secondoftwo
 \fi
}%
\providecommand \natexlab [1]{#1}%
\providecommand \enquote  [1]{``#1''}%
\providecommand \bibnamefont  [1]{#1}%
\providecommand \bibfnamefont [1]{#1}%
\providecommand \citenamefont [1]{#1}%
\providecommand \href@noop [0]{\@secondoftwo}%
\providecommand \href [0]{\begingroup \@sanitize@url \@href}%
\providecommand \@href[1]{\@@startlink{#1}\@@href}%
\providecommand \@@href[1]{\endgroup#1\@@endlink}%
\providecommand \@sanitize@url [0]{\catcode `\\12\catcode `\$12\catcode `\&12\catcode `\#12\catcode `\^12\catcode `\_12\catcode `\%12\relax}%
\providecommand \@@startlink[1]{}%
\providecommand \@@endlink[0]{}%
\providecommand \url  [0]{\begingroup\@sanitize@url \@url }%
\providecommand \@url [1]{\endgroup\@href {#1}{\urlprefix }}%
\providecommand \urlprefix  [0]{URL }%
\providecommand \Eprint [0]{\href }%
\providecommand \doibase [0]{https://doi.org/}%
\providecommand \selectlanguage [0]{\@gobble}%
\providecommand \bibinfo  [0]{\@secondoftwo}%
\providecommand \bibfield  [0]{\@secondoftwo}%
\providecommand \translation [1]{[#1]}%
\providecommand \BibitemOpen [0]{}%
\providecommand \bibitemStop [0]{}%
\providecommand \bibitemNoStop [0]{.\EOS\space}%
\providecommand \EOS [0]{\spacefactor3000\relax}%
\providecommand \BibitemShut  [1]{\csname bibitem#1\endcsname}%
\let\auto@bib@innerbib\@empty
\bibitem [{\citenamefont {Lukin}(2003)}]{Lukin2003}%
  \BibitemOpen
  \bibfield  {author} {\bibinfo {author} {\bibfnamefont {M.~D.}\ \bibnamefont {Lukin}},\ }\bibfield  {title} {\bibinfo {title} {{Colloquium: Trapping and manipulating photon states in atomic ensembles}},\ }\href {https://doi.org/10.1103/RevModPhys.75.457} {\bibfield  {journal} {\bibinfo  {journal} {Reviews of Modern Physics}\ }\textbf {\bibinfo {volume} {75}},\ \bibinfo {pages} {457} (\bibinfo {year} {2003})}\BibitemShut {NoStop}%
\bibitem [{\citenamefont {Imamo\u{g}lu}(2009)}]{Atac2009}%
  \BibitemOpen
  \bibfield  {author} {\bibinfo {author} {\bibfnamefont {A.}~\bibnamefont {Imamo\u{g}lu}},\ }\bibfield  {title} {\bibinfo {title} {{Cavity QED Based on Collective Magnetic Dipole Coupling: Spin Ensembles as Hybrid Two-Level Systems}},\ }\href {https://doi.org/10.1103/PhysRevLett.102.083602} {\bibfield  {journal} {\bibinfo  {journal} {Physical Review Letters}\ }\textbf {\bibinfo {volume} {102}},\ \bibinfo {pages} {083602} (\bibinfo {year} {2009})}\BibitemShut {NoStop}%
\bibitem [{\citenamefont {{Zare Rameshti}}\ \emph {et~al.}(2022)\citenamefont {{Zare Rameshti}}, \citenamefont {{Viola Kusminskiy}}, \citenamefont {Haigh}, \citenamefont {Usami}, \citenamefont {Lachance-Quirion}, \citenamefont {Nakamura}, \citenamefont {Hu}, \citenamefont {Tang}, \citenamefont {Bauer},\ and\ \citenamefont {Blanter}}]{ZARERAMESHTI20221}%
  \BibitemOpen
  \bibfield  {author} {\bibinfo {author} {\bibfnamefont {B.}~\bibnamefont {{Zare Rameshti}}}, \bibinfo {author} {\bibfnamefont {S.}~\bibnamefont {{Viola Kusminskiy}}}, \bibinfo {author} {\bibfnamefont {J.~A.}\ \bibnamefont {Haigh}}, \bibinfo {author} {\bibfnamefont {K.}~\bibnamefont {Usami}}, \bibinfo {author} {\bibfnamefont {D.}~\bibnamefont {Lachance-Quirion}}, \bibinfo {author} {\bibfnamefont {Y.}~\bibnamefont {Nakamura}}, \bibinfo {author} {\bibfnamefont {C.-M.}\ \bibnamefont {Hu}}, \bibinfo {author} {\bibfnamefont {H.~X.}\ \bibnamefont {Tang}}, \bibinfo {author} {\bibfnamefont {G.~E.}\ \bibnamefont {Bauer}},\ and\ \bibinfo {author} {\bibfnamefont {Y.~M.}\ \bibnamefont {Blanter}},\ }\bibfield  {title} {\bibinfo {title} {{Cavity magnonics}},\ }\href {https://doi.org/https://doi.org/10.1016/j.physrep.2022.06.001} {\bibfield  {journal} {\bibinfo  {journal} {Physics Reports}\ }\textbf {\bibinfo {volume} {979}},\ \bibinfo {pages} {1} (\bibinfo {year} {2022})}\BibitemShut {NoStop}%
\bibitem [{\citenamefont {Tabuchi}\ \emph {et~al.}(2014)\citenamefont {Tabuchi}, \citenamefont {Ishino}, \citenamefont {Ishikawa}, \citenamefont {Yamazaki}, \citenamefont {Usami},\ and\ \citenamefont {Nakamura}}]{Tabuchi2014}%
  \BibitemOpen
  \bibfield  {author} {\bibinfo {author} {\bibfnamefont {Y.}~\bibnamefont {Tabuchi}}, \bibinfo {author} {\bibfnamefont {S.}~\bibnamefont {Ishino}}, \bibinfo {author} {\bibfnamefont {T.}~\bibnamefont {Ishikawa}}, \bibinfo {author} {\bibfnamefont {R.}~\bibnamefont {Yamazaki}}, \bibinfo {author} {\bibfnamefont {K.}~\bibnamefont {Usami}},\ and\ \bibinfo {author} {\bibfnamefont {Y.}~\bibnamefont {Nakamura}},\ }\bibfield  {title} {\bibinfo {title} {{Hybridizing Ferromagnetic Magnons and Microwave Photons in the Quantum Limit}},\ }\href {https://doi.org/10.1103/PhysRevLett.113.083603} {\bibfield  {journal} {\bibinfo  {journal} {Physical Review Letters}\ }\textbf {\bibinfo {volume} {113}},\ \bibinfo {pages} {083603} (\bibinfo {year} {2014})}\BibitemShut {NoStop}%
\bibitem [{\citenamefont {Zhang}\ \emph {et~al.}(2014)\citenamefont {Zhang}, \citenamefont {Zou}, \citenamefont {Jiang},\ and\ \citenamefont {Tang}}]{Zhang2014}%
  \BibitemOpen
  \bibfield  {author} {\bibinfo {author} {\bibfnamefont {X.}~\bibnamefont {Zhang}}, \bibinfo {author} {\bibfnamefont {C.-L.}\ \bibnamefont {Zou}}, \bibinfo {author} {\bibfnamefont {L.}~\bibnamefont {Jiang}},\ and\ \bibinfo {author} {\bibfnamefont {H.~X.}\ \bibnamefont {Tang}},\ }\bibfield  {title} {\bibinfo {title} {{Strongly Coupled Magnons and Cavity Microwave Photons}},\ }\href {https://doi.org/10.1103/PhysRevLett.113.156401} {\bibfield  {journal} {\bibinfo  {journal} {Physical Review Letters}\ }\textbf {\bibinfo {volume} {113}},\ \bibinfo {pages} {156401} (\bibinfo {year} {2014})}\BibitemShut {NoStop}%
\bibitem [{\citenamefont {Goryachev}\ \emph {et~al.}(2014)\citenamefont {Goryachev}, \citenamefont {Farr}, \citenamefont {Creedon}, \citenamefont {Fan}, \citenamefont {Kostylev},\ and\ \citenamefont {Tobar}}]{Goryachev_2014}%
  \BibitemOpen
  \bibfield  {author} {\bibinfo {author} {\bibfnamefont {M.}~\bibnamefont {Goryachev}}, \bibinfo {author} {\bibfnamefont {W.~G.}\ \bibnamefont {Farr}}, \bibinfo {author} {\bibfnamefont {D.~L.}\ \bibnamefont {Creedon}}, \bibinfo {author} {\bibfnamefont {Y.}~\bibnamefont {Fan}}, \bibinfo {author} {\bibfnamefont {M.}~\bibnamefont {Kostylev}},\ and\ \bibinfo {author} {\bibfnamefont {M.~E.}\ \bibnamefont {Tobar}},\ }\bibfield  {title} {\bibinfo {title} {{High-Cooperativity Cavity QED with Magnons at Microwave Frequencies}},\ }\href {https://doi.org/10.1103/PhysRevApplied.2.054002} {\bibfield  {journal} {\bibinfo  {journal} {Physical Review Applied}\ }\textbf {\bibinfo {volume} {2}},\ \bibinfo {pages} {054002} (\bibinfo {year} {2014})}\BibitemShut {NoStop}%
\bibitem [{\citenamefont {Zhang}\ \emph {et~al.}(2015{\natexlab{a}})\citenamefont {Zhang}, \citenamefont {Wang}, \citenamefont {Li}, \citenamefont {Luo}, \citenamefont {Wu}, \citenamefont {Nori},\ and\ \citenamefont {You}}]{Zhang2015}%
  \BibitemOpen
  \bibfield  {author} {\bibinfo {author} {\bibfnamefont {D.}~\bibnamefont {Zhang}}, \bibinfo {author} {\bibfnamefont {X.-M.}\ \bibnamefont {Wang}}, \bibinfo {author} {\bibfnamefont {T.-F.}\ \bibnamefont {Li}}, \bibinfo {author} {\bibfnamefont {X.-Q.}\ \bibnamefont {Luo}}, \bibinfo {author} {\bibfnamefont {W.}~\bibnamefont {Wu}}, \bibinfo {author} {\bibfnamefont {F.}~\bibnamefont {Nori}},\ and\ \bibinfo {author} {\bibfnamefont {J.~Q.}\ \bibnamefont {You}},\ }\bibfield  {title} {\bibinfo {title} {{Cavity quantum electrodynamics with ferromagnetic magnons in a small yttrium-iron-garnet sphere}},\ }\href {https://doi.org/10.1038/npjqi.2015.14} {\bibfield  {journal} {\bibinfo  {journal} {npj Quantum Information}\ }\textbf {\bibinfo {volume} {1}},\ \bibinfo {pages} {15014} (\bibinfo {year} {2015}{\natexlab{a}})}\BibitemShut {NoStop}%
\bibitem [{\citenamefont {Bourhill}\ \emph {et~al.}(2016)\citenamefont {Bourhill}, \citenamefont {Kostylev}, \citenamefont {Goryachev}, \citenamefont {Creedon},\ and\ \citenamefont {Tobar}}]{Bourhill2016}%
  \BibitemOpen
  \bibfield  {author} {\bibinfo {author} {\bibfnamefont {J.}~\bibnamefont {Bourhill}}, \bibinfo {author} {\bibfnamefont {N.}~\bibnamefont {Kostylev}}, \bibinfo {author} {\bibfnamefont {M.}~\bibnamefont {Goryachev}}, \bibinfo {author} {\bibfnamefont {D.~L.}\ \bibnamefont {Creedon}},\ and\ \bibinfo {author} {\bibfnamefont {M.~E.}\ \bibnamefont {Tobar}},\ }\bibfield  {title} {\bibinfo {title} {{Ultrahigh cooperativity interactions between magnons and resonant photons in a YIG sphere}},\ }\href {https://doi.org/10.1103/PhysRevB.93.144420} {\bibfield  {journal} {\bibinfo  {journal} {Physical Review B}\ }\textbf {\bibinfo {volume} {93}},\ \bibinfo {pages} {144420} (\bibinfo {year} {2016})}\BibitemShut {NoStop}%
\bibitem [{\citenamefont {Kostylev}\ \emph {et~al.}(2016)\citenamefont {Kostylev}, \citenamefont {Goryachev},\ and\ \citenamefont {Tobar}}]{Kostylev2016}%
  \BibitemOpen
  \bibfield  {author} {\bibinfo {author} {\bibfnamefont {N.}~\bibnamefont {Kostylev}}, \bibinfo {author} {\bibfnamefont {M.}~\bibnamefont {Goryachev}},\ and\ \bibinfo {author} {\bibfnamefont {M.~E.}\ \bibnamefont {Tobar}},\ }\bibfield  {title} {\bibinfo {title} {{Superstrong coupling of a microwave cavity to yttrium iron garnet magnons}},\ }\href {https://doi.org/10.1063/1.4941730} {\bibfield  {journal} {\bibinfo  {journal} {Applied Physics Letters}\ }\textbf {\bibinfo {volume} {108}},\ \bibinfo {pages} {062402} (\bibinfo {year} {2016})}\BibitemShut {NoStop}%
\bibitem [{\citenamefont {Goryachev}\ \emph {et~al.}(2018)\citenamefont {Goryachev}, \citenamefont {Watt}, \citenamefont {Bourhill}, \citenamefont {Kostylev},\ and\ \citenamefont {Tobar}}]{Goryachev_2018}%
  \BibitemOpen
  \bibfield  {author} {\bibinfo {author} {\bibfnamefont {M.}~\bibnamefont {Goryachev}}, \bibinfo {author} {\bibfnamefont {S.}~\bibnamefont {Watt}}, \bibinfo {author} {\bibfnamefont {J.}~\bibnamefont {Bourhill}}, \bibinfo {author} {\bibfnamefont {M.}~\bibnamefont {Kostylev}},\ and\ \bibinfo {author} {\bibfnamefont {M.~E.}\ \bibnamefont {Tobar}},\ }\bibfield  {title} {\bibinfo {title} {{Cavity magnon polaritons with lithium ferrite and three-dimensional microwave resonators at millikelvin temperatures}},\ }\href {https://doi.org/10.1103/PhysRevB.97.155129} {\bibfield  {journal} {\bibinfo  {journal} {Physical Review B}\ }\textbf {\bibinfo {volume} {97}},\ \bibinfo {pages} {155129} (\bibinfo {year} {2018})}\BibitemShut {NoStop}%
\bibitem [{\citenamefont {Flower}\ \emph {et~al.}(2019)\citenamefont {Flower}, \citenamefont {Goryachev}, \citenamefont {Bourhill},\ and\ \citenamefont {Tobar}}]{Flower_2019}%
  \BibitemOpen
  \bibfield  {author} {\bibinfo {author} {\bibfnamefont {G.}~\bibnamefont {Flower}}, \bibinfo {author} {\bibfnamefont {M.}~\bibnamefont {Goryachev}}, \bibinfo {author} {\bibfnamefont {J.}~\bibnamefont {Bourhill}},\ and\ \bibinfo {author} {\bibfnamefont {M.~E.}\ \bibnamefont {Tobar}},\ }\bibfield  {title} {\bibinfo {title} {{Experimental implementations of cavity-magnon systems: from ultra strong coupling to applications in precision measurement}},\ }\href {https://doi.org/10.1088/1367-2630/ab3e1c} {\bibfield  {journal} {\bibinfo  {journal} {New Journal of Physics}\ }\textbf {\bibinfo {volume} {21}},\ \bibinfo {pages} {095004} (\bibinfo {year} {2019})}\BibitemShut {NoStop}%
\bibitem [{\citenamefont {Zhang}\ \emph {et~al.}(2016)\citenamefont {Zhang}, \citenamefont {Zou}, \citenamefont {Jiang},\ and\ \citenamefont {Tang}}]{Xufengscienceadvance2016}%
  \BibitemOpen
  \bibfield  {author} {\bibinfo {author} {\bibfnamefont {X.}~\bibnamefont {Zhang}}, \bibinfo {author} {\bibfnamefont {C.-L.}\ \bibnamefont {Zou}}, \bibinfo {author} {\bibfnamefont {L.}~\bibnamefont {Jiang}},\ and\ \bibinfo {author} {\bibfnamefont {H.~X.}\ \bibnamefont {Tang}},\ }\bibfield  {title} {\bibinfo {title} {{Cavity magnomechanics}},\ }\href {https://doi.org/10.1126/sciadv.1501286} {\bibfield  {journal} {\bibinfo  {journal} {Science Advances}\ }\textbf {\bibinfo {volume} {2}},\ \bibinfo {pages} {e1501286} (\bibinfo {year} {2016})}\BibitemShut {NoStop}%
\bibitem [{\citenamefont {Hisatomi}\ \emph {et~al.}(2016)\citenamefont {Hisatomi}, \citenamefont {Osada}, \citenamefont {Tabuchi}, \citenamefont {Ishikawa}, \citenamefont {Noguchi}, \citenamefont {Yamazaki}, \citenamefont {Usami},\ and\ \citenamefont {Nakamura}}]{Hisatomi2016}%
  \BibitemOpen
  \bibfield  {author} {\bibinfo {author} {\bibfnamefont {R.}~\bibnamefont {Hisatomi}}, \bibinfo {author} {\bibfnamefont {A.}~\bibnamefont {Osada}}, \bibinfo {author} {\bibfnamefont {Y.}~\bibnamefont {Tabuchi}}, \bibinfo {author} {\bibfnamefont {T.}~\bibnamefont {Ishikawa}}, \bibinfo {author} {\bibfnamefont {A.}~\bibnamefont {Noguchi}}, \bibinfo {author} {\bibfnamefont {R.}~\bibnamefont {Yamazaki}}, \bibinfo {author} {\bibfnamefont {K.}~\bibnamefont {Usami}},\ and\ \bibinfo {author} {\bibfnamefont {Y.}~\bibnamefont {Nakamura}},\ }\bibfield  {title} {\bibinfo {title} {{Bidirectional conversion between microwave and light via ferromagnetic magnons}},\ }\href {https://doi.org/10.1103/PhysRevB.93.174427} {\bibfield  {journal} {\bibinfo  {journal} {Physical Review B}\ }\textbf {\bibinfo {volume} {93}},\ \bibinfo {pages} {174427} (\bibinfo {year} {2016})}\BibitemShut {NoStop}%
\bibitem [{\citenamefont {Zhang}\ \emph {et~al.}(2015{\natexlab{b}})\citenamefont {Zhang}, \citenamefont {Zou}, \citenamefont {Zhu}, \citenamefont {Marquardt}, \citenamefont {Jiang},\ and\ \citenamefont {Tang}}]{Zhang2015NC}%
  \BibitemOpen
  \bibfield  {author} {\bibinfo {author} {\bibfnamefont {X.}~\bibnamefont {Zhang}}, \bibinfo {author} {\bibfnamefont {C.-L.}\ \bibnamefont {Zou}}, \bibinfo {author} {\bibfnamefont {N.}~\bibnamefont {Zhu}}, \bibinfo {author} {\bibfnamefont {F.}~\bibnamefont {Marquardt}}, \bibinfo {author} {\bibfnamefont {L.}~\bibnamefont {Jiang}},\ and\ \bibinfo {author} {\bibfnamefont {H.~X.}\ \bibnamefont {Tang}},\ }\bibfield  {title} {\bibinfo {title} {{Magnon dark modes and gradient memory}},\ }\href {https://doi.org/10.1038/ncomms9914} {\bibfield  {journal} {\bibinfo  {journal} {Nature Communications}\ }\textbf {\bibinfo {volume} {6}},\ \bibinfo {pages} {8914} (\bibinfo {year} {2015}{\natexlab{b}})}\BibitemShut {NoStop}%
\bibitem [{\citenamefont {Shen}\ \emph {et~al.}(2021)\citenamefont {Shen}, \citenamefont {Wang}, \citenamefont {Li}, \citenamefont {Zhu}, \citenamefont {Agarwal},\ and\ \citenamefont {You}}]{ShenPRL2021}%
  \BibitemOpen
  \bibfield  {author} {\bibinfo {author} {\bibfnamefont {R.-C.}\ \bibnamefont {Shen}}, \bibinfo {author} {\bibfnamefont {Y.-P.}\ \bibnamefont {Wang}}, \bibinfo {author} {\bibfnamefont {J.}~\bibnamefont {Li}}, \bibinfo {author} {\bibfnamefont {S.-Y.}\ \bibnamefont {Zhu}}, \bibinfo {author} {\bibfnamefont {G.~S.}\ \bibnamefont {Agarwal}},\ and\ \bibinfo {author} {\bibfnamefont {J.~Q.}\ \bibnamefont {You}},\ }\bibfield  {title} {\bibinfo {title} {{Long-Time Memory and Ternary Logic Gate Using a Multistable Cavity Magnonic System}},\ }\href {https://doi.org/10.1103/PhysRevLett.127.183202} {\bibfield  {journal} {\bibinfo  {journal} {Physical Review Letters}\ }\textbf {\bibinfo {volume} {127}},\ \bibinfo {pages} {183202} (\bibinfo {year} {2021})}\BibitemShut {NoStop}%
\bibitem [{\citenamefont {Osada}\ \emph {et~al.}(2016)\citenamefont {Osada}, \citenamefont {Hisatomi}, \citenamefont {Noguchi}, \citenamefont {Tabuchi}, \citenamefont {Yamazaki}, \citenamefont {Usami}, \citenamefont {Sadgrove}, \citenamefont {Yalla}, \citenamefont {Nomura},\ and\ \citenamefont {Nakamura}}]{Osada2016}%
  \BibitemOpen
  \bibfield  {author} {\bibinfo {author} {\bibfnamefont {A.}~\bibnamefont {Osada}}, \bibinfo {author} {\bibfnamefont {R.}~\bibnamefont {Hisatomi}}, \bibinfo {author} {\bibfnamefont {A.}~\bibnamefont {Noguchi}}, \bibinfo {author} {\bibfnamefont {Y.}~\bibnamefont {Tabuchi}}, \bibinfo {author} {\bibfnamefont {R.}~\bibnamefont {Yamazaki}}, \bibinfo {author} {\bibfnamefont {K.}~\bibnamefont {Usami}}, \bibinfo {author} {\bibfnamefont {M.}~\bibnamefont {Sadgrove}}, \bibinfo {author} {\bibfnamefont {R.}~\bibnamefont {Yalla}}, \bibinfo {author} {\bibfnamefont {M.}~\bibnamefont {Nomura}},\ and\ \bibinfo {author} {\bibfnamefont {Y.}~\bibnamefont {Nakamura}},\ }\bibfield  {title} {\bibinfo {title} {{Cavity Optomagnonics with Spin-Orbit Coupled Photons}},\ }\href {https://doi.org/10.1103/PhysRevLett.116.223601} {\bibfield  {journal} {\bibinfo  {journal} {Physical Review Letters}\ }\textbf {\bibinfo {volume} {116}},\ \bibinfo {pages} {223601} (\bibinfo {year} {2016})}\BibitemShut {NoStop}%
\bibitem [{\citenamefont {Wang}\ \emph {et~al.}(2019)\citenamefont {Wang}, \citenamefont {Rao}, \citenamefont {Yang}, \citenamefont {Xu}, \citenamefont {Gui}, \citenamefont {Yao}, \citenamefont {You},\ and\ \citenamefont {Hu}}]{Wangprl2019}%
  \BibitemOpen
  \bibfield  {author} {\bibinfo {author} {\bibfnamefont {Y.-P.}\ \bibnamefont {Wang}}, \bibinfo {author} {\bibfnamefont {J.~W.}\ \bibnamefont {Rao}}, \bibinfo {author} {\bibfnamefont {Y.}~\bibnamefont {Yang}}, \bibinfo {author} {\bibfnamefont {P.-C.}\ \bibnamefont {Xu}}, \bibinfo {author} {\bibfnamefont {Y.~S.}\ \bibnamefont {Gui}}, \bibinfo {author} {\bibfnamefont {B.~M.}\ \bibnamefont {Yao}}, \bibinfo {author} {\bibfnamefont {J.~Q.}\ \bibnamefont {You}},\ and\ \bibinfo {author} {\bibfnamefont {C.-M.}\ \bibnamefont {Hu}},\ }\bibfield  {title} {\bibinfo {title} {{Nonreciprocity and Unidirectional Invisibility in Cavity Magnonics}},\ }\href {https://doi.org/10.1103/PhysRevLett.123.127202} {\bibfield  {journal} {\bibinfo  {journal} {Physical Review Letters}\ }\textbf {\bibinfo {volume} {123}},\ \bibinfo {pages} {127202} (\bibinfo {year} {2019})}\BibitemShut {NoStop}%
\bibitem [{\citenamefont {Wang}\ \emph {et~al.}(2018)\citenamefont {Wang}, \citenamefont {Zhang}, \citenamefont {Zhang}, \citenamefont {Li}, \citenamefont {Hu},\ and\ \citenamefont {You}}]{Yi-Pu2018}%
  \BibitemOpen
  \bibfield  {author} {\bibinfo {author} {\bibfnamefont {Y.-P.}\ \bibnamefont {Wang}}, \bibinfo {author} {\bibfnamefont {G.-Q.}\ \bibnamefont {Zhang}}, \bibinfo {author} {\bibfnamefont {D.}~\bibnamefont {Zhang}}, \bibinfo {author} {\bibfnamefont {T.-F.}\ \bibnamefont {Li}}, \bibinfo {author} {\bibfnamefont {C.-M.}\ \bibnamefont {Hu}},\ and\ \bibinfo {author} {\bibfnamefont {J.~Q.}\ \bibnamefont {You}},\ }\bibfield  {title} {\bibinfo {title} {{Bistability of Cavity Magnon Polaritons}},\ }\href {https://doi.org/10.1103/PhysRevLett.120.057202} {\bibfield  {journal} {\bibinfo  {journal} {Physical Review Letters}\ }\textbf {\bibinfo {volume} {120}},\ \bibinfo {pages} {057202} (\bibinfo {year} {2018})}\BibitemShut {NoStop}%
\bibitem [{\citenamefont {Nair}\ \emph {et~al.}(2021)\citenamefont {Nair}, \citenamefont {Mukhopadhyay},\ and\ \citenamefont {Agarwal}}]{Nair_2021}%
  \BibitemOpen
  \bibfield  {author} {\bibinfo {author} {\bibfnamefont {J.~M.~P.}\ \bibnamefont {Nair}}, \bibinfo {author} {\bibfnamefont {D.}~\bibnamefont {Mukhopadhyay}},\ and\ \bibinfo {author} {\bibfnamefont {G.~S.}\ \bibnamefont {Agarwal}},\ }\bibfield  {title} {\bibinfo {title} {{Ultralow threshold bistability and generation of long-lived mode in a dissipatively coupled nonlinear system: Application to magnonics}},\ }\href {https://doi.org/10.1103/PhysRevB.103.224401} {\bibfield  {journal} {\bibinfo  {journal} {Physical Review B}\ }\textbf {\bibinfo {volume} {103}},\ \bibinfo {pages} {224401} (\bibinfo {year} {2021})}\BibitemShut {NoStop}%
\bibitem [{\citenamefont {Shen}\ \emph {et~al.}(2022)\citenamefont {Shen}, \citenamefont {Li}, \citenamefont {Fan}, \citenamefont {Wang},\ and\ \citenamefont {You}}]{Shen2022}%
  \BibitemOpen
  \bibfield  {author} {\bibinfo {author} {\bibfnamefont {R.-C.}\ \bibnamefont {Shen}}, \bibinfo {author} {\bibfnamefont {J.}~\bibnamefont {Li}}, \bibinfo {author} {\bibfnamefont {Z.-Y.}\ \bibnamefont {Fan}}, \bibinfo {author} {\bibfnamefont {Y.-P.}\ \bibnamefont {Wang}},\ and\ \bibinfo {author} {\bibfnamefont {J.~Q.}\ \bibnamefont {You}},\ }\bibfield  {title} {\bibinfo {title} {{Mechanical Bistability in Kerr-modified Cavity Magnomechanics}},\ }\href {https://doi.org/10.1103/PhysRevLett.129.123601} {\bibfield  {journal} {\bibinfo  {journal} {Physical Review Letters}\ }\textbf {\bibinfo {volume} {129}},\ \bibinfo {pages} {123601} (\bibinfo {year} {2022})}\BibitemShut {NoStop}%
\bibitem [{\citenamefont {Zhang}\ \emph {et~al.}(2017)\citenamefont {Zhang}, \citenamefont {Luo}, \citenamefont {Wang}, \citenamefont {Li},\ and\ \citenamefont {You}}]{ZhangNC2017}%
  \BibitemOpen
  \bibfield  {author} {\bibinfo {author} {\bibfnamefont {D.}~\bibnamefont {Zhang}}, \bibinfo {author} {\bibfnamefont {X.-Q.}\ \bibnamefont {Luo}}, \bibinfo {author} {\bibfnamefont {Y.-P.}\ \bibnamefont {Wang}}, \bibinfo {author} {\bibfnamefont {T.-F.}\ \bibnamefont {Li}},\ and\ \bibinfo {author} {\bibfnamefont {J.~Q.}\ \bibnamefont {You}},\ }\bibfield  {title} {\bibinfo {title} {{Observation of the exceptional point in cavity magnon-polaritons}},\ }\href {https://doi.org/10.1038/s41467-017-01634-w} {\bibfield  {journal} {\bibinfo  {journal} {Nature Communications}\ }\textbf {\bibinfo {volume} {8}},\ \bibinfo {pages} {1368} (\bibinfo {year} {2017})}\BibitemShut {NoStop}%
\bibitem [{\citenamefont {Zhang}\ and\ \citenamefont {You}(2019)}]{ZhangPRB2019}%
  \BibitemOpen
  \bibfield  {author} {\bibinfo {author} {\bibfnamefont {G.-Q.}\ \bibnamefont {Zhang}}\ and\ \bibinfo {author} {\bibfnamefont {J.~Q.}\ \bibnamefont {You}},\ }\bibfield  {title} {\bibinfo {title} {{Higher-order exceptional point in a cavity magnonics system}},\ }\href {https://doi.org/10.1103/PhysRevB.99.054404} {\bibfield  {journal} {\bibinfo  {journal} {Physical Review B}\ }\textbf {\bibinfo {volume} {99}},\ \bibinfo {pages} {054404} (\bibinfo {year} {2019})}\BibitemShut {NoStop}%
\bibitem [{\citenamefont {Li}\ \emph {et~al.}(2018)\citenamefont {Li}, \citenamefont {Zhu},\ and\ \citenamefont {Agarwal}}]{Li_PRL_2018}%
  \BibitemOpen
  \bibfield  {author} {\bibinfo {author} {\bibfnamefont {J.}~\bibnamefont {Li}}, \bibinfo {author} {\bibfnamefont {S.-Y.}\ \bibnamefont {Zhu}},\ and\ \bibinfo {author} {\bibfnamefont {G.~S.}\ \bibnamefont {Agarwal}},\ }\bibfield  {title} {\bibinfo {title} {{Magnon-Photon-Phonon Entanglement in Cavity Magnomechanics}},\ }\href {https://doi.org/10.1103/PhysRevLett.121.203601} {\bibfield  {journal} {\bibinfo  {journal} {Physical Review Letters}\ }\textbf {\bibinfo {volume} {121}},\ \bibinfo {pages} {203601} (\bibinfo {year} {2018})}\BibitemShut {NoStop}%
\bibitem [{\citenamefont {Zhang}\ \emph {et~al.}(2019)\citenamefont {Zhang}, \citenamefont {Scully},\ and\ \citenamefont {Agarwal}}]{Zhangzhedong_2019}%
  \BibitemOpen
  \bibfield  {author} {\bibinfo {author} {\bibfnamefont {Z.}~\bibnamefont {Zhang}}, \bibinfo {author} {\bibfnamefont {M.~O.}\ \bibnamefont {Scully}},\ and\ \bibinfo {author} {\bibfnamefont {G.~S.}\ \bibnamefont {Agarwal}},\ }\bibfield  {title} {\bibinfo {title} {{Quantum entanglement between two magnon modes via Kerr nonlinearity driven far from equilibrium}},\ }\href {https://doi.org/10.1103/PhysRevResearch.1.023021} {\bibfield  {journal} {\bibinfo  {journal} {Physical Review Research}\ }\textbf {\bibinfo {volume} {1}},\ \bibinfo {pages} {023021} (\bibinfo {year} {2019})}\BibitemShut {NoStop}%
\bibitem [{\citenamefont {Tabuchi}\ \emph {et~al.}(2015)\citenamefont {Tabuchi}, \citenamefont {Ishino}, \citenamefont {Noguchi}, \citenamefont {Ishikawa}, \citenamefont {Yamazaki}, \citenamefont {Usami},\ and\ \citenamefont {Nakamura}}]{TabuchiScience2015}%
  \BibitemOpen
  \bibfield  {author} {\bibinfo {author} {\bibfnamefont {Y.}~\bibnamefont {Tabuchi}}, \bibinfo {author} {\bibfnamefont {S.}~\bibnamefont {Ishino}}, \bibinfo {author} {\bibfnamefont {A.}~\bibnamefont {Noguchi}}, \bibinfo {author} {\bibfnamefont {T.}~\bibnamefont {Ishikawa}}, \bibinfo {author} {\bibfnamefont {R.}~\bibnamefont {Yamazaki}}, \bibinfo {author} {\bibfnamefont {K.}~\bibnamefont {Usami}},\ and\ \bibinfo {author} {\bibfnamefont {Y.}~\bibnamefont {Nakamura}},\ }\bibfield  {title} {\bibinfo {title} {{Coherent coupling between a ferromagnetic magnon and a superconducting qubit}},\ }\href {https://doi.org/10.1126/science.aaa3693} {\bibfield  {journal} {\bibinfo  {journal} {Science}\ }\textbf {\bibinfo {volume} {349}},\ \bibinfo {pages} {405} (\bibinfo {year} {2015})}\BibitemShut {NoStop}%
\bibitem [{\citenamefont {Lachance-Quirion}\ \emph {et~al.}(2017)\citenamefont {Lachance-Quirion}, \citenamefont {Tabuchi}, \citenamefont {Ishino}, \citenamefont {Noguchi}, \citenamefont {Ishikawa}, \citenamefont {Yamazaki},\ and\ \citenamefont {Nakamura}}]{Danysciadv2017}%
  \BibitemOpen
  \bibfield  {author} {\bibinfo {author} {\bibfnamefont {D.}~\bibnamefont {Lachance-Quirion}}, \bibinfo {author} {\bibfnamefont {Y.}~\bibnamefont {Tabuchi}}, \bibinfo {author} {\bibfnamefont {S.}~\bibnamefont {Ishino}}, \bibinfo {author} {\bibfnamefont {A.}~\bibnamefont {Noguchi}}, \bibinfo {author} {\bibfnamefont {T.}~\bibnamefont {Ishikawa}}, \bibinfo {author} {\bibfnamefont {R.}~\bibnamefont {Yamazaki}},\ and\ \bibinfo {author} {\bibfnamefont {Y.}~\bibnamefont {Nakamura}},\ }\bibfield  {title} {\bibinfo {title} {{Resolving quanta of collective spin excitations in a millimeter-sized ferromagnet}},\ }\href {https://doi.org/10.1126/sciadv.1603150} {\bibfield  {journal} {\bibinfo  {journal} {Science Advances}\ }\textbf {\bibinfo {volume} {3}},\ \bibinfo {pages} {e1603150} (\bibinfo {year} {2017})}\BibitemShut {NoStop}%
\bibitem [{\citenamefont {Lachance-Quirion}\ \emph {et~al.}(2020)\citenamefont {Lachance-Quirion}, \citenamefont {Wolski}, \citenamefont {Tabuchi}, \citenamefont {Kono}, \citenamefont {Usami},\ and\ \citenamefont {Nakamura}}]{DanyScience2020}%
  \BibitemOpen
  \bibfield  {author} {\bibinfo {author} {\bibfnamefont {D.}~\bibnamefont {Lachance-Quirion}}, \bibinfo {author} {\bibfnamefont {S.~P.}\ \bibnamefont {Wolski}}, \bibinfo {author} {\bibfnamefont {Y.}~\bibnamefont {Tabuchi}}, \bibinfo {author} {\bibfnamefont {S.}~\bibnamefont {Kono}}, \bibinfo {author} {\bibfnamefont {K.}~\bibnamefont {Usami}},\ and\ \bibinfo {author} {\bibfnamefont {Y.}~\bibnamefont {Nakamura}},\ }\bibfield  {title} {\bibinfo {title} {{Entanglement-based single-shot detection of a single magnon with a superconducting qubit}},\ }\href {https://doi.org/10.1126/science.aaz9236} {\bibfield  {journal} {\bibinfo  {journal} {Science}\ }\textbf {\bibinfo {volume} {367}},\ \bibinfo {pages} {425} (\bibinfo {year} {2020})}\BibitemShut {NoStop}%
\bibitem [{\citenamefont {Xu}\ \emph {et~al.}(2023)\citenamefont {Xu}, \citenamefont {Gu}, \citenamefont {Li}, \citenamefont {Weng}, \citenamefont {Wang}, \citenamefont {Li}, \citenamefont {Wang}, \citenamefont {Zhu},\ and\ \citenamefont {You}}]{XuPRL2023}%
  \BibitemOpen
  \bibfield  {author} {\bibinfo {author} {\bibfnamefont {D.}~\bibnamefont {Xu}}, \bibinfo {author} {\bibfnamefont {X.-K.}\ \bibnamefont {Gu}}, \bibinfo {author} {\bibfnamefont {H.-K.}\ \bibnamefont {Li}}, \bibinfo {author} {\bibfnamefont {Y.-C.}\ \bibnamefont {Weng}}, \bibinfo {author} {\bibfnamefont {Y.-P.}\ \bibnamefont {Wang}}, \bibinfo {author} {\bibfnamefont {J.}~\bibnamefont {Li}}, \bibinfo {author} {\bibfnamefont {H.}~\bibnamefont {Wang}}, \bibinfo {author} {\bibfnamefont {S.-Y.}\ \bibnamefont {Zhu}},\ and\ \bibinfo {author} {\bibfnamefont {J.~Q.}\ \bibnamefont {You}},\ }\bibfield  {title} {\bibinfo {title} {{Quantum Control of a Single Magnon in a Macroscopic Spin System}},\ }\href {https://doi.org/10.1103/PhysRevLett.130.193603} {\bibfield  {journal} {\bibinfo  {journal} {Physical Review Letters}\ }\textbf {\bibinfo {volume} {130}},\ \bibinfo {pages} {193603} (\bibinfo {year} {2023})}\BibitemShut {NoStop}%
\bibitem [{\citenamefont {Rao}\ \emph {et~al.}(2020)\citenamefont {Rao}, \citenamefont {Wang}, \citenamefont {Yang}, \citenamefont {Yu}, \citenamefont {Gui}, \citenamefont {Fan}, \citenamefont {Xue},\ and\ \citenamefont {Hu}}]{Rao2020}%
  \BibitemOpen
  \bibfield  {author} {\bibinfo {author} {\bibfnamefont {J.~W.}\ \bibnamefont {Rao}}, \bibinfo {author} {\bibfnamefont {Y.~P.}\ \bibnamefont {Wang}}, \bibinfo {author} {\bibfnamefont {Y.}~\bibnamefont {Yang}}, \bibinfo {author} {\bibfnamefont {T.}~\bibnamefont {Yu}}, \bibinfo {author} {\bibfnamefont {Y.~S.}\ \bibnamefont {Gui}}, \bibinfo {author} {\bibfnamefont {X.~L.}\ \bibnamefont {Fan}}, \bibinfo {author} {\bibfnamefont {D.~S.}\ \bibnamefont {Xue}},\ and\ \bibinfo {author} {\bibfnamefont {C.-M.}\ \bibnamefont {Hu}},\ }\bibfield  {title} {\bibinfo {title} {{Interactions between a magnon mode and a cavity photon mode mediated by traveling photons}},\ }\href {https://doi.org/10.1103/PhysRevB.101.064404} {\bibfield  {journal} {\bibinfo  {journal} {Physical Review B}\ }\textbf {\bibinfo {volume} {101}},\ \bibinfo {pages} {064404} (\bibinfo {year} {2020})}\BibitemShut {NoStop}%
\bibitem [{\citenamefont {Wang}\ \emph {et~al.}(2022)\citenamefont {Wang}, \citenamefont {Wang}, \citenamefont {Yao}, \citenamefont {Shen}, \citenamefont {Wu}, \citenamefont {Qian}, \citenamefont {Li}, \citenamefont {Zhu},\ and\ \citenamefont {You}}]{Wang2022}%
  \BibitemOpen
  \bibfield  {author} {\bibinfo {author} {\bibfnamefont {Z.-Q.}\ \bibnamefont {Wang}}, \bibinfo {author} {\bibfnamefont {Y.-P.}\ \bibnamefont {Wang}}, \bibinfo {author} {\bibfnamefont {J.}~\bibnamefont {Yao}}, \bibinfo {author} {\bibfnamefont {R.-C.}\ \bibnamefont {Shen}}, \bibinfo {author} {\bibfnamefont {W.-J.}\ \bibnamefont {Wu}}, \bibinfo {author} {\bibfnamefont {J.}~\bibnamefont {Qian}}, \bibinfo {author} {\bibfnamefont {J.}~\bibnamefont {Li}}, \bibinfo {author} {\bibfnamefont {S.-Y.}\ \bibnamefont {Zhu}},\ and\ \bibinfo {author} {\bibfnamefont {J.~Q.}\ \bibnamefont {You}},\ }\bibfield  {title} {\bibinfo {title} {{Giant spin ensembles in waveguide magnonics}},\ }\href {https://doi.org/10.1038/s41467-022-35174-9} {\bibfield  {journal} {\bibinfo  {journal} {Nature Communications}\ }\textbf {\bibinfo {volume} {13}},\ \bibinfo {pages} {7580} (\bibinfo {year} {2022})}\BibitemShut {NoStop}%
\bibitem [{\citenamefont {Qian}\ \emph {et~al.}(2023)\citenamefont {Qian}, \citenamefont {Meng}, \citenamefont {Rao}, \citenamefont {Rao}, \citenamefont {An}, \citenamefont {Gui},\ and\ \citenamefont {Hu}}]{Qian2023}%
  \BibitemOpen
  \bibfield  {author} {\bibinfo {author} {\bibfnamefont {J.}~\bibnamefont {Qian}}, \bibinfo {author} {\bibfnamefont {C.~H.}\ \bibnamefont {Meng}}, \bibinfo {author} {\bibfnamefont {J.~W.}\ \bibnamefont {Rao}}, \bibinfo {author} {\bibfnamefont {Z.~J.}\ \bibnamefont {Rao}}, \bibinfo {author} {\bibfnamefont {Z.}~\bibnamefont {An}}, \bibinfo {author} {\bibfnamefont {Y.}~\bibnamefont {Gui}},\ and\ \bibinfo {author} {\bibfnamefont {C.-M.}\ \bibnamefont {Hu}},\ }\bibfield  {title} {\bibinfo {title} {{Non-Hermitian control between absorption and transparency in perfect zero-reflection magnonics}},\ }\href {https://doi.org/10.1038/s41467-023-39102-3} {\bibfield  {journal} {\bibinfo  {journal} {Nature Communications}\ }\textbf {\bibinfo {volume} {14}},\ \bibinfo {pages} {3437} (\bibinfo {year} {2023})}\BibitemShut {NoStop}%
\bibitem [{\citenamefont {Rao}\ \emph {et~al.}(2023{\natexlab{a}})\citenamefont {Rao}, \citenamefont {Yao}, \citenamefont {Wang}, \citenamefont {Zhang}, \citenamefont {Yu},\ and\ \citenamefont {Lu}}]{Raoprl2023}%
  \BibitemOpen
  \bibfield  {author} {\bibinfo {author} {\bibfnamefont {J.~W.}\ \bibnamefont {Rao}}, \bibinfo {author} {\bibfnamefont {B.}~\bibnamefont {Yao}}, \bibinfo {author} {\bibfnamefont {C.~Y.}\ \bibnamefont {Wang}}, \bibinfo {author} {\bibfnamefont {C.}~\bibnamefont {Zhang}}, \bibinfo {author} {\bibfnamefont {T.}~\bibnamefont {Yu}},\ and\ \bibinfo {author} {\bibfnamefont {W.}~\bibnamefont {Lu}},\ }\bibfield  {title} {\bibinfo {title} {{Unveiling a Pump-Induced Magnon Mode via Its Strong Interaction with Walker Modes}},\ }\href {https://doi.org/10.1103/PhysRevLett.130.046705} {\bibfield  {journal} {\bibinfo  {journal} {Phys. Rev. Lett.}\ }\textbf {\bibinfo {volume} {130}},\ \bibinfo {pages} {046705} (\bibinfo {year} {2023}{\natexlab{a}})}\BibitemShut {NoStop}%
\bibitem [{\citenamefont {Wang}\ \emph {et~al.}(2024)\citenamefont {Wang}, \citenamefont {Rao}, \citenamefont {Chen}, \citenamefont {Zhao}, \citenamefont {Sun}, \citenamefont {Yao}, \citenamefont {Yu}, \citenamefont {Wang},\ and\ \citenamefont {Lu}}]{Wang_NP_2024}%
  \BibitemOpen
  \bibfield  {author} {\bibinfo {author} {\bibfnamefont {C.}~\bibnamefont {Wang}}, \bibinfo {author} {\bibfnamefont {J.}~\bibnamefont {Rao}}, \bibinfo {author} {\bibfnamefont {Z.}~\bibnamefont {Chen}}, \bibinfo {author} {\bibfnamefont {K.}~\bibnamefont {Zhao}}, \bibinfo {author} {\bibfnamefont {L.}~\bibnamefont {Sun}}, \bibinfo {author} {\bibfnamefont {B.}~\bibnamefont {Yao}}, \bibinfo {author} {\bibfnamefont {T.}~\bibnamefont {Yu}}, \bibinfo {author} {\bibfnamefont {Y.-P.}\ \bibnamefont {Wang}},\ and\ \bibinfo {author} {\bibfnamefont {W.}~\bibnamefont {Lu}},\ }\bibfield  {title} {\bibinfo {title} {{Enhancement of magnonic frequency combs by exceptional points}},\ }\href {https://doi.org/10.1038/s41567-024-02478-0} {\bibfield  {journal} {\bibinfo  {journal} {Nature Physics}\ }\textbf {\bibinfo {volume} {20}},\ \bibinfo {pages} {1139} (\bibinfo {year} {2024})}\BibitemShut {NoStop}%
\bibitem [{\citenamefont {Wu}\ \emph {et~al.}(2024)\citenamefont {Wu}, \citenamefont {Cheng}, \citenamefont {Lin}, \citenamefont {Aziz}, \citenamefont {Liu}, \citenamefont {Rangdhol}, \citenamefont {Yeung}, \citenamefont {Yang}, \citenamefont {Shao}, \citenamefont {Wang}, \citenamefont {Lin}, \citenamefont {Nori},\ and\ \citenamefont {Hoi}}]{wu2024arxiv}%
  \BibitemOpen
  \bibfield  {author} {\bibinfo {author} {\bibfnamefont {B.-Y.}\ \bibnamefont {Wu}}, \bibinfo {author} {\bibfnamefont {Y.-T.}\ \bibnamefont {Cheng}}, \bibinfo {author} {\bibfnamefont {K.-T.}\ \bibnamefont {Lin}}, \bibinfo {author} {\bibfnamefont {F.}~\bibnamefont {Aziz}}, \bibinfo {author} {\bibfnamefont {J.-C.}\ \bibnamefont {Liu}}, \bibinfo {author} {\bibfnamefont {K.-V.}\ \bibnamefont {Rangdhol}}, \bibinfo {author} {\bibfnamefont {Y.-Y.}\ \bibnamefont {Yeung}}, \bibinfo {author} {\bibfnamefont {S.}~\bibnamefont {Yang}}, \bibinfo {author} {\bibfnamefont {Q.}~\bibnamefont {Shao}}, \bibinfo {author} {\bibfnamefont {X.}~\bibnamefont {Wang}}, \bibinfo {author} {\bibfnamefont {G.-D.}\ \bibnamefont {Lin}}, \bibinfo {author} {\bibfnamefont {F.}~\bibnamefont {Nori}},\ and\ \bibinfo {author} {\bibfnamefont {I.-C.}\ \bibnamefont {Hoi}},\ }\href@noop {} {\bibinfo {title} {{Microwave interference from a spin ensemble and its mirror image in waveguide magnonics}}} (\bibinfo {year} {2024}),\ \Eprint
  {https://arxiv.org/abs/2409.17867} {arXiv:2409.17867} \BibitemShut {NoStop}%
\bibitem [{\citenamefont {Song}\ \emph {et~al.}(2019)\citenamefont {Song}, \citenamefont {Xu}, \citenamefont {Li}, \citenamefont {Zhang}, \citenamefont {Zhang}, \citenamefont {Liu}, \citenamefont {Guo}, \citenamefont {Wang}, \citenamefont {Ren}, \citenamefont {Hao}, \citenamefont {Feng}, \citenamefont {Fan}, \citenamefont {Zheng}, \citenamefont {Wang}, \citenamefont {Wang},\ and\ \citenamefont {Zhu}}]{songchaosci2019}%
  \BibitemOpen
  \bibfield  {author} {\bibinfo {author} {\bibfnamefont {C.}~\bibnamefont {Song}}, \bibinfo {author} {\bibfnamefont {K.}~\bibnamefont {Xu}}, \bibinfo {author} {\bibfnamefont {H.}~\bibnamefont {Li}}, \bibinfo {author} {\bibfnamefont {Y.-R.}\ \bibnamefont {Zhang}}, \bibinfo {author} {\bibfnamefont {X.}~\bibnamefont {Zhang}}, \bibinfo {author} {\bibfnamefont {W.}~\bibnamefont {Liu}}, \bibinfo {author} {\bibfnamefont {Q.}~\bibnamefont {Guo}}, \bibinfo {author} {\bibfnamefont {Z.}~\bibnamefont {Wang}}, \bibinfo {author} {\bibfnamefont {W.}~\bibnamefont {Ren}}, \bibinfo {author} {\bibfnamefont {J.}~\bibnamefont {Hao}}, \bibinfo {author} {\bibfnamefont {H.}~\bibnamefont {Feng}}, \bibinfo {author} {\bibfnamefont {H.}~\bibnamefont {Fan}}, \bibinfo {author} {\bibfnamefont {D.}~\bibnamefont {Zheng}}, \bibinfo {author} {\bibfnamefont {D.-W.}\ \bibnamefont {Wang}}, \bibinfo {author} {\bibfnamefont {H.}~\bibnamefont {Wang}},\ and\ \bibinfo {author} {\bibfnamefont {S.-Y.}\ \bibnamefont {Zhu}},\ }\bibfield  {title} {\bibinfo
  {title} {{Generation of multicomponent atomic {Schr$\ddot{\rm{o}}$dinger} cat states of up to 20 qubits}},\ }\href {https://doi.org/10.1126/science.aay0600} {\bibfield  {journal} {\bibinfo  {journal} {Science}\ }\textbf {\bibinfo {volume} {365}},\ \bibinfo {pages} {574} (\bibinfo {year} {2019})}\BibitemShut {NoStop}%
\bibitem [{\citenamefont {Stassi}\ \emph {et~al.}(2020)\citenamefont {Stassi}, \citenamefont {Cirio},\ and\ \citenamefont {Nori}}]{stassi2020scalable}%
  \BibitemOpen
  \bibfield  {author} {\bibinfo {author} {\bibfnamefont {R.}~\bibnamefont {Stassi}}, \bibinfo {author} {\bibfnamefont {M.}~\bibnamefont {Cirio}},\ and\ \bibinfo {author} {\bibfnamefont {F.}~\bibnamefont {Nori}},\ }\bibfield  {title} {\bibinfo {title} {Scalable quantum computer with superconducting circuits in the ultrastrong coupling regime},\ }\href {https://doi.org/10.1038/s41534-020-00294-x} {\bibfield  {journal} {\bibinfo  {journal} {npj Quantum Information}\ }\textbf {\bibinfo {volume} {6}},\ \bibinfo {pages} {67} (\bibinfo {year} {2020})}\BibitemShut {NoStop}%
\bibitem [{\citenamefont {Bhattacharjee}\ \emph {et~al.}(2025)\citenamefont {Bhattacharjee}, \citenamefont {Jain}, \citenamefont {Deshmukh}, \citenamefont {Das}, \citenamefont {Chand}, \citenamefont {Patankar},\ and\ \citenamefont {Vijay}}]{bhattacharjee2025}%
  \BibitemOpen
  \bibfield  {author} {\bibinfo {author} {\bibfnamefont {A.}~\bibnamefont {Bhattacharjee}}, \bibinfo {author} {\bibfnamefont {P.}~\bibnamefont {Jain}}, \bibinfo {author} {\bibfnamefont {J.}~\bibnamefont {Deshmukh}}, \bibinfo {author} {\bibfnamefont {S.}~\bibnamefont {Das}}, \bibinfo {author} {\bibfnamefont {M.}~\bibnamefont {Chand}}, \bibinfo {author} {\bibfnamefont {M.~P.}\ \bibnamefont {Patankar}},\ and\ \bibinfo {author} {\bibfnamefont {R.}~\bibnamefont {Vijay}},\ }\bibfield  {title} {\bibinfo {title} {{Demonstration of two qubit entangling gates in a 2D ring resonator based coupler architecture}},\ }\href {https://doi.org/10.1038/s41598-025-87410-z} {\bibfield  {journal} {\bibinfo  {journal} {Scientific Reports}\ }\textbf {\bibinfo {volume} {15}},\ \bibinfo {pages} {4426} (\bibinfo {year} {2025})}\BibitemShut {NoStop}%
\bibitem [{\citenamefont {Arute}\ \emph {et~al.}(2019)\citenamefont {Arute} \emph {et~al.}}]{Arute2019}%
  \BibitemOpen
  \bibfield  {author} {\bibinfo {author} {\bibfnamefont {F.}~\bibnamefont {Arute}} \emph {et~al.},\ }\bibfield  {title} {\bibinfo {title} {{Quantum supremacy using a programmable superconducting processor}},\ }\href {https://doi.org/10.1038/s41586-019-1666-5} {\bibfield  {journal} {\bibinfo  {journal} {Nature}\ }\textbf {\bibinfo {volume} {574}},\ \bibinfo {pages} {505} (\bibinfo {year} {2019})}\BibitemShut {NoStop}%
\bibitem [{\citenamefont {Mirhosseini}\ \emph {et~al.}(2019)\citenamefont {Mirhosseini}, \citenamefont {Kim}, \citenamefont {Zhang}, \citenamefont {Sipahigil}, \citenamefont {Dieterle}, \citenamefont {Keller}, \citenamefont {Asenjo-Garcia}, \citenamefont {Chang},\ and\ \citenamefont {Painter}}]{Mir2019nat}%
  \BibitemOpen
  \bibfield  {author} {\bibinfo {author} {\bibfnamefont {M.}~\bibnamefont {Mirhosseini}}, \bibinfo {author} {\bibfnamefont {E.}~\bibnamefont {Kim}}, \bibinfo {author} {\bibfnamefont {X.}~\bibnamefont {Zhang}}, \bibinfo {author} {\bibfnamefont {A.}~\bibnamefont {Sipahigil}}, \bibinfo {author} {\bibfnamefont {P.~B.}\ \bibnamefont {Dieterle}}, \bibinfo {author} {\bibfnamefont {A.~J.}\ \bibnamefont {Keller}}, \bibinfo {author} {\bibfnamefont {A.}~\bibnamefont {Asenjo-Garcia}}, \bibinfo {author} {\bibfnamefont {D.~E.}\ \bibnamefont {Chang}},\ and\ \bibinfo {author} {\bibfnamefont {O.}~\bibnamefont {Painter}},\ }\bibfield  {title} {\bibinfo {title} {{Cavity quantum electrodynamics with atom-like mirrors}},\ }\href {https://doi.org/10.1038/s41586-019-1196-1} {\bibfield  {journal} {\bibinfo  {journal} {Nature}\ }\textbf {\bibinfo {volume} {569}},\ \bibinfo {pages} {692} (\bibinfo {year} {2019})}\BibitemShut {NoStop}%
\bibitem [{\citenamefont {Wen}\ \emph {et~al.}(2019)\citenamefont {Wen}, \citenamefont {Lin}, \citenamefont {Kockum}, \citenamefont {Suri}, \citenamefont {Ian}, \citenamefont {Chen}, \citenamefont {Mao}, \citenamefont {Chiu}, \citenamefont {Delsing}, \citenamefont {Nori}, \citenamefont {Lin},\ and\ \citenamefont {Hoi}}]{Wen2019}%
  \BibitemOpen
  \bibfield  {author} {\bibinfo {author} {\bibfnamefont {P.~Y.}\ \bibnamefont {Wen}}, \bibinfo {author} {\bibfnamefont {K.-T.}\ \bibnamefont {Lin}}, \bibinfo {author} {\bibfnamefont {A.~F.}\ \bibnamefont {Kockum}}, \bibinfo {author} {\bibfnamefont {B.}~\bibnamefont {Suri}}, \bibinfo {author} {\bibfnamefont {H.}~\bibnamefont {Ian}}, \bibinfo {author} {\bibfnamefont {J.~C.}\ \bibnamefont {Chen}}, \bibinfo {author} {\bibfnamefont {S.~Y.}\ \bibnamefont {Mao}}, \bibinfo {author} {\bibfnamefont {C.~C.}\ \bibnamefont {Chiu}}, \bibinfo {author} {\bibfnamefont {P.}~\bibnamefont {Delsing}}, \bibinfo {author} {\bibfnamefont {F.}~\bibnamefont {Nori}}, \bibinfo {author} {\bibfnamefont {G.-D.}\ \bibnamefont {Lin}},\ and\ \bibinfo {author} {\bibfnamefont {I.-C.}\ \bibnamefont {Hoi}},\ }\bibfield  {title} {\bibinfo {title} {{Large Collective Lamb Shift of Two Distant Superconducting Artificial Atoms}},\ }\href {https://doi.org/10.1103/PhysRevLett.123.233602} {\bibfield  {journal} {\bibinfo  {journal} {Physical Review
  Letters}\ }\textbf {\bibinfo {volume} {123}},\ \bibinfo {pages} {233602} (\bibinfo {year} {2019})}\BibitemShut {NoStop}%
\bibitem [{\citenamefont {Lin}\ \emph {et~al.}(2019)\citenamefont {Lin}, \citenamefont {Hsu}, \citenamefont {Lee}, \citenamefont {Hoi},\ and\ \citenamefont {Lin}}]{Lin2019}%
  \BibitemOpen
  \bibfield  {author} {\bibinfo {author} {\bibfnamefont {K.-T.}\ \bibnamefont {Lin}}, \bibinfo {author} {\bibfnamefont {T.}~\bibnamefont {Hsu}}, \bibinfo {author} {\bibfnamefont {C.-Y.}\ \bibnamefont {Lee}}, \bibinfo {author} {\bibfnamefont {I.-C.}\ \bibnamefont {Hoi}},\ and\ \bibinfo {author} {\bibfnamefont {G.-D.}\ \bibnamefont {Lin}},\ }\bibfield  {title} {\bibinfo {title} {{Scalable collective Lamb shift of a 1D superconducting qubit array in front of a mirror}},\ }\href {https://doi.org/10.1038/s41598-019-55545-5} {\bibfield  {journal} {\bibinfo  {journal} {Scientific Reports}\ }\textbf {\bibinfo {volume} {9}},\ \bibinfo {pages} {19175} (\bibinfo {year} {2019})}\BibitemShut {NoStop}%
\bibitem [{\citenamefont {Li}\ \emph {et~al.}(2022)\citenamefont {Li}, \citenamefont {Yefremenko}, \citenamefont {Lisovenko}, \citenamefont {Trevillian}, \citenamefont {Polakovic}, \citenamefont {Cecil}, \citenamefont {Barry}, \citenamefont {Pearson}, \citenamefont {Divan}, \citenamefont {Tyberkevych}, \citenamefont {Chang}, \citenamefont {Welp}, \citenamefont {Kwok},\ and\ \citenamefont {Novosad}}]{li2022coherent}%
  \BibitemOpen
  \bibfield  {author} {\bibinfo {author} {\bibfnamefont {Y.}~\bibnamefont {Li}}, \bibinfo {author} {\bibfnamefont {V.~G.}\ \bibnamefont {Yefremenko}}, \bibinfo {author} {\bibfnamefont {M.}~\bibnamefont {Lisovenko}}, \bibinfo {author} {\bibfnamefont {C.}~\bibnamefont {Trevillian}}, \bibinfo {author} {\bibfnamefont {T.}~\bibnamefont {Polakovic}}, \bibinfo {author} {\bibfnamefont {T.~W.}\ \bibnamefont {Cecil}}, \bibinfo {author} {\bibfnamefont {P.~S.}\ \bibnamefont {Barry}}, \bibinfo {author} {\bibfnamefont {J.}~\bibnamefont {Pearson}}, \bibinfo {author} {\bibfnamefont {R.}~\bibnamefont {Divan}}, \bibinfo {author} {\bibfnamefont {V.}~\bibnamefont {Tyberkevych}}, \bibinfo {author} {\bibfnamefont {C.~L.}\ \bibnamefont {Chang}}, \bibinfo {author} {\bibfnamefont {U.}~\bibnamefont {Welp}}, \bibinfo {author} {\bibfnamefont {W.-K.}\ \bibnamefont {Kwok}},\ and\ \bibinfo {author} {\bibfnamefont {V.}~\bibnamefont {Novosad}},\ }\bibfield  {title} {\bibinfo {title} {Coherent coupling of two remote magnonic resonators
  mediated by superconducting circuits},\ }\href {https://doi.org/10.1103/PhysRevLett.128.047701} {\bibfield  {journal} {\bibinfo  {journal} {Physical Review Letters}\ }\textbf {\bibinfo {volume} {128}},\ \bibinfo {pages} {047701} (\bibinfo {year} {2022})}\BibitemShut {NoStop}%
\bibitem [{\citenamefont {Kockum}\ \emph {et~al.}(2018)\citenamefont {Kockum}, \citenamefont {Johansson},\ and\ \citenamefont {Nori}}]{kockprl2018}%
  \BibitemOpen
  \bibfield  {author} {\bibinfo {author} {\bibfnamefont {A.~F.}\ \bibnamefont {Kockum}}, \bibinfo {author} {\bibfnamefont {G.}~\bibnamefont {Johansson}},\ and\ \bibinfo {author} {\bibfnamefont {F.}~\bibnamefont {Nori}},\ }\bibfield  {title} {\bibinfo {title} {{Decoherence-Free Interaction between Giant Atoms in Waveguide Quantum Electrodynamics}},\ }\href {https://doi.org/10.1103/PhysRevLett.120.140404} {\bibfield  {journal} {\bibinfo  {journal} {Physical Review Letters}\ }\textbf {\bibinfo {volume} {120}},\ \bibinfo {pages} {140404} (\bibinfo {year} {2018})}\BibitemShut {NoStop}%
\bibitem [{\citenamefont {Kannan}\ \emph {et~al.}(2020)\citenamefont {Kannan}, \citenamefont {Ruckriegel}, \citenamefont {Campbell}, \citenamefont {Frisk~Kockum}, \citenamefont {Braum{\"u}ller}, \citenamefont {Kim}, \citenamefont {Kjaergaard}, \citenamefont {Krantz}, \citenamefont {Melville}, \citenamefont {Niedzielski}, \citenamefont {Veps{\"a}l{\"a}inen}, \citenamefont {Winik}, \citenamefont {Yoder}, \citenamefont {Nori}, \citenamefont {Orlando}, \citenamefont {Gustavsson},\ and\ \citenamefont {Oliver}}]{Kannan2020}%
  \BibitemOpen
  \bibfield  {author} {\bibinfo {author} {\bibfnamefont {B.}~\bibnamefont {Kannan}}, \bibinfo {author} {\bibfnamefont {M.~J.}\ \bibnamefont {Ruckriegel}}, \bibinfo {author} {\bibfnamefont {D.~L.}\ \bibnamefont {Campbell}}, \bibinfo {author} {\bibfnamefont {A.}~\bibnamefont {Frisk~Kockum}}, \bibinfo {author} {\bibfnamefont {J.}~\bibnamefont {Braum{\"u}ller}}, \bibinfo {author} {\bibfnamefont {D.~K.}\ \bibnamefont {Kim}}, \bibinfo {author} {\bibfnamefont {M.}~\bibnamefont {Kjaergaard}}, \bibinfo {author} {\bibfnamefont {P.}~\bibnamefont {Krantz}}, \bibinfo {author} {\bibfnamefont {A.}~\bibnamefont {Melville}}, \bibinfo {author} {\bibfnamefont {B.~M.}\ \bibnamefont {Niedzielski}}, \bibinfo {author} {\bibfnamefont {A.}~\bibnamefont {Veps{\"a}l{\"a}inen}}, \bibinfo {author} {\bibfnamefont {R.}~\bibnamefont {Winik}}, \bibinfo {author} {\bibfnamefont {J.~L.}\ \bibnamefont {Yoder}}, \bibinfo {author} {\bibfnamefont {F.}~\bibnamefont {Nori}}, \bibinfo {author} {\bibfnamefont {T.~P.}\ \bibnamefont {Orlando}}, \bibinfo
  {author} {\bibfnamefont {S.}~\bibnamefont {Gustavsson}},\ and\ \bibinfo {author} {\bibfnamefont {W.~D.}\ \bibnamefont {Oliver}},\ }\bibfield  {title} {\bibinfo {title} {{Waveguide quantum electrodynamics with superconducting artificial giant atoms}},\ }\href {https://doi.org/10.1038/s41586-020-2529-9} {\bibfield  {journal} {\bibinfo  {journal} {Nature}\ }\textbf {\bibinfo {volume} {583}},\ \bibinfo {pages} {775} (\bibinfo {year} {2020})}\BibitemShut {NoStop}%
\bibitem [{\citenamefont {Almanakly}\ \emph {et~al.}(2025)\citenamefont {Almanakly}, \citenamefont {Yankelevich}, \citenamefont {Hays}, \citenamefont {Kannan}, \citenamefont {Assouly}, \citenamefont {Greene}, \citenamefont {Gingras}, \citenamefont {Niedzielski}, \citenamefont {Stickler}, \citenamefont {Schwartz}, \citenamefont {Serniak}, \citenamefont {Wang}, \citenamefont {Orlando}, \citenamefont {Gustavsson}, \citenamefont {Grover},\ and\ \citenamefont {Oliver}}]{Almanakly2025}%
  \BibitemOpen
  \bibfield  {author} {\bibinfo {author} {\bibfnamefont {A.}~\bibnamefont {Almanakly}}, \bibinfo {author} {\bibfnamefont {B.}~\bibnamefont {Yankelevich}}, \bibinfo {author} {\bibfnamefont {M.}~\bibnamefont {Hays}}, \bibinfo {author} {\bibfnamefont {B.}~\bibnamefont {Kannan}}, \bibinfo {author} {\bibfnamefont {R.}~\bibnamefont {Assouly}}, \bibinfo {author} {\bibfnamefont {A.}~\bibnamefont {Greene}}, \bibinfo {author} {\bibfnamefont {M.}~\bibnamefont {Gingras}}, \bibinfo {author} {\bibfnamefont {B.~M.}\ \bibnamefont {Niedzielski}}, \bibinfo {author} {\bibfnamefont {H.}~\bibnamefont {Stickler}}, \bibinfo {author} {\bibfnamefont {M.~E.}\ \bibnamefont {Schwartz}}, \bibinfo {author} {\bibfnamefont {K.}~\bibnamefont {Serniak}}, \bibinfo {author} {\bibfnamefont {J.~{\^{I}}.-j.}\ \bibnamefont {Wang}}, \bibinfo {author} {\bibfnamefont {T.~P.}\ \bibnamefont {Orlando}}, \bibinfo {author} {\bibfnamefont {S.}~\bibnamefont {Gustavsson}}, \bibinfo {author} {\bibfnamefont {J.~A.}\ \bibnamefont {Grover}},\ and\ \bibinfo
  {author} {\bibfnamefont {W.~D.}\ \bibnamefont {Oliver}},\ }\bibfield  {title} {\bibinfo {title} {{Deterministic remote entanglement using a chiral quantum interconnect}},\ }\href {https://doi.org/10.1038/s41567-025-02811-1} {\bibfield  {journal} {\bibinfo  {journal} {Nature Physics}\ }\textbf {\bibinfo {volume} {21}},\ \bibinfo {pages} {825} (\bibinfo {year} {2025})}\BibitemShut {NoStop}%
\bibitem [{\citenamefont {Karg}\ \emph {et~al.}(2019)\citenamefont {Karg}, \citenamefont {Gouraud}, \citenamefont {Treutlein},\ and\ \citenamefont {Hammerer}}]{karg2019}%
  \BibitemOpen
  \bibfield  {author} {\bibinfo {author} {\bibfnamefont {T.~M.}\ \bibnamefont {Karg}}, \bibinfo {author} {\bibfnamefont {B.}~\bibnamefont {Gouraud}}, \bibinfo {author} {\bibfnamefont {P.}~\bibnamefont {Treutlein}},\ and\ \bibinfo {author} {\bibfnamefont {K.}~\bibnamefont {Hammerer}},\ }\bibfield  {title} {\bibinfo {title} {{Remote Hamiltonian interactions mediated by light}},\ }\href {https://doi.org/10.1103/PhysRevA.99.063829} {\bibfield  {journal} {\bibinfo  {journal} {Physical Review A}\ }\textbf {\bibinfo {volume} {99}},\ \bibinfo {pages} {063829} (\bibinfo {year} {2019})}\BibitemShut {NoStop}%
\bibitem [{\citenamefont {Karg}\ \emph {et~al.}(2020)\citenamefont {Karg}, \citenamefont {Gouraud}, \citenamefont {Ngai}, \citenamefont {Schmid}, \citenamefont {Hammerer},\ and\ \citenamefont {Treutlein}}]{Thomssci2020}%
  \BibitemOpen
  \bibfield  {author} {\bibinfo {author} {\bibfnamefont {T.~M.}\ \bibnamefont {Karg}}, \bibinfo {author} {\bibfnamefont {B.}~\bibnamefont {Gouraud}}, \bibinfo {author} {\bibfnamefont {C.~T.}\ \bibnamefont {Ngai}}, \bibinfo {author} {\bibfnamefont {G.-L.}\ \bibnamefont {Schmid}}, \bibinfo {author} {\bibfnamefont {K.}~\bibnamefont {Hammerer}},\ and\ \bibinfo {author} {\bibfnamefont {P.}~\bibnamefont {Treutlein}},\ }\bibfield  {title} {\bibinfo {title} {{Light-mediated strong coupling between a mechanical oscillator and atomic spins 1 meter apart}},\ }\href {https://doi.org/10.1126/science.abb0328} {\bibfield  {journal} {\bibinfo  {journal} {Science}\ }\textbf {\bibinfo {volume} {369}},\ \bibinfo {pages} {174} (\bibinfo {year} {2020})}\BibitemShut {NoStop}%
\bibitem [{\citenamefont {Zhong}\ \emph {et~al.}(2021)\citenamefont {Zhong}, \citenamefont {Chang}, \citenamefont {Bienfait}, \citenamefont {Dumur}, \citenamefont {Chou}, \citenamefont {Conner}, \citenamefont {Grebel}, \citenamefont {Povey}, \citenamefont {Yan}, \citenamefont {Schuster},\ and\ \citenamefont {Cleland}}]{zhong2021deterministic}%
  \BibitemOpen
  \bibfield  {author} {\bibinfo {author} {\bibfnamefont {Y.}~\bibnamefont {Zhong}}, \bibinfo {author} {\bibfnamefont {H.-S.}\ \bibnamefont {Chang}}, \bibinfo {author} {\bibfnamefont {A.}~\bibnamefont {Bienfait}}, \bibinfo {author} {\bibfnamefont {{\'E}.}~\bibnamefont {Dumur}}, \bibinfo {author} {\bibfnamefont {M.-H.}\ \bibnamefont {Chou}}, \bibinfo {author} {\bibfnamefont {C.~R.}\ \bibnamefont {Conner}}, \bibinfo {author} {\bibfnamefont {J.}~\bibnamefont {Grebel}}, \bibinfo {author} {\bibfnamefont {R.~G.}\ \bibnamefont {Povey}}, \bibinfo {author} {\bibfnamefont {H.}~\bibnamefont {Yan}}, \bibinfo {author} {\bibfnamefont {D.~I.}\ \bibnamefont {Schuster}},\ and\ \bibinfo {author} {\bibfnamefont {A.~N.}\ \bibnamefont {Cleland}},\ }\bibfield  {title} {\bibinfo {title} {Deterministic multi-qubit entanglement in a quantum network},\ }\href {https://doi.org/10.1038/s41586-021-03288-7} {\bibfield  {journal} {\bibinfo  {journal} {Nature}\ }\textbf {\bibinfo {volume} {590}},\ \bibinfo {pages} {571} (\bibinfo {year}
  {2021})}\BibitemShut {NoStop}%
\bibitem [{\citenamefont {Li}\ \emph {et~al.}(2021)\citenamefont {Li}, \citenamefont {Wang}, \citenamefont {Wu}, \citenamefont {Zhu},\ and\ \citenamefont {You}}]{Liprxq2021}%
  \BibitemOpen
  \bibfield  {author} {\bibinfo {author} {\bibfnamefont {J.}~\bibnamefont {Li}}, \bibinfo {author} {\bibfnamefont {Y.-P.}\ \bibnamefont {Wang}}, \bibinfo {author} {\bibfnamefont {W.-J.}\ \bibnamefont {Wu}}, \bibinfo {author} {\bibfnamefont {S.-Y.}\ \bibnamefont {Zhu}},\ and\ \bibinfo {author} {\bibfnamefont {J.}~\bibnamefont {You}},\ }\bibfield  {title} {\bibinfo {title} {{Quantum Network with Magnonic and Mechanical Nodes}},\ }\href {https://doi.org/10.1103/PRXQuantum.2.040344} {\bibfield  {journal} {\bibinfo  {journal} {PRX Quantum}\ }\textbf {\bibinfo {volume} {2}},\ \bibinfo {pages} {040344} (\bibinfo {year} {2021})}\BibitemShut {NoStop}%
\bibitem [{\citenamefont {Hollerith}\ \emph {et~al.}(2022)\citenamefont {Hollerith}, \citenamefont {Srakaew}, \citenamefont {Wei}, \citenamefont {Rubio-Abadal}, \citenamefont {Adler}, \citenamefont {Weckesser}, \citenamefont {Kruckenhauser}, \citenamefont {Walther}, \citenamefont {van Bijnen}, \citenamefont {Rui}, \citenamefont {Gross}, \citenamefont {Bloch},\ and\ \citenamefont {Zeiher}}]{HOLLER2022}%
  \BibitemOpen
  \bibfield  {author} {\bibinfo {author} {\bibfnamefont {S.}~\bibnamefont {Hollerith}}, \bibinfo {author} {\bibfnamefont {K.}~\bibnamefont {Srakaew}}, \bibinfo {author} {\bibfnamefont {D.}~\bibnamefont {Wei}}, \bibinfo {author} {\bibfnamefont {A.}~\bibnamefont {Rubio-Abadal}}, \bibinfo {author} {\bibfnamefont {D.}~\bibnamefont {Adler}}, \bibinfo {author} {\bibfnamefont {P.}~\bibnamefont {Weckesser}}, \bibinfo {author} {\bibfnamefont {A.}~\bibnamefont {Kruckenhauser}}, \bibinfo {author} {\bibfnamefont {V.}~\bibnamefont {Walther}}, \bibinfo {author} {\bibfnamefont {R.}~\bibnamefont {van Bijnen}}, \bibinfo {author} {\bibfnamefont {J.}~\bibnamefont {Rui}}, \bibinfo {author} {\bibfnamefont {C.}~\bibnamefont {Gross}}, \bibinfo {author} {\bibfnamefont {I.}~\bibnamefont {Bloch}},\ and\ \bibinfo {author} {\bibfnamefont {J.}~\bibnamefont {Zeiher}},\ }\bibfield  {title} {\bibinfo {title} {{Realizing Distance-Selective Interactions in a Rydberg-Dressed Atom Array}},\ }\href
  {https://doi.org/10.1103/PhysRevLett.128.113602} {\bibfield  {journal} {\bibinfo  {journal} {Physical Review Letters}\ }\textbf {\bibinfo {volume} {128}},\ \bibinfo {pages} {113602} (\bibinfo {year} {2022})}\BibitemShut {NoStop}%
\bibitem [{\citenamefont {Cheung}\ \emph {et~al.}(2024)\citenamefont {Cheung}, \citenamefont {Haller}, \citenamefont {Kononov}, \citenamefont {Ciaccia}, \citenamefont {Ungerer}, \citenamefont {Kanne}, \citenamefont {Nyg{\aa}rd}, \citenamefont {Winkel}, \citenamefont {Reisinger}, \citenamefont {Pop}, \citenamefont {Baumgartner},\ and\ \citenamefont {Sch\"onenberger}}]{cheung2024photon}%
  \BibitemOpen
  \bibfield  {author} {\bibinfo {author} {\bibfnamefont {L.}~\bibnamefont {Cheung}}, \bibinfo {author} {\bibfnamefont {R.}~\bibnamefont {Haller}}, \bibinfo {author} {\bibfnamefont {A.}~\bibnamefont {Kononov}}, \bibinfo {author} {\bibfnamefont {C.}~\bibnamefont {Ciaccia}}, \bibinfo {author} {\bibfnamefont {J.}~\bibnamefont {Ungerer}}, \bibinfo {author} {\bibfnamefont {T.}~\bibnamefont {Kanne}}, \bibinfo {author} {\bibfnamefont {J.}~\bibnamefont {Nyg{\aa}rd}}, \bibinfo {author} {\bibfnamefont {P.}~\bibnamefont {Winkel}}, \bibinfo {author} {\bibfnamefont {T.}~\bibnamefont {Reisinger}}, \bibinfo {author} {\bibfnamefont {I.}~\bibnamefont {Pop}}, \bibinfo {author} {\bibfnamefont {A.}~\bibnamefont {Baumgartner}},\ and\ \bibinfo {author} {\bibfnamefont {C.}~\bibnamefont {Sch\"onenberger}},\ }\bibfield  {title} {\bibinfo {title} {{Photon-mediated long-range coupling of two Andreev pair qubits}},\ }\href {https://doi.org/10.1038/s41567-024-02630-w} {\bibfield  {journal} {\bibinfo  {journal} {Nature Physics}\ }\textbf
  {\bibinfo {volume} {20}},\ \bibinfo {pages} {1793} (\bibinfo {year} {2024})}\BibitemShut {NoStop}%
\bibitem [{\citenamefont {Zhou}\ \emph {et~al.}(2023)\citenamefont {Zhou}, \citenamefont {Lu}, \citenamefont {Praquin}, \citenamefont {Chien}, \citenamefont {Kaufman}, \citenamefont {Cao}, \citenamefont {Xia}, \citenamefont {Mong}, \citenamefont {Pfaff}, \citenamefont {Pekker},\ and\ \citenamefont {Hatridge}}]{Zhou2023}%
  \BibitemOpen
  \bibfield  {author} {\bibinfo {author} {\bibfnamefont {C.}~\bibnamefont {Zhou}}, \bibinfo {author} {\bibfnamefont {P.}~\bibnamefont {Lu}}, \bibinfo {author} {\bibfnamefont {M.}~\bibnamefont {Praquin}}, \bibinfo {author} {\bibfnamefont {T.-C.}\ \bibnamefont {Chien}}, \bibinfo {author} {\bibfnamefont {R.}~\bibnamefont {Kaufman}}, \bibinfo {author} {\bibfnamefont {X.}~\bibnamefont {Cao}}, \bibinfo {author} {\bibfnamefont {M.}~\bibnamefont {Xia}}, \bibinfo {author} {\bibfnamefont {R.~S.~K.}\ \bibnamefont {Mong}}, \bibinfo {author} {\bibfnamefont {W.}~\bibnamefont {Pfaff}}, \bibinfo {author} {\bibfnamefont {D.}~\bibnamefont {Pekker}},\ and\ \bibinfo {author} {\bibfnamefont {M.}~\bibnamefont {Hatridge}},\ }\bibfield  {title} {\bibinfo {title} {{Realizing all-to-all couplings among detachable quantum modules using a microwave quantum state router}},\ }\href {https://doi.org/10.1038/s41534-023-00723-7} {\bibfield  {journal} {\bibinfo  {journal} {npj Quantum Information}\ }\textbf {\bibinfo {volume} {9}},\ \bibinfo
  {pages} {54} (\bibinfo {year} {2023})}\BibitemShut {NoStop}%
\bibitem [{\citenamefont {Rao}\ \emph {et~al.}(2023{\natexlab{b}})\citenamefont {Rao}, \citenamefont {Wang}, \citenamefont {Yao}, \citenamefont {Chen}, \citenamefont {Zhao},\ and\ \citenamefont {Lu}}]{RAO2023PRL}%
  \BibitemOpen
  \bibfield  {author} {\bibinfo {author} {\bibfnamefont {J.}~\bibnamefont {Rao}}, \bibinfo {author} {\bibfnamefont {C.~Y.}\ \bibnamefont {Wang}}, \bibinfo {author} {\bibfnamefont {B.}~\bibnamefont {Yao}}, \bibinfo {author} {\bibfnamefont {Z.~J.}\ \bibnamefont {Chen}}, \bibinfo {author} {\bibfnamefont {K.~X.}\ \bibnamefont {Zhao}},\ and\ \bibinfo {author} {\bibfnamefont {W.}~\bibnamefont {Lu}},\ }\bibfield  {title} {\bibinfo {title} {{Meterscale Strong Coupling between Magnons and Photons}},\ }\href {https://doi.org/10.1103/PhysRevLett.131.106702} {\bibfield  {journal} {\bibinfo  {journal} {Physical Review Letters}\ }\textbf {\bibinfo {volume} {131}},\ \bibinfo {pages} {106702} (\bibinfo {year} {2023}{\natexlab{b}})}\BibitemShut {NoStop}%
\bibitem [{\citenamefont {Yang}\ \emph {et~al.}(2024)\citenamefont {Yang}, \citenamefont {Yao}, \citenamefont {Xiao}, \citenamefont {Fong}, \citenamefont {Lau},\ and\ \citenamefont {Hu}}]{YANG2024PRL}%
  \BibitemOpen
  \bibfield  {author} {\bibinfo {author} {\bibfnamefont {Y.}~\bibnamefont {Yang}}, \bibinfo {author} {\bibfnamefont {J.}~\bibnamefont {Yao}}, \bibinfo {author} {\bibfnamefont {Y.}~\bibnamefont {Xiao}}, \bibinfo {author} {\bibfnamefont {P.-T.}\ \bibnamefont {Fong}}, \bibinfo {author} {\bibfnamefont {H.-K.}\ \bibnamefont {Lau}},\ and\ \bibinfo {author} {\bibfnamefont {C.-M.}\ \bibnamefont {Hu}},\ }\bibfield  {title} {\bibinfo {title} {{Anomalous Long-Distance Coherence in Critically Driven Cavity Magnonics}},\ }\href {https://doi.org/10.1103/PhysRevLett.132.206902} {\bibfield  {journal} {\bibinfo  {journal} {Physical Review Letters}\ }\textbf {\bibinfo {volume} {132}},\ \bibinfo {pages} {206902} (\bibinfo {year} {2024})}\BibitemShut {NoStop}%
\bibitem [{\citenamefont {Gilleo}\ and\ \citenamefont {Geller}(1958)}]{Gilleo1958}%
  \BibitemOpen
  \bibfield  {author} {\bibinfo {author} {\bibfnamefont {M.~A.}\ \bibnamefont {Gilleo}}\ and\ \bibinfo {author} {\bibfnamefont {S.}~\bibnamefont {Geller}},\ }\bibfield  {title} {\bibinfo {title} {{Magnetic and Crystallographic Properties of Substituted {Yttrium-Iron Garnet}, $\rm3{{Y}}_{2}{{O}}_{3}\ifmmode\cdot\else\textperiodcentered\fi{}x{{M}}_{2}{{O}}_{3}\ifmmode\cdot\else\textperiodcentered\fi{}(5\ensuremath{-}x){{Fe}}_{2}{{O}}_{3}$}},\ }\href {https://doi.org/10.1103/PhysRev.110.73} {\bibfield  {journal} {\bibinfo  {journal} {Physical Review}\ }\textbf {\bibinfo {volume} {110}},\ \bibinfo {pages} {73} (\bibinfo {year} {1958})}\BibitemShut {NoStop}%
\bibitem [{\citenamefont {Huebl}\ \emph {et~al.}(2013)\citenamefont {Huebl}, \citenamefont {Zollitsch}, \citenamefont {Lotze}, \citenamefont {Hocke}, \citenamefont {Greifenstein}, \citenamefont {Marx}, \citenamefont {Gross},\ and\ \citenamefont {Goennenwein}}]{Huebl2013}%
  \BibitemOpen
  \bibfield  {author} {\bibinfo {author} {\bibfnamefont {H.}~\bibnamefont {Huebl}}, \bibinfo {author} {\bibfnamefont {C.~W.}\ \bibnamefont {Zollitsch}}, \bibinfo {author} {\bibfnamefont {J.}~\bibnamefont {Lotze}}, \bibinfo {author} {\bibfnamefont {F.}~\bibnamefont {Hocke}}, \bibinfo {author} {\bibfnamefont {M.}~\bibnamefont {Greifenstein}}, \bibinfo {author} {\bibfnamefont {A.}~\bibnamefont {Marx}}, \bibinfo {author} {\bibfnamefont {R.}~\bibnamefont {Gross}},\ and\ \bibinfo {author} {\bibfnamefont {S.~T.~B.}\ \bibnamefont {Goennenwein}},\ }\bibfield  {title} {\bibinfo {title} {{High Cooperativity in Coupled Microwave Resonator Ferrimagnetic Insulator Hybrids}},\ }\href {https://doi.org/10.1103/PhysRevLett.111.127003} {\bibfield  {journal} {\bibinfo  {journal} {Physical Review Letters}\ }\textbf {\bibinfo {volume} {111}},\ \bibinfo {pages} {127003} (\bibinfo {year} {2013})}\BibitemShut {NoStop}%
\bibitem [{\citenamefont {Astafiev}\ \emph {et~al.}(2010)\citenamefont {Astafiev}, \citenamefont {Zagoskin}, \citenamefont {Abdumalikov}, \citenamefont {Pashkin}, \citenamefont {Yamamoto}, \citenamefont {Inomata}, \citenamefont {Nakamura},\ and\ \citenamefont {Tsai}}]{Astafiev2010}%
  \BibitemOpen
  \bibfield  {author} {\bibinfo {author} {\bibfnamefont {O.}~\bibnamefont {Astafiev}}, \bibinfo {author} {\bibfnamefont {A.~M.}\ \bibnamefont {Zagoskin}}, \bibinfo {author} {\bibfnamefont {A.~A.}\ \bibnamefont {Abdumalikov}}, \bibinfo {author} {\bibfnamefont {Y.~A.}\ \bibnamefont {Pashkin}}, \bibinfo {author} {\bibfnamefont {T.}~\bibnamefont {Yamamoto}}, \bibinfo {author} {\bibfnamefont {K.}~\bibnamefont {Inomata}}, \bibinfo {author} {\bibfnamefont {Y.}~\bibnamefont {Nakamura}},\ and\ \bibinfo {author} {\bibfnamefont {J.~S.}\ \bibnamefont {Tsai}},\ }\bibfield  {title} {\bibinfo {title} {{Resonance Fluorescence of a Single Artificial Atom}},\ }\href {https://doi.org/10.1126/science.1181918} {\bibfield  {journal} {\bibinfo  {journal} {Science}\ }\textbf {\bibinfo {volume} {327}},\ \bibinfo {pages} {840} (\bibinfo {year} {2010})}\BibitemShut {NoStop}%
\bibitem [{\citenamefont {You}\ and\ \citenamefont {Nori}(2011)}]{You2011}%
  \BibitemOpen
  \bibfield  {author} {\bibinfo {author} {\bibfnamefont {J.~Q.}\ \bibnamefont {You}}\ and\ \bibinfo {author} {\bibfnamefont {F.}~\bibnamefont {Nori}},\ }\bibfield  {title} {\bibinfo {title} {{Atomic physics and quantum optics using superconducting circuits}},\ }\href {https://doi.org/10.1038/nature10122} {\bibfield  {journal} {\bibinfo  {journal} {Nature}\ }\textbf {\bibinfo {volume} {474}},\ \bibinfo {pages} {589} (\bibinfo {year} {2011})}\BibitemShut {NoStop}%
\bibitem [{\citenamefont {Hoi}\ \emph {et~al.}(2011)\citenamefont {Hoi}, \citenamefont {Wilson}, \citenamefont {Johansson}, \citenamefont {Palomaki}, \citenamefont {Peropadre},\ and\ \citenamefont {Delsing}}]{Hoi2011}%
  \BibitemOpen
  \bibfield  {author} {\bibinfo {author} {\bibfnamefont {I.-C.}\ \bibnamefont {Hoi}}, \bibinfo {author} {\bibfnamefont {C.~M.}\ \bibnamefont {Wilson}}, \bibinfo {author} {\bibfnamefont {G.}~\bibnamefont {Johansson}}, \bibinfo {author} {\bibfnamefont {T.}~\bibnamefont {Palomaki}}, \bibinfo {author} {\bibfnamefont {B.}~\bibnamefont {Peropadre}},\ and\ \bibinfo {author} {\bibfnamefont {P.}~\bibnamefont {Delsing}},\ }\bibfield  {title} {\bibinfo {title} {{Demonstration of a Single-Photon Router in the Microwave Regime}},\ }\href {https://doi.org/10.1103/PhysRevLett.107.073601} {\bibfield  {journal} {\bibinfo  {journal} {Physical Review Letters}\ }\textbf {\bibinfo {volume} {107}},\ \bibinfo {pages} {073601} (\bibinfo {year} {2011})}\BibitemShut {NoStop}%
\bibitem [{\citenamefont {Hoi}\ \emph {et~al.}(2012)\citenamefont {Hoi}, \citenamefont {Palomaki}, \citenamefont {Lindkvist}, \citenamefont {Johansson}, \citenamefont {Delsing},\ and\ \citenamefont {Wilson}}]{Hoi2012}%
  \BibitemOpen
  \bibfield  {author} {\bibinfo {author} {\bibfnamefont {I.-C.}\ \bibnamefont {Hoi}}, \bibinfo {author} {\bibfnamefont {T.}~\bibnamefont {Palomaki}}, \bibinfo {author} {\bibfnamefont {J.}~\bibnamefont {Lindkvist}}, \bibinfo {author} {\bibfnamefont {G.}~\bibnamefont {Johansson}}, \bibinfo {author} {\bibfnamefont {P.}~\bibnamefont {Delsing}},\ and\ \bibinfo {author} {\bibfnamefont {C.~M.}\ \bibnamefont {Wilson}},\ }\bibfield  {title} {\bibinfo {title} {{Generation of Nonclassical Microwave States Using an Artificial Atom in 1D Open Space}},\ }\href {https://doi.org/10.1103/PhysRevLett.108.263601} {\bibfield  {journal} {\bibinfo  {journal} {Physical Review Letters}\ }\textbf {\bibinfo {volume} {108}},\ \bibinfo {pages} {263601} (\bibinfo {year} {2012})}\BibitemShut {NoStop}%
\bibitem [{\citenamefont {Hoi}\ \emph {et~al.}(2013)\citenamefont {Hoi}, \citenamefont {Kockum}, \citenamefont {Palomaki}, \citenamefont {Stace}, \citenamefont {Fan}, \citenamefont {Tornberg}, \citenamefont {Sathyamoorthy}, \citenamefont {Johansson}, \citenamefont {Delsing},\ and\ \citenamefont {Wilson}}]{Hoi2013}%
  \BibitemOpen
  \bibfield  {author} {\bibinfo {author} {\bibfnamefont {I.-C.}\ \bibnamefont {Hoi}}, \bibinfo {author} {\bibfnamefont {A.~F.}\ \bibnamefont {Kockum}}, \bibinfo {author} {\bibfnamefont {T.}~\bibnamefont {Palomaki}}, \bibinfo {author} {\bibfnamefont {T.~M.}\ \bibnamefont {Stace}}, \bibinfo {author} {\bibfnamefont {B.}~\bibnamefont {Fan}}, \bibinfo {author} {\bibfnamefont {L.}~\bibnamefont {Tornberg}}, \bibinfo {author} {\bibfnamefont {S.~R.}\ \bibnamefont {Sathyamoorthy}}, \bibinfo {author} {\bibfnamefont {G.}~\bibnamefont {Johansson}}, \bibinfo {author} {\bibfnamefont {P.}~\bibnamefont {Delsing}},\ and\ \bibinfo {author} {\bibfnamefont {C.~M.}\ \bibnamefont {Wilson}},\ }\bibfield  {title} {\bibinfo {title} {{Giant Cross-Kerr Effect for Propagating Microwaves Induced by an Artificial Atom}},\ }\href {https://doi.org/10.1103/PhysRevLett.111.053601} {\bibfield  {journal} {\bibinfo  {journal} {Physical Review Letters}\ }\textbf {\bibinfo {volume} {111}},\ \bibinfo {pages} {053601} (\bibinfo {year}
  {2013})}\BibitemShut {NoStop}%
\bibitem [{\citenamefont {Gu}\ \emph {et~al.}(2017)\citenamefont {Gu}, \citenamefont {Kockum}, \citenamefont {Miranowicz}, \citenamefont {Liu},\ and\ \citenamefont {Nori}}]{GU20171}%
  \BibitemOpen
  \bibfield  {author} {\bibinfo {author} {\bibfnamefont {X.}~\bibnamefont {Gu}}, \bibinfo {author} {\bibfnamefont {A.~F.}\ \bibnamefont {Kockum}}, \bibinfo {author} {\bibfnamefont {A.}~\bibnamefont {Miranowicz}}, \bibinfo {author} {\bibfnamefont {Y.-X.}\ \bibnamefont {Liu}},\ and\ \bibinfo {author} {\bibfnamefont {F.}~\bibnamefont {Nori}},\ }\bibfield  {title} {\bibinfo {title} {{Microwave photonics with superconducting quantum circuits}},\ }\href {https://doi.org/https://doi.org/10.1016/j.physrep.2017.10.002} {\bibfield  {journal} {\bibinfo  {journal} {Physics Reports}\ }\textbf {\bibinfo {volume} {718--719}},\ \bibinfo {pages} {1} (\bibinfo {year} {2017})}\BibitemShut {NoStop}%
\bibitem [{\citenamefont {Kockum}\ and\ \citenamefont {Nori}(2019)}]{Kockum_2019}%
  \BibitemOpen
  \bibfield  {author} {\bibinfo {author} {\bibfnamefont {A.~F.}\ \bibnamefont {Kockum}}\ and\ \bibinfo {author} {\bibfnamefont {F.}~\bibnamefont {Nori}},\ }\bibinfo {title} {{Quantum Bits with Josephson Junctions}},\ in\ \href {https://doi.org/10.1007/978-3-030-20726-7_17} {\emph {\bibinfo {booktitle} {{Fundamentals and Frontiers of the Josephson Effect}}}},\ \bibinfo {editor} {edited by\ \bibinfo {editor} {\bibfnamefont {F.}~\bibnamefont {Tafuri}}}\ (\bibinfo  {publisher} {Springer International Publishing},\ \bibinfo {year} {2019})\BibitemShut {NoStop}%
\bibitem [{\citenamefont {Wen}\ \emph {et~al.}(2018)\citenamefont {Wen}, \citenamefont {Kockum}, \citenamefont {Ian}, \citenamefont {Chen}, \citenamefont {Nori},\ and\ \citenamefont {Hoi}}]{Wen_PRL_2018}%
  \BibitemOpen
  \bibfield  {author} {\bibinfo {author} {\bibfnamefont {P.~Y.}\ \bibnamefont {Wen}}, \bibinfo {author} {\bibfnamefont {A.~F.}\ \bibnamefont {Kockum}}, \bibinfo {author} {\bibfnamefont {H.}~\bibnamefont {Ian}}, \bibinfo {author} {\bibfnamefont {J.~C.}\ \bibnamefont {Chen}}, \bibinfo {author} {\bibfnamefont {F.}~\bibnamefont {Nori}},\ and\ \bibinfo {author} {\bibfnamefont {I.-C.}\ \bibnamefont {Hoi}},\ }\bibfield  {title} {\bibinfo {title} {{Reflective Amplification without Population Inversion from a Strongly Driven Superconducting Qubit}},\ }\href {https://doi.org/10.1103/PhysRevLett.120.063603} {\bibfield  {journal} {\bibinfo  {journal} {Physical Review Letters}\ }\textbf {\bibinfo {volume} {120}},\ \bibinfo {pages} {063603} (\bibinfo {year} {2018})}\BibitemShut {NoStop}%
\bibitem [{Sup()}]{SupMat}%
  \BibitemOpen
  \href@noop {} {\bibinfo {title} {{See the Supplemental Material for the experimental setup, the theory of dipole-dipole interaction of YIG spheres in front of a mirror, all extracted and calculated parameters, additional experimental details regarding frequency-domain and time-domain measurements with corresponding simulation results, comparisons with other systems, and discussions on the downscaling and scalability of our device.}}}\BibitemShut {Stop}%
\bibitem [{\citenamefont {Hoi}\ \emph {et~al.}(2015)\citenamefont {Hoi}, \citenamefont {Kockum}, \citenamefont {Tornberg}, \citenamefont {Pourkabirian}, \citenamefont {Johansson}, \citenamefont {Delsing},\ and\ \citenamefont {Wilson}}]{Hoi2015}%
  \BibitemOpen
  \bibfield  {author} {\bibinfo {author} {\bibfnamefont {I.-C.}\ \bibnamefont {Hoi}}, \bibinfo {author} {\bibfnamefont {A.~F.}\ \bibnamefont {Kockum}}, \bibinfo {author} {\bibfnamefont {L.}~\bibnamefont {Tornberg}}, \bibinfo {author} {\bibfnamefont {A.}~\bibnamefont {Pourkabirian}}, \bibinfo {author} {\bibfnamefont {G.}~\bibnamefont {Johansson}}, \bibinfo {author} {\bibfnamefont {P.}~\bibnamefont {Delsing}},\ and\ \bibinfo {author} {\bibfnamefont {C.~M.}\ \bibnamefont {Wilson}},\ }\bibfield  {title} {\bibinfo {title} {{Probing the quantum vacuum with an artificial atom in front of a mirror}},\ }\href {https://doi.org/10.1038/nphys3484} {\bibfield  {journal} {\bibinfo  {journal} {Nature Physics}\ }\textbf {\bibinfo {volume} {11}},\ \bibinfo {pages} {1045} (\bibinfo {year} {2015})}\BibitemShut {NoStop}%
\end{thebibliography}%

\end{document}


\title{Supplementary Material for \\``Realizing on-demand all-to-all selective interactions between distant spin ensembles"}

\author{C.-X.~Run}
\thanks{These authors contributed equally}
\affiliation{Department of Physics, City University of Hong Kong, Kowloon, Hong Kong SAR, China}

\author{K.-T.~Lin}
\thanks{These authors contributed equally}
\affiliation{Trapped-Ion Quantum Computing Laboratory, Hon Hai Research Institute, Taipei 11492, Taiwan}

\author{K.-M.~Hsieh}
\thanks{These authors contributed equally}
\affiliation{Department of Physics, City University of Hong Kong, Kowloon, Hong Kong SAR, China} 

\author{B.-Y.~Wu}
\affiliation{Department of Physics, City University of Hong Kong, Kowloon, Hong Kong SAR, China}

\author{W.-M.~Zhou}
\affiliation{Department of Physics, City University of Hong Kong, Kowloon, Hong Kong SAR, China}

\author{G.-D.~Lin}
\affiliation{Trapped-Ion Quantum Computing Laboratory, Hon Hai Research Institute, Taipei 11492, Taiwan}
\affiliation{Department of Physics and Center for Quantum Science and Engineering, National Taiwan University, Taipei 10617, Taiwan}

\author{A. F.~Kockum}
\affiliation{Department of Microtechnology and Nanoscience, Chalmers University of Technology, 412 96 Gothenburg, Sweden}

\author{I.-C.~Hoi}
\email[e-mail:]{iochoi@cityu.edu.hk}
\affiliation{Department of Physics, City University of Hong Kong, Kowloon, Hong Kong SAR, China}

\date{\today}

\maketitle

\tableofcontents


\section{Experimental setup}

\begin{figure}
\includegraphics[width=0.7\textwidth]{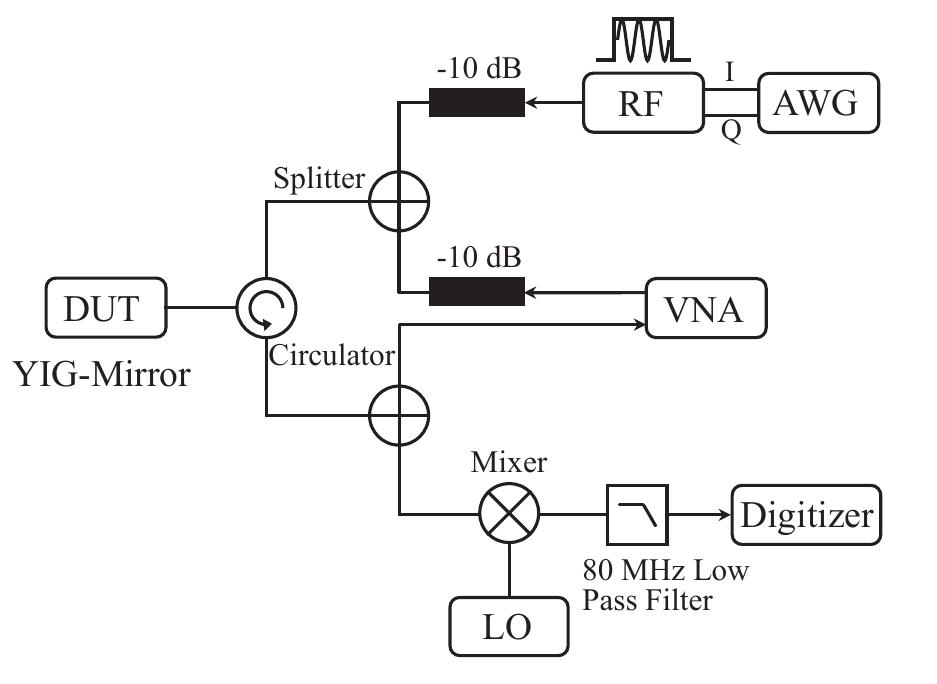}
\caption{Experimental setup for both frequency- and time-domain measurements. The device under test (DUT) here is four YIG spheres placed on the coplanar waveguide (CPW) shorted to ground, see Fig.~1(b) in the main text. The CPW is fabricated on a $\unit[14 \times 440]{mm^2}$ RO4003C board, with a center conductor width of \unit[1.5]{mm}, a gap of \unit[0.4]{mm} between the center conductor and the two ground planes, and a characteristic impedance of $Z_0 \approx \unit[50]{\Omega}$.}
\label{fig:S1}
\end{figure}

Figure~\ref{fig:S1} illustrates the experimental setup used for both frequency- and time-domain measurements. A vector network analyzer (VNA), employed for reflection spectroscopy, is connected in parallel with the time-domain setup, which consists of an arbitrary waveform generator (AWG), a radio-frequency (RF) source, and a digitizer. A coherent microwave square pulse is generated by the RF source, modulated in-phase and quadrature (IQ) by the AWG, and then transmitted into the system via an attenuator, a splitter, and a circulator. The probe power used on the yttrium iron garnet (YIG) spheres is approximately $\unit[-30]{dBm}$ to ensure that excitation of the Kittel mode (KM) remains within the linear regime. After interacting with the system, the reflected microwaves are down-converted by a mixer using a signal generated by a local oscillator (LO), and the encoded system information is recorded by the digitizer.


\section{Dipole-dipole interaction}

In this section, we summarize the derivation of the resonant dipole-dipole interaction (RDDI) in a system consisting of an array of $N$ YIG spheres coupled to a one-dimensional (1D) coplanar waveguide (CPW), which is terminated at one end by an antinode mirror~\cite{Wen2019, Wu2024}. 
All YIG spheres are positioned in front of the mirror, and their locations can be adjusted to enable all-to-all connectivity. 
The total Hamiltonian of the system is given by~\cite{Wang2022, Zhan2022}
\begin{equation}
H = H_m + H_b + V ,
\end{equation}
where $H_m$ and $H_b$ represent the magnonic and field energies, respectively, and are expressed as 
\begin{align}
H_m &= \hbar \sum_i \omega_{m_i} a_i^\dag a_i , \\
H_b &= \hbar \int_0^\infty d\omega \: \omega b_\omega^\dag b_\omega .
\end{align}
Here, the operator $a_i^\dag$ ($a_i$) denotes the creation (annihilation) operator of a magnon with energy $\hbar \omega_{m_i}$ for the $i$th YIG, and $b_\omega^\dag$ ($b_\omega$) creates (annihilates) a waveguide photon of frequency $\omega$. 

The interaction between the YIG spheres and the waveguide is described by
\begin{equation}
V = \hbar \sum_i \int_0^\infty d\omega \: g_i (\omega) \cos \mleft( k_\omega x_i \mright) a_i^\dag b_\omega + {\rm H.c.}, 
\label{eq: interaction H'}
\end{equation}
where $x_i$ is the position of the $i$th YIG sphere. The position-dependent coupling strength $g_i (\omega) \cos \mleft( k_\omega x_i \mright)$ reflects the boundary condition imposed by the antinode at $x = 0$. The wavenumber is given by $k_\omega = \omega / v$, with $v$ denoting the speed of light, and H.c.~denotes the Hermitian conjugate.

The equations of motion of the photonic and magnonic annihilation operators are
\begin{align}
\frac{d b_\omega}{dt} &= - i \omega b_\omega - i \sum_i g_i (\omega) \cos \mleft( k_\omega x_i \mright) a_i,
\label{eq: b dynamics}
\\
\frac{d a_i}{dt} &= - i \omega_{m_i} a_i - i \int_0^\infty d\omega \: g_i (\omega) \cos \mleft(k_\omega x_i \mright) b_\omega.
\label{eq: a dynamics}
\end{align}
By integrating \eqref{eq: b dynamics} and applying the Born--Markov approximation~\cite{Lehmberg1970}, we obtain
\begin{equation}
b_\omega (t) = b_\omega (0) e^{- i \omega t} - \sum_i \frac{g_i (\omega) \cos \mleft( k_\omega x_i \mright)}{\mleft( \omega - \omega_{m_i} \mright) - i \epsilon} a_i,
\label{eq: b solution}
\end{equation}
where a small positive quantity $\epsilon$ ensures convergence of the integral. Substituting \eqref{eq: b solution} into \eqref{eq: a dynamics}, we obtain
\begin{equation}
\frac{d a_i}{dt} = - i \omega_{m_i} a_i - i \sqrt{\kappa_i (m_i)} \cos \mleft( k_i x_i \mright) b_{\rm in} (t) - \sum_j F (x_i, x_j) a_j,
\label{eq: traced a dynamics}
\end{equation}
where $b_{\rm in} (t)$ represents the photonic input field~\cite{Meystre2021}, and $k_i = \omega_{m_i} / v$ is the wavenumber corresponding to the $i$th YIG sphere. The function $F (x_i, x_j)$ captures the collective interaction between YIG spheres; it is given by
\begin{equation}
F (x_i, x_j ) = \int_0^\infty d\omega \frac{g_i (\omega) g_j (\omega) \cos \mleft( k_\omega x_i \mright) \cos \mleft( k_\omega x_j \mright)}{\epsilon + i \mleft(\omega - \omega_{m_i} \mright)} \equiv \frac{\gamma_{ij}}{2} + i \Delta_{ij},
\label{eq: F function}
\end{equation}
where $\gamma_{ij}$ and $\Delta_{ij}$ denote collective dissipative and coherent coupling strengths, respectively~\cite{Lin2019}.
Using contour-integration techniques, we finally obtain
\begin{align}
\gamma_{ij} = \sqrt{\kappa_i (\omega_{m_i}) \kappa_j (\omega_{m_i})} \mleft[ \cos k_i \mleft( x_i + x_j \mright) + \cos k_i \mleft| x_i - x_j \mright| \mright],
\label{eq: collective decay}
\\
\Delta_{ij} = \frac{\sqrt{\kappa_i (\omega_{m_i}) \kappa_j (\omega_{m_i})}}{2} \mleft[ \sin k_i \mleft (x_i + x_j \mright) + \sin k_i \mleft | x_i - x_j \mright| \mright],
\label{eq: collective energy shift}
\end{align}
where the bare radiative damping rate of the $i$th YIG is given by
\begin{equation}
\kappa_i (\omega_{m_i}) = \pi g_i^2 (\omega_{m_i}) ,
\end{equation}
and the radiative damping rate $\kappa_r$ can be expressed as
\begin{equation}
\kappa_r = 2 \kappa_i \cos^2 \mleft( k_i x_i \mright).
\label{eq: kr}
\end{equation}

When YIG sphere $i$ is at an antinode ($k_i x_i = n_i \pi$, $n_i = 0, 1, 2, \ldots$) and YIG sphere $j$ is at a node ($k_j x_j  = n_j \pi + \frac{\pi}{2}$, $n_j = 0, 1, 2, \ldots$, $n_j \geq n_i$), the dissipative coupling vanishes ($\gamma_{ij}=0$) while the coherent interaction reaches its maximum, given by $\Delta_{ij} = \kappa_{ij} \equiv \sqrt{\kappa_i(\omega_{m_i})\kappa_j(\omega_{m_i})}$, resulting in purely coherent coupling between the two YIG spheres. 
Remarkably, when both YIG spheres are located at nodes, both $\gamma_{ij}$ and $\Delta_{ij}$ vanish ($\gamma_{ij}=\Delta_{ij}=0$), leading to a complete suppression of the magnon-magnon interaction, which cannot be achieved in the open-CPW configuration~\cite{Li2022}. 
This mirror-assisted configuration therefore offers a robust method for achieving full decoupling simply by positioning the YIG spheres at different nodes of the standing wave corresponding to the resonance frequency.


\section{Frequency-domain response}

To probe the system, we send a weak coherent probe signal with frequency $\omega_p$ into the shorted end of the waveguide, allowing it to interact with the YIG-sphere array, as illustrated in Fig.~1 of the main text. 
We then measure the reflection coefficient, defined as $|r|=\mleft| \mleft \langle b_{\rm out} \mright \rangle / \mleft \langle b_{\rm in} \mright \rangle \mright| $, where the output field $b_{\rm out}$ is determined by both the input signal and the collective response of the YIG spheres via the input-output relation~\cite{Gardiner1985, Lin2025}
\begin{equation}
b_{\rm out} (t) = b_{\rm in } (t) - i \sum_i \sqrt{\kappa_i (\omega_{m_i})} \cos \mleft( k_i x_i \mright) a_i (t).
\label{eq: input-output relation}
\end{equation}
Using Eqs.~(\ref{eq: traced a dynamics}) and (\ref{eq: input-output relation}), the reflection coefficient $|r|$ in the rotating frame of the probe frequency $\omega_p$ is given by
\begin{equation}
|r| = \mleft| 1 + c^T M^{-1} c \mright|,
\label{eq: reflection r}
\end{equation}
with the vector
\begin{equation}
c^T = \begin{bmatrix}
\sqrt{\kappa_1 (\omega_{m_1})} \cos \mleft(k_1 x_1 \mright), & 
\sqrt{\kappa_2 (\omega_{m_2})} \cos \mleft( k_2 x_2 \mright), & 
\cdots, & 
\sqrt{\kappa_N (\omega_{m_3})} \cos \mleft( k_N x_N \mright)
\end{bmatrix}
\end{equation}
and the coefficient matrix
\begin{equation}
M = \begin{bmatrix}
\frac{\gamma_{11}}{2} + \frac{\alpha_1}{2} + i \Delta_{11} - i \delta_1 & 
\frac{\gamma_{12}}{2} + i \Delta_{12} & 
\cdots & 
\frac{\gamma_{1N}}{2} + i \Delta_{1N} \\
\frac{\gamma_{21}}{2} + i \Delta_{21} & 
\frac{\gamma_{22}}{2} + \frac{\alpha_2}{2} + i \Delta_{22} - i \delta_2 & 
\ldots & 
\frac{\gamma_{2N}}{2} + i \Delta_{2N} \\
\vdots & \vdots & \ddots & \vdots \\
\frac{\gamma_{N1}}{2} + i \Delta_{N1} & 
\frac{\gamma_{N2}}{2} + i \Delta_{N2} & 
\cdots & 
\frac{\gamma_{NN}}{2} + \frac{\alpha_N}{2} + i \Delta_{NN} - i \delta_N
\end{bmatrix} .
\label{eq: coefficient matrix}
\end{equation}
Here $\delta_i = \omega_p - \omega_{m_i}$ is the detuning of the probe from the resonance frequency of the $i$th YIG sphere and $\alpha_i$, included manually, denotes the non-radiative decay rate of the $i$th YIG sphere~\cite{Koshino2012}.


\section{Time-domain dynamics}

To study temporal evolution, we excite the YIG array using a microwave square pulse of frequency $\omega_p$ and duration $t_0$.
The time-domain dynamics, using \eqref{eq: traced a dynamics}, are then governed by
\begin{align}
\frac{d \tilde{A}_i}{dt} = & - \mleft( \frac{\gamma_{ii} + \alpha_i}{2} \mright) \tilde{A}_i + i \mleft( \omega_p - \omega_{m_i} - \Delta_{ii} \mright) \tilde{A}_n - \sum_{j \neq i} \mleft( \frac{\gamma_{ij}}{2} + i \Delta_{ij} \mright) \tilde{A}_j \nonumber \\
& - i \sqrt{\kappa_i (\omega_{m_i})} \cos \mleft( k_i x_i \mright) \mleft( u (t) - u (t - t_0) \mright) ,
\label{eq: A dynamics}
\end{align}
where $\tilde{A}_i = \mleft\langle a_i \mright\rangle e^{- i \omega_p t} / \beta$ is the normalized magnon amplitude for the $i$th YIG sphere in the frame rotating at frequency $\omega_p$, $\beta$ is the input field amplitude, and $u (t)$ denotes the unit step function. By solving \eqref{eq: A dynamics}, the time evolution of the YIG-sphere array can be characterized through the normalized reflected power in log scale~\cite{Zhang2014}, defined as
\begin{equation}
\tilde{P} \propto 20 \log \mleft| \tilde{V}_{\rm out} \mright| ,
\label{eq: reflect power}
\end{equation}
where the normalized output voltage is given by
\begin{equation}
\mleft| \tilde{V}_{\rm out} \mright| \propto \mleft| \mleft( u (t) - u (t - t_0) \mright) - i \sum_i \sqrt{\kappa_i (\omega_{m_i})} \cos \mleft( k_i x_i \mright) \tilde{A}_i (t) \mright| . 
\label{eq: Vout}
\end{equation}
This framework captures the time-dependent exchange of excitations mediated by the dipole-dipole interactions and is essential for analyzing coherent dynamics within the array.


\section{All extracted parameters of the YIG spheres}

In the main text, we have two configurations of YIG-sphere arrangements, as illustrated in Fig.~2(a) and Fig.~3(a) there. The only difference between them is the location of YIG C, which in the second configuration is moved \unit[2]{cm} away from its original position in the direction opposite to the mirror. This section aims to characterize the parameters of all YIG spheres at specific positions and corresponding KM frequencies $\omega_m$.

\begin{table}
\centering
\begin{tabular}{|c|c|c|c|c|c|c|}
\hline
~ & $x$ & $\omega_m / 2\pi$ & $\kappa_r / 2\pi$ & $\Gamma / 2\pi$ & $\alpha / 2\pi$ & $v$  \\ \hline
~ & cm & GHz & MHz & MHz & MHz & $\unit[10^8]{m/s}$ \\ \hline
YIG A & 0 & 2.0926 (antinode) & 1.383 & 1.812 & 2.242 & 1.86 \\ \cline{1-6}
YIG A & 0 & 2.3244 (antinode) & 1.640 & 1.817 & 1.995 &  \\ \cline{1-6}
YIG B & 10 & 2.0142 & 8.877 & 5.427 & 1.976 & \\ \cline{1-6}
YIG B & 10 & 2.0928 & 3.871 & 3.058 & 2.245 & \\ \cline{1-6}
YIG B & 10 & 2.7900 (antinode) & 9.845 & 5.971 & 2.097 & \\ \cline{1-6}
YIG C & 20 & 2.3255 (antinode) & 12.779 & 7.715 & 2.651 & \\ \cline{1-6}
YIG C & 22 & 2.1138 (antinode) & 12.691 & 7.514 & 2.338 & \\ \cline{1-6}
YIG D & 30 & 2.1692 (antinode) & 13.982 & 7.827 & 1.671 & \\ \hline
\end{tabular}
\caption{Summary of extracted parameters of the YIG spheres. $x$ is the distance of the YIG spheres from the mirror. Note that due to their non-identical nature and spatial arrangement, different YIG spheres exhibit various $\kappa_r$.}
\label{tab:S1}
\end{table}

We first conduct reflection spectroscopy with a weak probe tone on each YIG sphere. When focusing on one YIG sphere, the remaining spheres are significantly detuned. The reflection coefficient $r$ is recorded by VNA as a function of both $\omega_p$ and $\omega_m$, which is modulated by the current $I$. In the steady state, the reflection coefficient is given by~\cite{Wu2024}
\begin{equation}
r = 1 - \frac{\kappa_r}{\Gamma - i (\omega_p - \omega_m)} ,
\label{eq:r}
\end{equation}
where $\Gamma = (\kappa_r + \alpha) / 2$ is the overall damping rate of the KM, with radiative damping rate $\kappa_r$ and non-radiative decay rate $\alpha$. The measured data are fitted by \eqref{eq:r} to extract the parameters as summarized in \tabref{tab:S1}. 
The microwave velocity $v$ within the CPW is extracted by finding multiple nodes of the field for YIG D following the method introduced in Ref.~\cite{Wu2024}.


\section{Detailed data for frequency-domain results}

\begin{table}
\centering
\begin{tabular}{|c|c|c|c|c|}
\hline
Pair & ~ & $\omega_m / 2\pi$ & $\kappa_i / 2\pi$ & $\alpha / 2\pi$ \\ \hline
~ & ~ & GHz & MHz & MHz \\ \hline
A\&B & YIG A & 2.3249 & 0.7307 & 2.453 \\ \hline
~ & YIG B & 2.3251 & 5.1276 & 1.5728 \\ \hline
A\&C & YIG A & 2.0939 & 0.6536 & 2.4681 \\ \hline
~ & YIG C & 2.0920 & 7.1867 & 3.2784 \\ \hline
A\&D & YIG A & 2.3242 & 0.6152 & 3.5922 \\ \hline
~ & YIG D & 2.3253 & 8.8241 & 2.0752 \\ \hline
B\&C & YIG B & 2.0932 & 3.2735 & 2.8632 \\ \hline
~ & YIG C & 2.0918 & 6.8204 & 3.0760 \\ \hline
B\&D & YIG B & 2.0138 & 4.698 & 3.0977 \\ \hline
~ & YIG D & 2.0148 & 5.8921 & 3.4641 \\ \hline
C\&D & YIG C & 2.3249 & 4.9116 & 4.2947 \\ \hline
~ & YIG D & 2.3250 & 9.3355 & 2.1120 \\ \hline
\end{tabular}
\caption{Parameters used for fitting all-to-all interaction between YIG pairs in \figref{fig:S2}. Note that $\kappa_i$ is the bare radiative damping rate, and the radiative damping rate is related to $\kappa_i$ through \eqref{eq: kr}.}
\label{tab:S2}
\end{table}

\begin{table}
\centering
\begin{tabular}{|c|c|c|c|}
\hline
~ & $\omega_m / 2\pi$ & $\kappa_i / 2\pi$ & $\alpha / 2\pi$ \\ \hline
~ & GHz & MHz & MHz  \\ \hline
YIG A & 2.325 & 0.8202 & 1.995 \\ \hline
YIG B & 2.325 & 4.1202 & 2.097 \\ \hline
YIG C & 2.325 & 6.9795 & 2.338 \\ \hline
YIG D & 2.325 & 7.4931 & 1.671 \\ \hline
\end{tabular}
\caption{Parameters used for fitting the interactions between multiple YIG spheres in Figs.~\ref{fig:S3} and \ref{fig:S5}.}
\label{tab:S3}
\end{table}


\subsection{Multi-sphere interactions}

This section presents simulation results for the all-to-all selective network shown in Fig.~2 of the main text, as well as for multiple YIG spheres configured as in Fig.~3(a), based on \eqref{eq: reflection r} and the fitting parameters summarized in \tabref{tab:S2} and \tabref{tab:S3}, respectively.

\begin{figure}
\includegraphics[width=\linewidth]{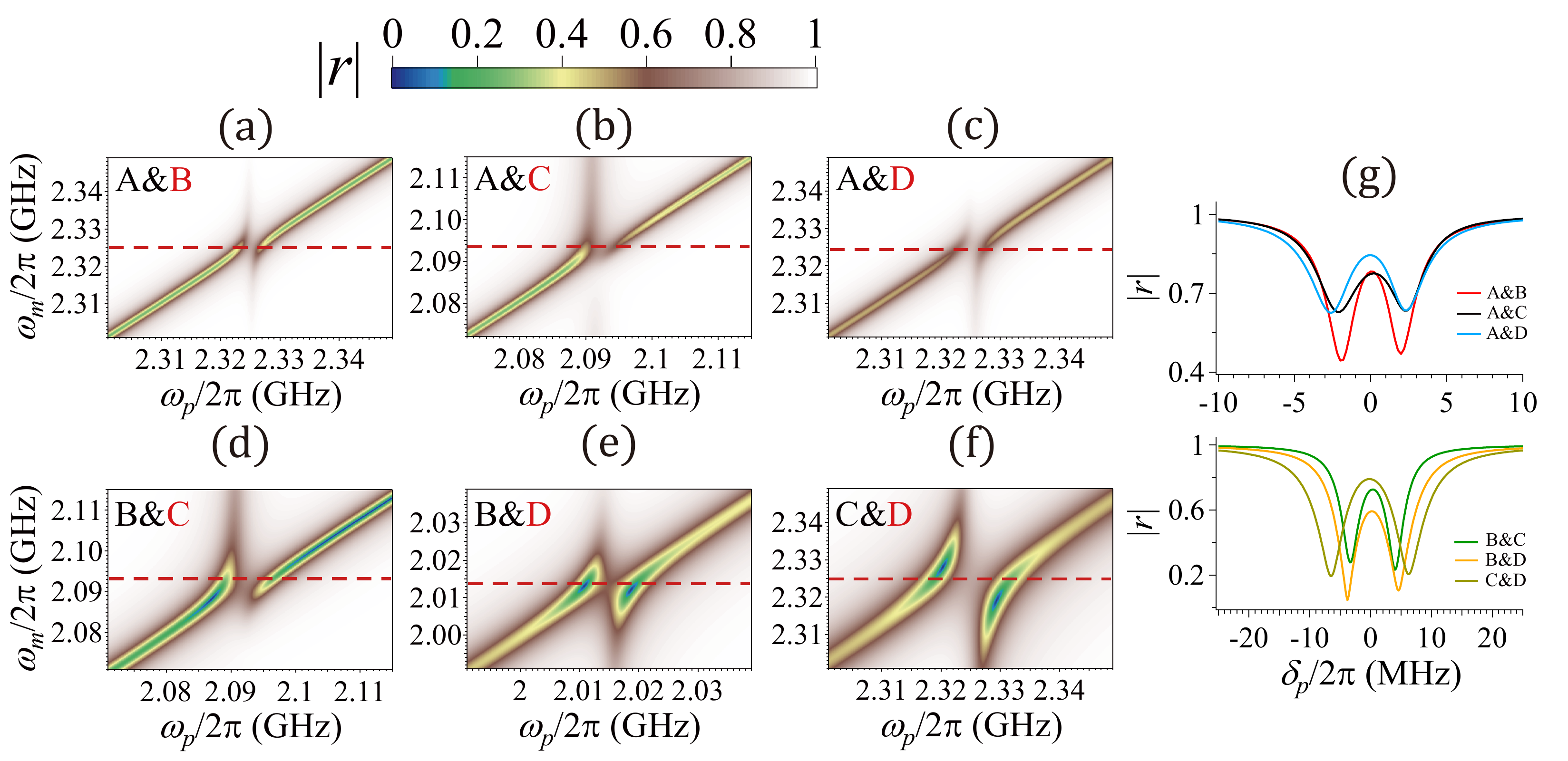}
\caption{Simulation results for selective all-to-all interaction between YIG-sphere pairs in Fig.~2 of the main text.
(a)--(f) Simulated amplitude reflection coefficient $|r|$ for a weak probe as a function of the probe frequency $\omega_p$ and the KM transition frequency $\omega_m$ of one YIG sphere in the pair. In each panel, $\omega_m$ of the YIG sphere located at the node is fixed (highlighted in red font), while $\omega_m$ of the other is tuned through resonance.
(g) Line cuts from panels (a)--(f), taken at resonance (dashed lines), show the interaction profiles for each YIG pair.}
\label{fig:S2}
\end{figure}

\begin{figure}
\includegraphics[width=0.85\linewidth]{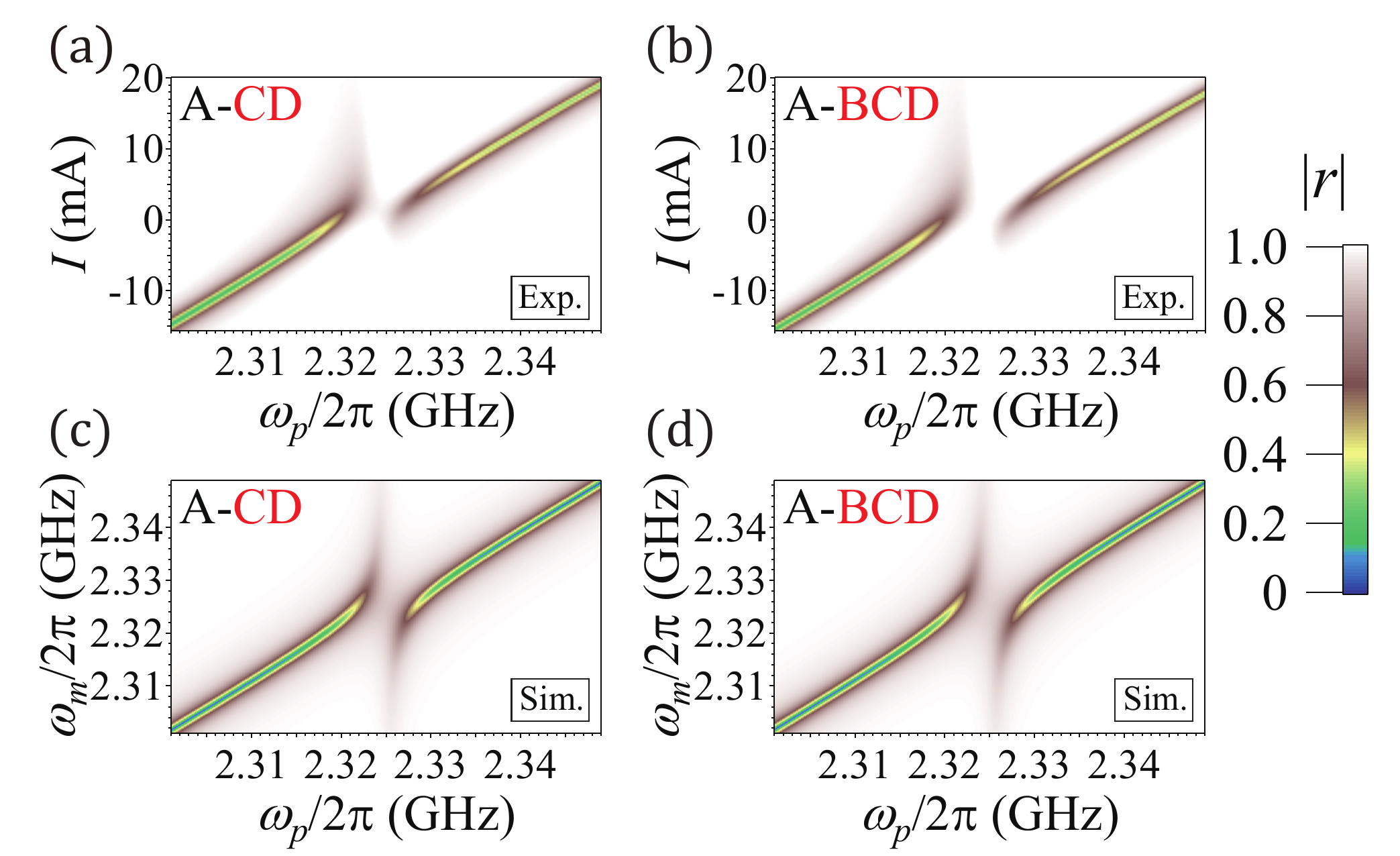}
\caption{Interaction between multiple YIG spheres. Amplitude reflection coefficient $|r|$ for a weak probe as a function of $\omega_p$ and $I$ ($\omega_m$) of YIG A.
(a) $\omega_m$ of YIG spheres C and D are fixed at \unit[2.325]{GHz}, and $\omega_m$ of YIG A is tuned through resonance with this value.
(b) $\omega_m$ of YIG spheres B, C, and D are fixed at \unit[2.325]{GHz}, and $\omega_m$ of YIG A is tuned through resonance with this value.
(c), (d) Theoretical simulations corresponding to (a) and (b), respectively, based on \eqref{eq: reflection r} and fitting parameters listed in \tabref{tab:S3}.}
\label{fig:S3}
\end{figure}

Figure~\ref{fig:S2} displays theoretical simulations of all-to-all pairwise interactions between YIG spheres A\&B, A\&C, A\&D, B\&C, B\&D, and C\&D, corresponding to Fig.~2 of the main text. In each panel, the $\omega_m$ of the YIG sphere positioned closer to the mirror is varied, while the $\omega_m$ of the other YIG sphere in the pair (marked in red font) is fixed at a node. 

In \figpanel{fig:S3}{a}, three YIG spheres, A, C, and D, are coupled by fixing the $\omega_m$ of YIG spheres C and D at \unit[2.325]{GHz}, while varying the $\omega_m$ of YIG A by sweeping the current $I$ to observe their interaction. Similarly, in \figpanel{fig:S3}{b}, four YIG spheres are coupled together by fixing the $\omega_m$ of YIG spheres B, C, and D at \unit[2.325]{GHz} and sweeping the $\omega_m$ of YIG A. As mentioned in the main text, when they are resonant at $\omega_r / 2\pi = \unit[2.325]{GHz}$, YIG spheres B, C, and D are positioned at the nodes, while YIG A consistently remains at the antinode of the microwave field within the CPW due to its position at the mirror. The corresponding simulations are shown in \figpanels{fig:S3}{c}{d}. 

Our observations indicate that the coupling strength $\rm{\Delta}$ is proportional to the number of YIG spheres located at the nodes, as shown in Fig.~3(d) of the main text.


\subsection{Selective coupling between distant YIG spheres}

\begin{figure}
\includegraphics[width=0.6\textwidth]{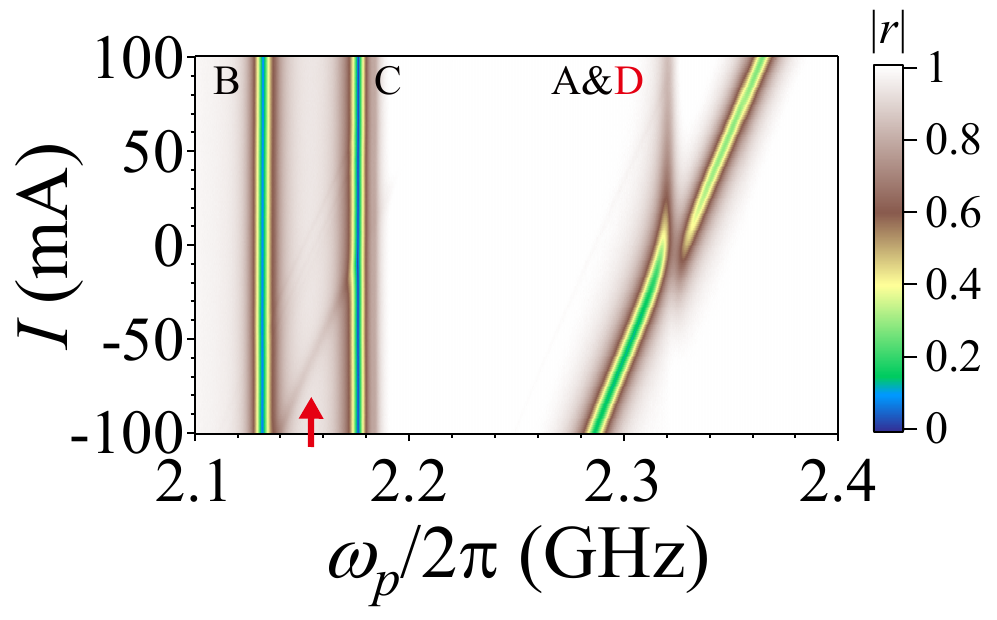}
\caption{Selective coupling between YIG A and YIG D. Only YIG A and D are tuned on resonance, while the remaining spheres are far detuned and act as spectators. A clear avoided crossing between YIG A and D is observed, indicating selective pairwise interaction. The red marker indicates another magnetostatic mode of YIG A.}
\label{fig:S4}
\end{figure}

To further illustrate selective control over pairwise interactions, we present in \figref{fig:S4} experimental result where only YIG A and YIG D are tuned into resonance, while YIG B and C remain far detuned, serving as non-interacting spectators. When YIG A and D are brought into resonance at \unit[2.325]{GHz}, a distinct avoided crossing appears in the reflection spectrum, indicating coherent interaction between them.

The absence of additional splitting or interference confirms that YIG B and C do not participate in the interaction, thereby validating the system's capability to achieve selective coupling between arbitrary pairs of YIG spheres. This experimental setup aligns with the configuration employed in Fig.~2 of the main text, where a specific pair of YIG spheres is tuned to a designated frequency while all other spheres remain far detuned. Through this approach, an all-to-all coherent network featuring selective interactions among distant YIG pairs is successfully established.

This experiment underscores the platform's reconfigurability, as adjusting the magnetic bias of individual YIG spheres enables the selective activation of specific couplings between distant spin ensembles.

\section{Detailed data for time-domain results}

\begin{figure}
\includegraphics[width=\linewidth]{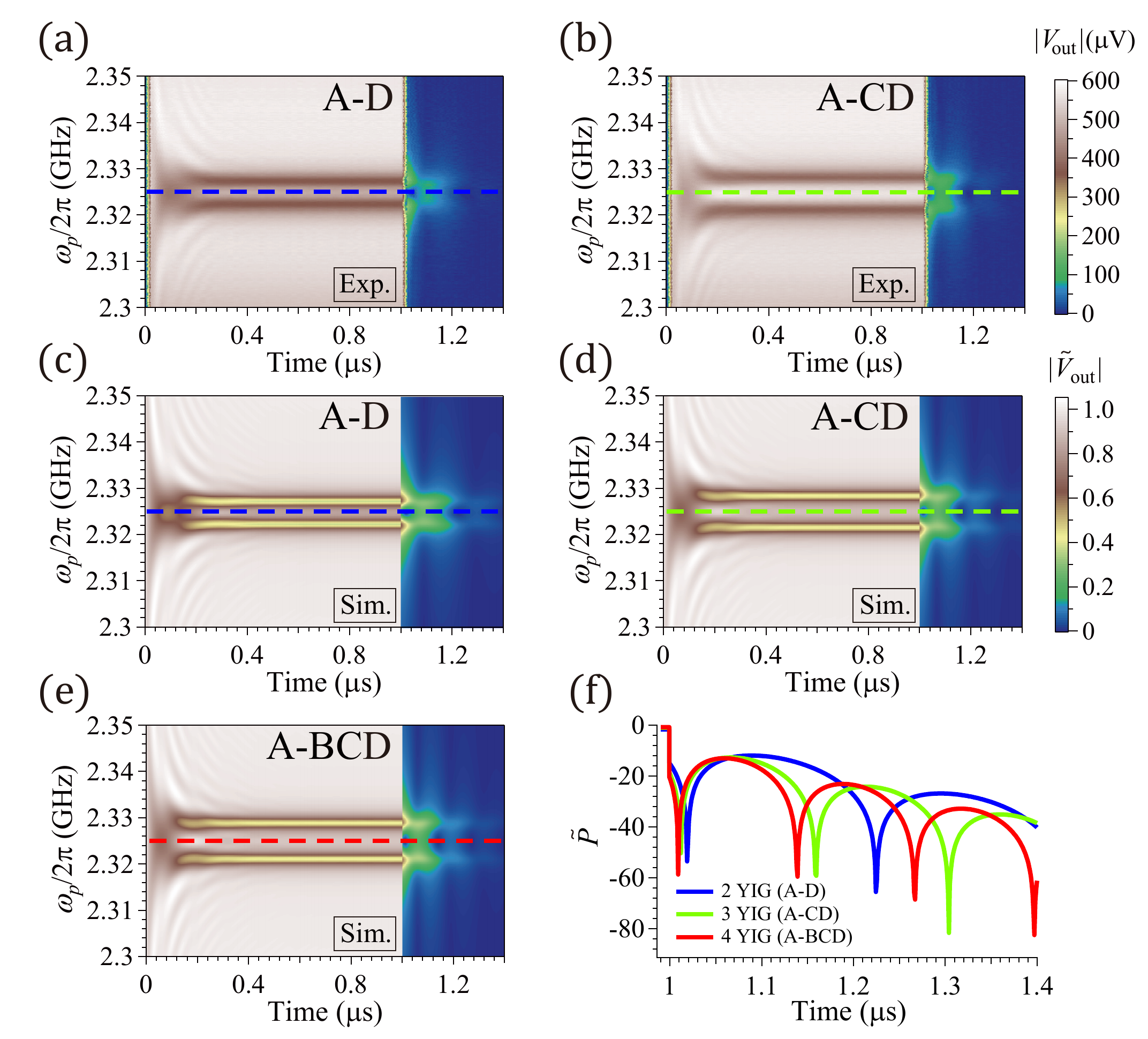}
\caption{Time-resolved energy oscillations for multiple YIG spheres. After applying a microwave square pulse into the system, the reflected signal $|V_{\rm out}|$ is recorded as a function of $\omega_p$ and evolution time.
(a) With two YIG spheres, A and D, tuned on resonance at \unit[2.325]{GHz}.
(b) With three YIG spheres, A, C, and D, tuned on resonance at \unit[2.325]{GHz}. The pulse is on from \unit[0]{$\mu$s} to \unit[1]{$\mu$s}, and off for the remaining durations. We can observe energy oscillations shortly after the pulse is activated and deactivated.
(c), (d) Theoretical simulations corresponding to the experiments in panels (a) and (b), respectively, based on \eqref{eq: Vout}.
(e) Theoretical simulation corresponding to Fig.~3(c) in the main text.
(f) Theoretical simulation corresponding to Fig.~3(b) in the main text. The time traces correspond to the line cuts of the data in panels (c)--(e), presented for configurations with two (blue), three (green), and four (red) YIG spheres, each tuned on resonance at $\omega_r / 2\pi = \unit[2.325]{GHz}$, following the termination of a long resonant microwave square pulse. To enhance visibility of the oscillations, $|\tilde{V}_{\rm out}|$ is converted to the normalized reflected power $\tilde{P}$ using \eqref{eq: reflect power}.}
\label{fig:S5}
\end{figure}


\subsection{Coherent energy oscillations}

In this section, we provide the detailed data corresponding to Fig.~3 in the main text. Using the same configuration as depicted in Fig.~3(a), we investigate the coherent interaction involving two to four YIG spheres to observe time-resolved energy oscillations. In \figpanel{fig:S5}{a}, two YIG spheres, A and D, are biased at \unit[2.325]{GHz}, and $|V_{\rm out}|$ is plotted as a function of probe frequency $\omega_p$ and evolution time, following the procedure described in the main text. In \figpanel{fig:S5}{b}, three YIG spheres, A, C, and D, are also biased at \unit[2.325]{GHz}, and $|V_{\rm out}|$ is presented. Based on \eqref{eq: Vout} and the fitting parameters summarized in \tabref{tab:S3}, the simulation results for interactions involving two to four YIG spheres are shown in \figpanels{fig:S5}{c}{e}. These simulation results are normalized, denoted as $|\tilde{V}_{\rm out}|$ and $\tilde{P}$, to eliminate the influence of non-ideal background signals from the overall measurement setup.

With the `joint spin ensemble' (YIG spheres at nodes) dynamically coupled to YIG A, we directly observe the system evolving into a steady state characterized by two split modes when the pulse is activated. Moreover, the relationship between coupling strength $\Delta$ and the number of YIG spheres influences the width of these split modes; a larger joint spin ensemble, comprising more YIG spheres at nodes, results in broader splitting of modes, as well as faster energy oscillation, as illustrated in \figpanel{fig:S5}{f}. 
Through time-domain measurements, we have thus successfully demonstrated coherent energy exchange.


\subsection{Dynamic coupling control using a detuning magnetic pulse}

\begin{figure}
\includegraphics[width=\linewidth]{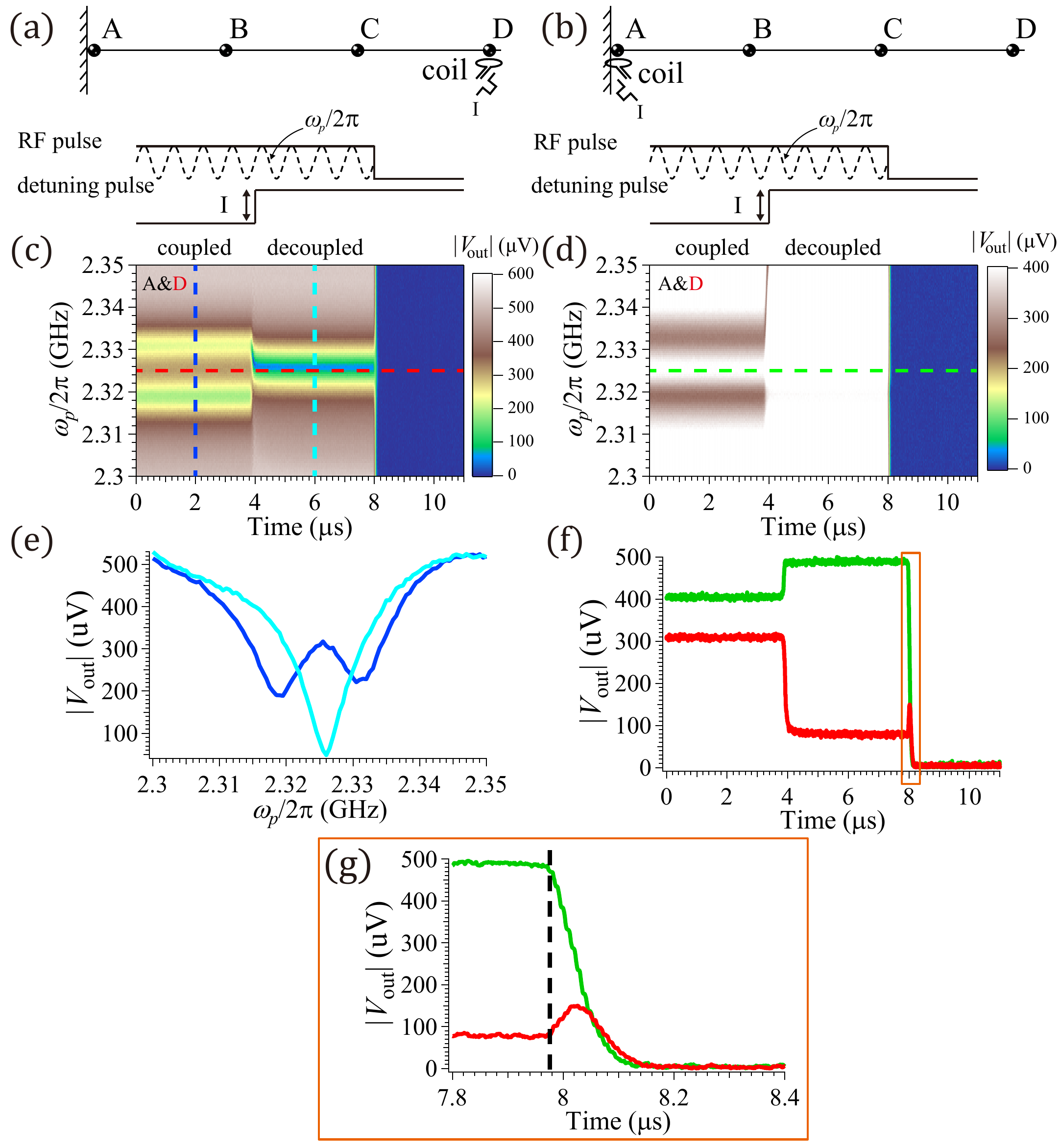}
\caption{Dynamic decoupling of YIG A and YIG D using a detuning magnetic pulse. 
(a), (b) Schematic of the experimental setup showing the RF excitation pulse with carrier frequency $\omega_p$ and the magnetic-field detuning pulse that detunes the YIG sphere via the coil. 
(c), (d) Time-resolved reflection signal $|{V}_{\rm out}|$ as a function of $\omega_p$ and evolution time when detuning (c) YIG D or (d) YIG A. The KM transition frequency $\omega_m$ of YIG D is fixed at the node (highlighted in red font).
(e) Vertical line cuts from (c) demonstrate the coupled (blue) and decoupled (cyan) states. The Lorentzian shape indicates the response of YIG A alone.
(f) Horizontal line cuts extracted from (c) and (d) illustrate the temporal transition from the coupled to the decoupled states. 
(g) Zoomed-in view corresponding to the orange box highlighted in (f), where the tail of the RF pulse (green) is a Gaussian square edge with $\sigma = \unit[50]{ns}$. The black dashed line marks the flat end of this Gaussian square RF pulse. An emission signal is observed from YIG~A (red) but not from YIG~D (green), since YIG~D is located at a node.
}
\label{fig:S6}
\end{figure}

To demonstrate real-time control of the magnon-magnon interaction, we apply a fast detuning magnetic pulse to dynamically switch the coupling between YIG A and YIG D, as shown in \figref{fig:S6}. The pulse sequence, illustrated in the lower part of \figpanels{fig:S6}{a}{b}, consists of a long-duration RF excitation pulse (from $0$ to $\unit[8]{\mu s}$) and a magnetic-field detuning pulse (from $\unit[4]{\mu s}$ to $\unit[12]{\mu s}$) that transiently detunes either YIG A [\figpanel{fig:S6}{b}] or YIG D [\figpanel{fig:S6}{a}] via the coil.

When two spheres resonate, the time-resolved reflection signal $|{V}_{\rm out}|$ exhibits two split modes, indicating coherent interaction between them, as shown in \figpanels{fig:S6}{c}{d} (from $0$ to $\unit[4]{\mu s}$). Upon activating the detuning pulse at $\unit[4]{\mu s}$, these split modes are rapidly suppressed, signaling the effective termination of the coherent coupling. The transition is distinctly observed through vertical line cuts extracted from \figpanel{fig:S6}{c} at $\unit[2]{\mu s}$ and $\unit[6]{\mu s}$, summarized in \figpanel{fig:S6}{e}, where the spectrum evolves from two distinct modes (blue) to a single mode (cyan). Additionally, the horizontal red and green line cuts, derived from \figpanels{fig:S6}{c}{d} and summarized in \figpanel{fig:S6}{f}, reveal the temporal transition from the coupled to the decoupled states. The decoupling time is $\sim\unit[200]{ns}$. A zoom-in view in \figpanel{fig:S6}{g} further demonstrates that, upon deactivating the RF pulse, an emission is observed from YIG A (red), whereas YIG D shows no emission (green), as it resides at a node along the CPW.

This experiment delivers a direct demonstration of rapid, on-demand decoupling of two distant YIG spheres. Such dynamic modulation of dipole-dipole interactions opens new avenues for developing time-dependent magnonic circuits and reconfigurable coherent networks based on waveguide magnonic systems.


\section{Comparison with other coupling schemes}

To underscore the differences between our mirror-induced standing-wave system and other representative platforms, we outline key distinctions in Table~\ref{tab:S4}. The mirror-terminated CPW generates standing-wave patterns along the waveguide, facilitating all-to-all connectivity among distant YIG spheres. This configuration provides two primary advantages over traditional cavity-based and open-waveguide systems: frequency selectivity and positional flexibility.

Unlike conventional cavity magnonic systems (e.g., Ref.~\cite{Zhang2015NC}), our scheme does not require all spin ensembles to couple to a single, fixed cavity resonance. Instead, any chosen pair of YIG spheres can be individually tuned into resonance at a designated frequency, while the remaining spheres remain far detuned and effectively decoupled. As shown in Fig.~2(b)--(g) of the main text and further validated in \figref{fig:S4}, each data set corresponds to a specific YIG pair (e.g., A\&B, A\&C, etc.) exhibiting a clear avoided crossing, with all other spheres acting as non-interacting spectators. This capability enables precise, frequency-selective control over pairwise magnonic interactions.

Moreover, YIG spheres in our mirror-terminated CPW can, in principle, be put at any position along the waveguide, with their interaction strength determined by their relative positions with respect to the standing-wave nodes and antinodes, as described by Eq.~(1) in the main text. This level of spatial tunability starkly contrasts with conventional cavity systems, where strong coupling occurs only near the fixed antinodes of the cavity field. 

The ability to independently control both frequency selectivity and spatial placement, resulting in controllable interference-enhanced or suppressed dipolar coupling as described by Eq.~(1) in the main text, provides a versatile and scalable approach for engineering reconfigurable magnonic networks.

\begin{table}[]
\begin{tabular}{|l|l|l|l|l|}
\hline
& \begin{tabular}[c]{@{}l@{}}Mirror-terminated \\ CPW \end{tabular} & Cavity & Open CPW & \begin{tabular}[c]{@{}l@{}}Dipolar coupled \\ spins in close \\ proximity\end{tabular}   \\ \hline
Field modes  & \begin{tabular}[c]{@{}l@{}}Continuum modes \\ exhibiting \\ mirror-induced \\ standing-wave \\ patterns at specific \\ frequencies, with \\ microwave nodes \\ present\end{tabular}                                       & \begin{tabular}[c]{@{}l@{}}Single-mode \\ standing wave, \\ with microwave \\ nodes present\end{tabular}                      & \begin{tabular}[c]{@{}l@{}}Continuum\\ modes, without \\ microwave nodes\end{tabular}                                    & \begin{tabular}[c]{@{}l@{}}Continuum \\ modes in three\\ dimensions,\\  without \\ microwave nodes\end{tabular}                                                                                  \\ \hline
\begin{tabular}[c]{@{}l@{}}Coupling objects\end{tabular}         & \begin{tabular}[c]{@{}l@{}}YIG \\ spheres\end{tabular}                                                                                                                                                            & \begin{tabular}[c]{@{}l@{}}Hybridization \\ modes (magnon \\ modes hybridized \\ with the \\ cavity mode)\end{tabular} & \begin{tabular}[c]{@{}l@{}}Two \\ atoms\end{tabular}                                                             & \begin{tabular}[c]{@{}l@{}}Dipoles\end{tabular}                                                                                              \\ \hline
\begin{tabular}[c]{@{}l@{}}Coupling mediator\end{tabular}        & \begin{tabular}[c]{@{}l@{}}Traveling photons in \\ the CPW (reflected \\ by the mirror)\end{tabular}                                                                                                                       & Cavity photons                                                                                                                 & \begin{tabular}[c]{@{}l@{}}Traveling photons \\ in the CPW\end{tabular}                                                   & \begin{tabular}[c]{@{}l@{}}Direct \\ dipole-dipole \\ exchange\end{tabular}                                                                                             \\ \hline
\begin{tabular}[c]{@{}l@{}}Coupling \\ characteristics\end{tabular} & \begin{tabular}[c]{@{}l@{}} We demonstrated: \\ \ding{172} Selective mutual \\ coupling between \\ arbitrary pairs in the \\ frequency domain \\ (Fig.~2, \figref{fig:S4});\\\ding{173} Simultaneous \\ coupling of multiple \\ spheres in the time\\ domain (Fig.~3);\\\ding{174} Dynamic on-\\ demand decoupling\\ between two YIG\\ spheres within \unit[200]{ns}\\ (\figref{fig:S6}).\end{tabular} & \begin{tabular}[c]{@{}l@{}}All spheres (e.g.,\\  8 in Ref.~\cite{Zhang2015NC}) \\ are coupled \\ collectively via \\ the cavity mode.\end{tabular} & \begin{tabular}[c]{@{}l@{}}The coupling \\ is weak in the \\ absence of \\ mirror-induced \\ standing waves.\end{tabular} & \begin{tabular}[c]{@{}l@{}}The coupling is \\ effective only at \\ very short \\ distances, with \\ coupling strength \\ diminishing \\ rapidly as \\ separation \\ increases.\end{tabular} \\ \hline
References                                                          & This work                                                                                                                                                                                                                 & \begin{tabular}[c]{@{}l@{}}Ref.~\cite{Zhang2015NC}\end{tabular}                                                         & \begin{tabular}[c]{@{}l@{}}Ref.~\cite{LalumiPRA2013}\end{tabular}                                                & \begin{tabular}[c]{@{}l@{}}Ref.~\cite{Lehmberg1970}\end{tabular}                                                                                                   \\ \hline
\end{tabular}
\caption{
Comparison of our mirror-terminated system with other representative coupling schemes. 
}
\label{tab:S4}
\end{table}

\section{Downscaling and scalability of our system}

The scalability of our mirror-terminated CPW platform is a key consideration for realizing dense, all-to-all magnonic networks in future quantum and hybrid information systems. In the present experiment, the YIG spheres have diameters of \unit[1.2]{mm}, typical for current magnonics research (\unit[0.25--2]{mm}). This size primarily reflects fabrication constraints, as conventional polishing limits the production of smaller spheres with high crystalline quality and low damping~\cite{zhang2016magnon}. Sub-millimeter spheres are, in principle, feasible and would enable compact, chip-scale integration; however, the magnon-photon coupling strength decreases with the square root of the sphere volume ($g\propto\sqrt{V}$)~\cite{Wu2024}, making strong coupling more difficult to maintain. Consequently, size reduction weakens the photon-mediated coherent coupling between YIG spheres, according to~\eqref{eq: collective energy shift}, and may compromise the system's performance in achieving long-range interactions.

The reduced coupling of smaller YIG spheres can be mitigated through several approaches. Positioning YIG spheres closer to the CPW surface enhances the local microwave magnetic field and increases field overlap. Adopting loop-type transmission lines that confine the field around the sphere, as demonstrated in Ref.~\cite{Li2022}, can further strengthen the magnon-photon coupling. In addition, advances in YIG fabrication techniques, such as chemical etching or laser-assisted polishing, may enable the realization of smaller spheres while maintaining high crystalline quality and coherence~\cite{zhang2016magnon}.

The microwave wavelength ($\lambda=v/f$) in the CPW also constrains the device size. This constraint can be mitigated by operating at higher frequencies or employing substrates with larger relative dielectric constants $\epsilon_{r}$ to reduce the phase velocity $v$. Compact waveguide geometries, including meandering or folded CPW~\cite{Wang2022}, can further minimize the footprint without degrading the coupling.

Alternative approaches for coupling spin ensembles on more compact scales have also been explored in hybrid magnonic systems. For instance, as shown in Ref.~\cite{Li2022}, coherent coupling between two remote YIG spheres can be achieved in a superconducting circuit by integrating them with resonators in the dispersive regime, where the resonator mode is detuned from the magnon frequencies. This configuration yields coherent interaction with coupling strengths of \unit[10--40]{MHz} over a \unit[7]{mm} separation via cavity-mediated interactions. Additionally, the same paper demonstrates that two YIG spheres mounted in circular CPW loops can enable coherent coupling mediated by propagating microwave photons, achieving coupling strengths up to \unit[8]{MHz} over a \unit[7.2]{mm} separation in certain frequency regimes, where the waveguide phase difference influences the coupling dynamics. Furthermore, as shown in Ref.~\cite{Jun2023}, microwave-mediated magnon-atom interactions couple two YIG spheres using a superconducting artificial atom in a hybrid ferromagnet-superconductor system, establishing coherent coupling via virtual photon exchange in microwave cavities. These schemes require operation in a millikelvin environment, while superconducting circuits commonly encounter fabrication challenges, which can be circumvented in our device. 


\bibliography{SI}